\documentclass[fleqn,usenatbib]{mnras}

\usepackage{newtxtext,newtxmath}

\usepackage[T2A]{fontenc}
\usepackage{ae,aecompl}

\usepackage[utf8]{inputenc}
\usepackage[english]{babel}

\hypersetup{draft}
\usepackage{color}
\usepackage{xcolor}
\usepackage{soul}
\sethlcolor{yellow}
\newcommand{\corr}[1]{#1}

\usepackage{graphicx}   
\usepackage{amsmath}    
\usepackage{amssymb}    
\binoppenalty=\maxdimen
\relpenalty=\maxdimen

\renewcommand{\C}{\text{c}}
\renewcommand{\it}{\textit}
\renewcommand{\vec}{\mathbfit}

\title[Core shift variability in AGN jets]{Significant core shift variability in parsec-scale jets of active galactic nuclei}

\author[Plavin et al.]{
A.~V.~Plavin,$^{1,2}$\thanks{E-mail: alexander@plav.in}
Y.~Y.~Kovalev,$^{1,2,3}$
A.~B.~Pushkarev$^{4,1}$
and A.~P.~Lobanov$^{3,5}$
\\
$^1$Astro Space Center of Lebedev Physical Institute, Profsoyuznaya 84/32, 117997 Moscow, Russia\\
$^2$Moscow Institute of Physics and Technology, Institutsky per. 9, Dolgoprudny 141700, Russia\\
$^3$Max-Planck-Institut f\"ur Radioastronomie, Auf dem H\"ugel 69, 53121 Bonn, Germany\\
$^4$Crimean Astrophysical Observatory, Nauchny 298409, Crimea, Russia\\
$^5$Institut f\"ur Experimentalphysik, Universit\"at Hamburg, Luruper Chaussee 149, 22761 Hamburg, Germany
}

\date{Accepted XXX. Received YYY; in original form 06.11.2018}

\pubyear{2018}

\begin{document}
\label{firstpage}
\pagerange{\pageref{firstpage}--\pageref{lastpage}}
\maketitle

\begin{abstract}
The apparent position of jet base (core) in radio-loud active galactic nuclei changes with frequency because of synchrotron self-absorption. Studying this `core shift` effect enables us to reconstruct properties of the jet regions close to the central engine.
We report here results from core shift measurements in AGNs observed with global VLBI at 2 and 8 GHz at epochs from 1994 to 2016. Our sample contains 40 objects observed at least 10 times during that period. The core shift is determined using a new automatic procedure introduced to minimize possible biases. The resulting multiple epoch measurements of the core position are employed for examining temporal variability of the core shift.
We argue that the core shift variability is a common phenomenon, as established for 33 of 40 AGNs we study. Our analysis shows that the typical offsets between the core positions at 2 and 8 GHz are about 0.5 mas and they vary in time. Typical variability of the individual core positions is about 0.3 mas. The measurements show a strong dependence between the core position and its flux density, suggesting that changes in both are likely related to the nuclear flares injecting denser plasma into the flow. We determine that density of emitting relativistic particles significantly increases during these flares, while relative magnetic field changes less and in the opposite direction.
\end{abstract}



\section{Introduction}

In images of relativistic AGN jets, the `core' is typically identified with the most compact component closest to the jet base.
It is usually the region with optical depth $\tau_\nu \approx 1$ (photosphere), which position depends on observing frequency $\nu$ \citep{1979ApJ...232...34B}.
Observations confirm that the core is located farther from the jet base and has larger apparent size at lower frequencies \citep[e.g.][]{2011A&A...532A..38S, 2012A&A...545A.113P}.

The most likely mechanisms causing the observable effects in compact jets are synchrotron emission and synchrotron self-absorption \citep{1981ApJ...243..700K}.
Under the assumption that the jet is freely expanding and there is an equipartition between the particle kinetic energy and magnetic field energy, the distance $r_\mathrm{c}(\nu)$ of the core from the jet base is expected to scale as $r_\mathrm{c}(\nu) \propto \nu^{-1}$.
While it is debatable whether these assumptions always hold, it has been shown \citep[e.g.][]{2011A&A...532A..38S} that the core distance follows $r_\mathrm{c}(\nu) \propto \nu^{-1}$ in many cases.

Investigations of the core shift effect and its variability are important both for astrometric \citep[cf.][]{2008A&A...483..759K} and astrophysical \citep[cf.][]{1998A&A...330...79L,2000ApJ...545..100H,2005ApJ...619...73H} applications.
This effect impacts precise radio Very Long Baseline Interferometry (VLBI) astrometric measurements as discussed by \citet{2005astro.ph..5475R} and \citet{2009A&A...505L...1P}.
It also introduces systematic offsets when aligning radio and optical reference frames \citep{2008A&A...483..759K, 2017A&A...598L...1K,r:gaia5}.
Thus, unaccounted core shift variability causes these offsets to change over time, further complicating long-term precise astrometry and reference frames alignment.
As for astrophysics, measuring how core position changes during flares and between them may provide more insight on origin of outbursts and their propagation along the jet.

For sources experiencing flares it has been shown \citep{1998A&A...330...79L,2008A&A...483..759K} that if the only changing parameter in the compact jet is emitting particles density, then the core shift magnitude is closely related to the core flux density: $r_\C \propto S_\C^{2/3}$.
\cite{1999ApJ...521..509L} showed that during nuclear flares jet speed changes only weakly, so it can be expected that this proportionality between core shift and flux density usually holds.
Some observational results \citep[e.g.][]{2008A&A...483..759K} also suggest that there is such a dependence, but they had a limited number of sources to make any robust conclusion.
Later, \citet{2011A&A...532A..38S} found no significant relation between core shift and flux density of the core region. \corr{Recent studies of individual sources also connect changes in the VLBI core position to X-ray and $\gamma$-ray flares \citep{2015ApJ...807L..14N,2017MNRAS.468.4478L}.}
Clearly, having a large enough dataset of core shift measurements, covering multiple flaring events in sources experiencing nuclear flares, is crucial for \corr{evaluating and studying the possible VLBI core flux-shift} relationship.

Earlier measurements of the core shift effect \citep[e.g.][]{2008A&A...483..759K,2009MNRAS.400...26O,2011A&A...532A..38S,2012A&A...545A.113P,2016MNRAS.462.2747K,2018arXiv180806138P} contained no more than a few epochs per source (with most objects measured only at a single epoch), and hence could not study effectively how the core shift changes with time.
This paper presents for the first time multi-epoch core shift measurements for 40 AGNs, with each of the targets having from 10 to 70 observing epochs, and apply these data to investigate in detail the core shift variability.
In \autoref{s:obsdata}, we describe VLBI data used for our analysis, followed by \autoref{s:measuring} presenting methods used for measuring core shifts.
Then, in \autoref{s:results}, we present results of these measurements and in \autoref{s:discussion} discuss their implications for astrophysical studies of AGN jets and astrometric measurements made with VLBI observations of compact extragalactic radio sources.
We summarize our results in \autoref{s:summary}.

Throughout the paper, we use the spectral index $\alpha$ defined as $S\propto\nu^\alpha$ and adopt the $\Lambda$CDM cosmology with $\Omega_m=0.287$, $\Omega_\Lambda=0.7185$ and $H_0=69.3$~km~s$^{-1}$~Mpc$^{-1}$ \citep{2013ApJS..208...19H}. All position angles are given in degrees east of north.

\section{Observational data}
\label{s:obsdata}

For our analysis we used simultaneous \it{S} and \it{X}-band (2 and 8~GHz) VLBI observations compiled in the \it{Astrogeo}\footnote{\url{ http://astrogeo.org/vlbi_images/}} \corr{VLBI FITS image} database, comprising the visibility and imaging data acquired from geodetic VLBI observations \citep{2009JGeod..83..859P,2012A&A...544A..34P,2012ApJ...758...84P} and the VLBA\footnote{Very Long Baseline Array of the National Radio Astronomy Observatory, Socorro, NM, USA} calibrator surveys (VCS; \citealt{2002ApJS..141...13B,2003AJ....126.2562F,2005AJ....129.1163P,2006AJ....131.1872P,2007AJ....133.1236K,2008AJ....136..580P}). \corr{For details on amplitude calibration of geodetic observations see \citet{2012A&A...544A..34P}.} These programs suit the core shift measurement problem very well, because the two observing frequencies are well-separated and low enough, whereas at higher frequencies the magnitude of core shift could be less pronounced. In total, the dataset contains 4143 sources observed in the time period from 1994 to 2016. We selected 40 sources which show extended structure warranting core shift measurements and have been observed at least at ten epochs at both frequencies. The cadence of these observations is highly nonuniform and typically ranges from a month to a few years. Basic information about these selected sources is given in \autoref{t:basicinfo}.

In order to provide a reference direction for the core shift measurements, the overall jet position angle was determined for each of the target objects, using the reconstructed 8~GHz images. For this, we took the average angle of significant emission (over 5 times the image noise) located farther than 3 beam sizes from the center. If more than one image was available for a given source, the position angle was averaged over all epochs. The resulting jet position angles are listed in column 6 of \autoref{t:basicinfo}.

The \it{Astrogeo} database provides calibrated visibility data (Fourier-space measurements) and hybrid images reconstructed from these data. The images \corr{collected in the database are not directly suited} for our purposes, as image alignment and core shift measurements require images restored with the same beam and pixel size at both frequencies. We therefore re-imaged the visibility data using the CLEAN algorithm \citep{1974A&AS...15..417H} as implemented in \textsc{difmap} \citep{1994BAAS...26..987S}, and convolved the resulting images with the average beam size (calculated individually for each given object) between 2 and 8~GHz. We applied this approach to all of the targets studied and produced a homogeneous set of images with dimensions of $2048 \times 2048$~pixels and the pixel size of $0.05$~mas. Potential systematic effects owing to different ($u$,$v$)-plane coverages at different frequencies \citep[e.g.][]{2014AJ....147..143H} should not affect our measurements, as they become significant at signal-to-noise ratios (SNR) of $\lesssim 7$ \citep{1998A&AS..132..261L} which is significantly lower the SNR levels of the jet regions we \corr{use} in our core shift measurements.

\begin{table*}
\caption[]{Source properties. Columns are as follows:
(1)~--~J2000 name;
(2)~--~B1950 name;
(3), (4)~--~J2000 coordinates from the Radio Fundamental Catalogue (RFC, \url{http://astrogeo.org/rfc}) version 2017c;
(5)~--~redshift from \cite{2009AJ....137.3718L};
(6)~--`jet positional angle estimated as described in \autoref{s:obsdata};
(7)~--~number of observational epochs used in our analysis.
}
\label{t:basicinfo}
\begin{centering}
\begin{tabular}{llllllc}
\hline
\multicolumn{2}{c}{Source Name} & \multicolumn{1}{c}{R.A. (J2000)} & \multicolumn{1}{c}{DEC. (J2000)} & $z$ &  P.A. & $N_\mathrm{obs}$ \\
\multicolumn{1}{c}{J2000} & \multicolumn{1}{c}{B1950} &  hh:mm:ss.ssss & \phantom{+}dd:mm:ss.sss &  & [$^\circ$] & \\
\multicolumn{1}{c}{(1)} & \multicolumn{1}{c}{(2)} & \multicolumn{1}{c}{(3)} & \multicolumn{1}{c}{(4)} & (5) & (6) & (7) \\
\hline
J0006$-$0623 & 0003$-$066 & 00:06:13.8929 & $-$06:23:35.3353 & 0.347 & 280.8 & 51 \\
J0017$+$8135 & 0014$+$813 & 00:17:08.4749 & $+$81:35:08.1365 & 3.384 & 190.7 & 71 \\
J0102$+$5824 & 0059$+$581 & 01:02:45.7624 & $+$58:24:11.1367 & 0.644 & 247.7 & 78 \\
J0136$+$4751 & 0133$+$476 & 01:36:58.5948 & $+$47:51:29.1001 & 0.859 & 327.9 & 69 \\
J0204$+$1514 & 0202$+$149 & 02:04:50.4139 & $+$15:14:11.0437 & 0.405 & 305.1 & 45 \\
J0237$+$2848 & 0234$+$285 & 02:37:52.4057 & $+$28:48:08.9901 & 1.206 & 350.4 & 49 \\
J0339$-$0146 & 0336$-$019 & 03:39:30.9378 & $-$01:46:35.8041 & 0.852 & 063.3 & 61 \\
J0501$-$0159 & 0458$-$020 & 05:01:12.8099 & $-$01:59:14.2564 & 2.286 & 304.3 & 43 \\
J0530$+$1331 & 0528$+$134 & 05:30:56.4167 & $+$13:31:55.1495 & 2.070 & 029.5 & 63 \\
J0609$-$1542 & 0607$-$157 & 06:09:40.9495 & $-$15:42:40.6727 & 0.323 & 054.1 & 17 \\
J0808$+$4052 & 0805$+$410 & 08:08:56.6520 & $+$40:52:44.8889 & 1.419 & 051.7 & 22 \\
J0808$+$4950 & 0804$+$499 & 08:08:39.6663 & $+$49:50:36.5304 & 1.436 & 138.5 & 59 \\
J0825$+$0309 & 0823$+$033 & 08:25:50.3384 & $+$03:09:24.5200 & 0.505 & 025.1 & 52 \\
J0854$+$2006 & 0851$+$202 & 08:54:48.8749 & $+$20:06:30.6408 & 0.306 & 247.4 & 74 \\
J0927$+$3902 & 0923$+$392 & 09:27:03.0139 & $+$39:02:20.8518 & 0.695 & 282.0 & 61 \\
J1007$+$1356 & 1004$+$141 & 10:07:41.4981 & $+$13:56:29.6008 & 2.715 & 131.1 & 15 \\
J1058$+$0133 & 1055$+$018 & 10:58:29.6052 & $+$01:33:58.8237 & 0.888 & 305.1 & 13 \\
J1104$+$3812 & 1101$+$384 & 11:04:27.3139 & $+$38:12:31.7990 & 0.031 & 319.9 & 48 \\
J1147$-$0724 & 1145$-$071 & 11:47:51.5540 & $-$07:24:41.1412 & 1.342 & 292.7 & 40 \\
J1159$+$2914 & 1156$+$295 & 11:59:31.8339 & $+$29:14:43.8268 & 0.725 & 024.7 & 56 \\
J1230$+$1223 & 1228$+$126 & 12:30:49.4234 & $+$12:23:28.0437 & 0.004 & 285.4 & 53 \\
J1310$+$3220 & 1308$+$326 & 13:10:28.6639 & $+$32:20:43.7828 & 0.997 & 287.5 & 59 \\
J1316$-$3338 & 1313$-$333 & 13:16:07.9859 & $-$33:38:59.1727 & 1.210 & 276.9 & 25 \\
J1419$+$5423 & 1418$+$546 & 14:19:46.5974 & $+$54:23:14.7871 & 0.153 & 127.9 & 31 \\
J1517$-$2422 & 1514$-$241 & 15:17:41.8131 & $-$24:22:19.4761 & 0.049 & 156.0 & 31 \\
J1608$+$1029 & 1606$+$106 & 16:08:46.2032 & $+$10:29:07.7757 & 1.232 & 314.2 & 53 \\
J1632$+$8232 & 1637$+$826 & 16:32:31.9699 & $+$82:32:16.3999 & 0.024 & 297.4 & 13 \\
J1638$+$5720 & 1637$+$574 & 16:38:13.4563 & $+$57:20:23.9790 & 0.751 & 203.4 & 11 \\
J1642$+$6856 & 1642$+$690 & 16:42:07.8485 & $+$68:56:39.7564 & 0.751 & 194.4 & 35 \\
J1727$+$4530 & 1726$+$455 & 17:27:27.6508 & $+$45:30:39.7313 & 0.717 & 262.0 & 29 \\
J1800$+$7828 & 1803$+$784 & 18:00:45.6839 & $+$78:28:04.0184 & 0.680 & 265.5 & 70 \\
J1911$-$2006 & 1908$-$201 & 19:11:09.6529 & $-$20:06:55.1091 & 1.119 & 048.0 & 40 \\
J2038$+$5119 & 2037$+$511 & 20:38:37.0347 & $+$51:19:12.6626 & 1.686 & 219.6 & 20 \\
J2115$+$2933 & 2113$+$293 & 21:15:29.4135 & $+$29:33:38.3670 & 1.514 & 183.4 & 15 \\
J2148$+$0657 & 2145$+$067 & 21:48:05.4587 & $+$06:57:38.6042 & 0.999 & 137.0 & 51 \\
J2202$+$4216 & 2200$+$420 & 22:02:43.2914 & $+$42:16:39.9800 & 0.069 & 189.9 & 53 \\
J2203$+$3145 & 2201$+$315 & 22:03:14.9758 & $+$31:45:38.2700 & 0.295 & 220.7 & 16 \\
J2225$-$0457 & 2223$-$052 & 22:25:47.2593 & $-$04:57:01.3907 & 1.404 & 103.3 & 29 \\
J2246$-$1206 & 2243$-$123 & 22:46:18.2320 & $-$12:06:51.2776 & 0.632 & 027.5 & 43 \\
J2258$-$2758 & 2255$-$282 & 22:58:05.9629 & $-$27:58:21.2568 & 0.927 & 226.9 & 27 \\
\hline
\end{tabular}
\end{centering}
\end{table*}

\section{Core shift measurements}
\label{s:measuring}

\subsection{Methodology of core shift measurements}

There are several methods and approaches developed for measuring core shifts in radio jets. All these methods must address the primary issues of identifying the core region at each individual observing frequency and cross-referencing the positions of the core registered at different frequencies.

Assuming that the core location is determined by the opacity $\tau$ in the emitting jet plasma, it is typically argued that the $\tau=1$ surface representing the core is the most compact and bright feature observed at the narrow end of the jet, close to the true jet origin. The robustness of such an identification may nevertheless be affected by potential blending of the core emission with the emission downstream the jet \citep{2012A&A...545A.113P} or by confusing the true, partially opaque core with a stationary shock downstream of it \citep{2016ApJ...817...96G}.
A stationary recollimation shock confused with the core could be recognized by its optically thin spectrum, lack of observable frequency dependence of its position, and recurrent appearances of a weaker feature upstream from it, resulting from flaring activity of the source.

Cross-referencing of the core positions registered at different frequencies (or {\em image alignment}) is, in principle, an astrometric task. In the absence of absolute or relative astrometric measurements, this task may become non-trivial, because absolute position information is lost during self-calibration part of imaging process \citep{1958MNRAS.118..276J}.

\subsubsection{Identification of the parsec-scale core region}

Approaches suitable for identifying the core region in jets depend on the complexity of the source structure and the number of sources included in a study with large samples requiring automated and unsupervised methods for this process.

In the simplest approach, the brightness peak in an image can be used as a proxy for the core. However, the core may not necessarily be the brightest region in the jet \citep[see, e.g., 0923+392 spectral index map in][]{2014AJ....147..143H} and even when it is the brightest region in the jet, the position of the peak of brightness can be influenced by the blending effect.

It is common to model the jet structure with Gaussian patterns ({\em Gaussian components}) and identify the component nearest to the jet origin~--- in most cases the brightest one~--- as the core \citep[e.g.][]{2008A&A...483..759K}. This method provides relatively accurate core identification, but is difficult to be implemented in an automated, unsupervised routine. In a more simplified automated approach \citep[e.g.][]{2005AJ....130.2473K,2015MNRAS.452.4274P}, the core was modeled with one Gaussian component and the rest of the jet with another. This approach yields reasonable results for sources in which the structure is dominated by the core region and the core itself is not blended significantly with the emission downstream.

\subsubsection{Image alignment}

Once the core region has been identified in images at different frequencies, the respective images must be cross-referenced (or {\em aligned}) in order to enable measurements of relative offsets of the core position.

One way to achieve this alignment is to use phase-referencing to a nearby calibrator source \citep{1984ApJ...276...56M,1995ASPC...82..363F,voitsik2018} resolving the core shift for targets and calibrators. The position of a target source can then be robustly located relative to the calibrator position. This method provides image alignment at different frequencies, but cannot be applied to the data used in this paper.

In the absence of astrometric information, image alignment can be made using some parts of the observed structure as a reference.
If the jet has clearly separated emitting regions (or {\em jet components}) in its optically thin part, these components can be assumed to be at the same location at both frequencies. Then they can explicitly be used as reference points and the core location can be determined relative to them \citep{2008A&A...483..759K,2011A&A...532A..38S}. However, this requires a detailed model of the source structure (usually in a form of several Gaussians) and selection of corresponding components for each observation. This is done manually, which leads to possible biases and is difficult to automate.

For objects with smooth structure which cannot be uniquely decomposed into Gaussian components, masked 2D cross-correlation was applied for the purpose of image alignment \citep{2000ApJ...530..233W}. The algorithm was also discussed by \cite{2008MNRAS.386..619C} and applied by \cite{2009MNRAS.400...26O,2014AJ....147..143H}. In this approach, a rectangular mask is manually selected to cover an optically thin jet region in one image, and the corresponding region of the second image is registered by finding the maximum value of cross-correlation. This implicitly assumes that the reference region is optically thin and lacks significant gradients of spectral index. Despite the necessity to select the reference region manually, this method is considered as one of the most accurate \citep{2012A&A...545A.113P} for object with pronounced extended emission.

Comparison of the alignments obtained from the modelfit based and the 2D cross-correlation methods shows that both methods yield consistent results \citep{2011A&A...532A..38S, 2012A&A...545A.113P}.

\subsection{Automated procedure for core shift measurements}

In this paper, we aim at providing a robust, automated, and, whenever possible, unsupervised procedure for locating the core region in VLBI images of compact jets and cross-referencing its position between individual multi-frequency images of the same object.

\subsubsection{Determining the core position form structure-subtracted data}
\label{s:method_corepos}

For the purpose of determining the core location, a detailed model of the entire source structure is not required. For this purpose, we adopt a method, in which the core region is represented by a single Gaussian component, after subtracting the extended source structure~(Homan et al., in prep). The method requires access to visibility data and a CLEAN component model obtained during hybrid imaging of the source structure. The algorithm for locating the core region includes the following steps:
\begin{enumerate}
\item Estimate approximate core position in a very simple way, e.g. using the brightest image pixel.
\item Subtract from the interferometric visibilities the contribution of CLEAN components located farther than some threshold distance from this position.
\item Fit a single circular Gaussian component to the residual visibilities.
\end{enumerate}

The threshold distance is a free parameter to be chosen. To assess how stable the results are with respect to this choice, we evaluated the described process for different threshold values ranging from $0.5$ to $1.5$ times the beam size. For 8~GHz data, the resulting Gaussian core component fitted to the visibility data after the extended structure subtraction is very close to the phase center, with offsets $\le 0.04$~mas in 90\% of cases. This corresponds to our expectations for the optically thin jet emission to decrease with frequency and to affect the core position estimation less. In the 2~GHz band the core component position is often significantly offset from zero, but also insensitive to the threshold value. Standard deviation of its position for the specified range of thresholds is less than $0.06$~mas for 90\% of observations. For further analysis we keep core component position estimates obtained using all thresholds from the range $0.5$ to $1.5$ beam sizes and use them to represent the measurement uncertainty.

\subsubsection{Aligning images using masked cross-correlation}
\label{s:align-ccorr}

\begin{figure}
    \centering
    \includegraphics[width=\columnwidth,trim=0cm 0.5cm 0cm 0cm]{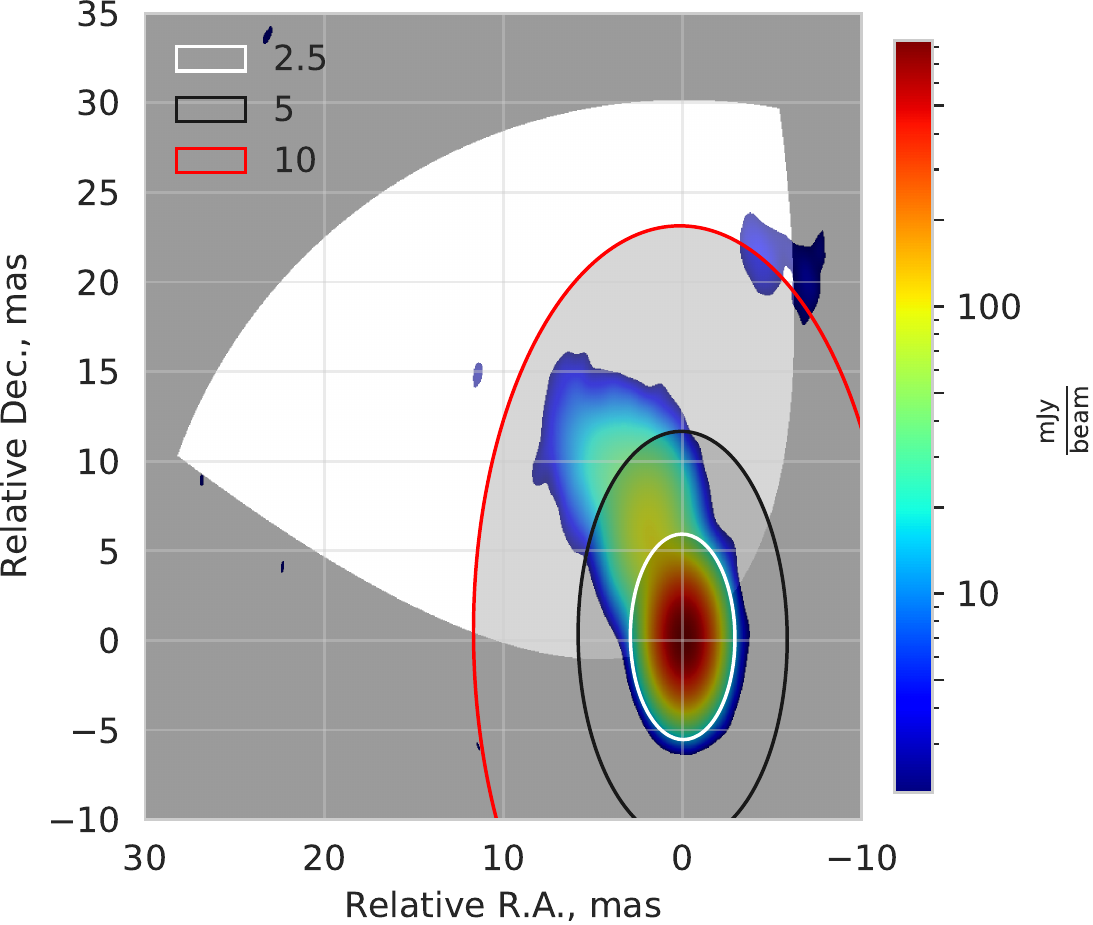}
    \caption{Example of cross-correlation mask for J1118+1234, with a parabolic mask and different elliptical exclusion regions for the core, corresponding to $m=2.5,\,5,\,10$ (shown with different colours). Regions which are ignored when using the corresponding mask are shaded.}
    \label{f:ccs_mask_sample}
\end{figure}

In order to automatically align multi-frequency images of the same object, we apply cross-correlation of optically thin {\em reference} regions detected in its structure. We also use a mask to identify the reference regions, but our implementation does not require the mask to be rectangular. For the mask, we use a parabolic shape, $|y| < a \sqrt{x}$ where $x$ is the distance from the core along the jet direction, and $y$ is the coordinate transverse to it. The value of $a$ was set to $\sqrt{20 \text{mas}}$ so that the parabola covers even the widest jets. The core emission is excluded from the cross-correlation by applying a beam-shaped elliptical exclusion region centered at the core position. The extend of the excluded core region is regulated by a parameter $m$, describing the size of the ellipse in the units of the restoring beam size. Application of the masking procedure is illustrated in \autoref{f:ccs_mask_sample}. We apply the mask to 8 GHz images only which gives equivalent results to masking images at both frequencies. Anything outside the mask is completely ignored, so as long as the mask applied to 2 GHz image contains the resulting match region, it would not change the highest-correlation alignment.

The parameter $m$ was varied between 2.5 and 10 in steps of 0.1. Not all sources have significant emission detected far from the core, and for large $m$ the mask may contain mostly noise. To filter such cases we set an empirical threshold and limit the maximum $m$ by requiring that the resulting mask area with brightness above 3 times the noise level is more than 4 times the area of the restoring beam. We keep image alignment estimates corresponding to all $m$ values to represent the measurement uncertainty.

There are pairs of images for which the masked cross-correlation gives obviously incorrect results, which can be inferred from strong gradients and non-physical values of spectral index in the spectral index maps obtained using this alignment procedure. Such cases have to be dealt with manually, to be either excluded from further consideration or re-analyzed using a different mask.

\subsubsection{Aligning images based on spectral index properties}
\label{s:method_align}

In an attempt to overcome the difficulties of the masked cross-correlation procedure and to reduce the amount of manual interventions required, we have devised a new alignment method based on machine learning, with a classifier algorithm trained on a relatively small amount of manually prepared data. The idea of the method is based on the observation that it is often very easy to choose the correct alignment of images from visual inspection of spectral index maps. In this case, the optimal alignment decision is based on a priori expected properties on a spectral index map of a compact jet: the spectral index distribution should be mostly symmetric perpendicular to the jet direction and approaching a uniform distribution in the optically thin jet regions, far from the core.

Hence we have compiled the following classification features for the classifier algorithm:
\begin{enumerate}
\item Symmetry perpendicular to the jet direction. This condition is implemented in form of average pairwise difference of spectral index values located at the same distances from the jet axis on both sides of it. The distance ranges selected were 0 to 1, 1 to 2, 2 to 3, 3 to 4 and 4 to 5 mas. The jet axis is defined as a straight line starting at the brightest image pixel and directed at the jet angle given in \autoref{t:basicinfo}.
\item Concentration of negative spectral index values, which represents how uniform the spectral index is in the optically thin regions. This condition is quantified by the number of pixels having spectral index equal (within its error) to the negative peak of spectral index distribution within the image.
\end{enumerate}

During the manual selection process, the algorithm is trained by the user who selects the best alignment from a grid of spectral index maps corresponding to different coordinate shifts of images from the same observation. The step in both coordinates is set to $0.1$~mas and the user is required to choose no more than two correctly-looking maps. Specific cases for which the selection is ambiguous and the user cannot choose an acceptable alignment are not used in the classifier training. This way we select optimal alignments for 50 random pairs of images and use them as the \emph{ground truth} answers for the classifier. For the purpose of classification we consider the user-selected shifts as correct, and shifts differing by more than $0.15$~mas from the selected ones as incorrect.

\begin{figure}
    \centering
    \includegraphics[width=\columnwidth,trim=0cm 1cm 0cm 0.1cm]{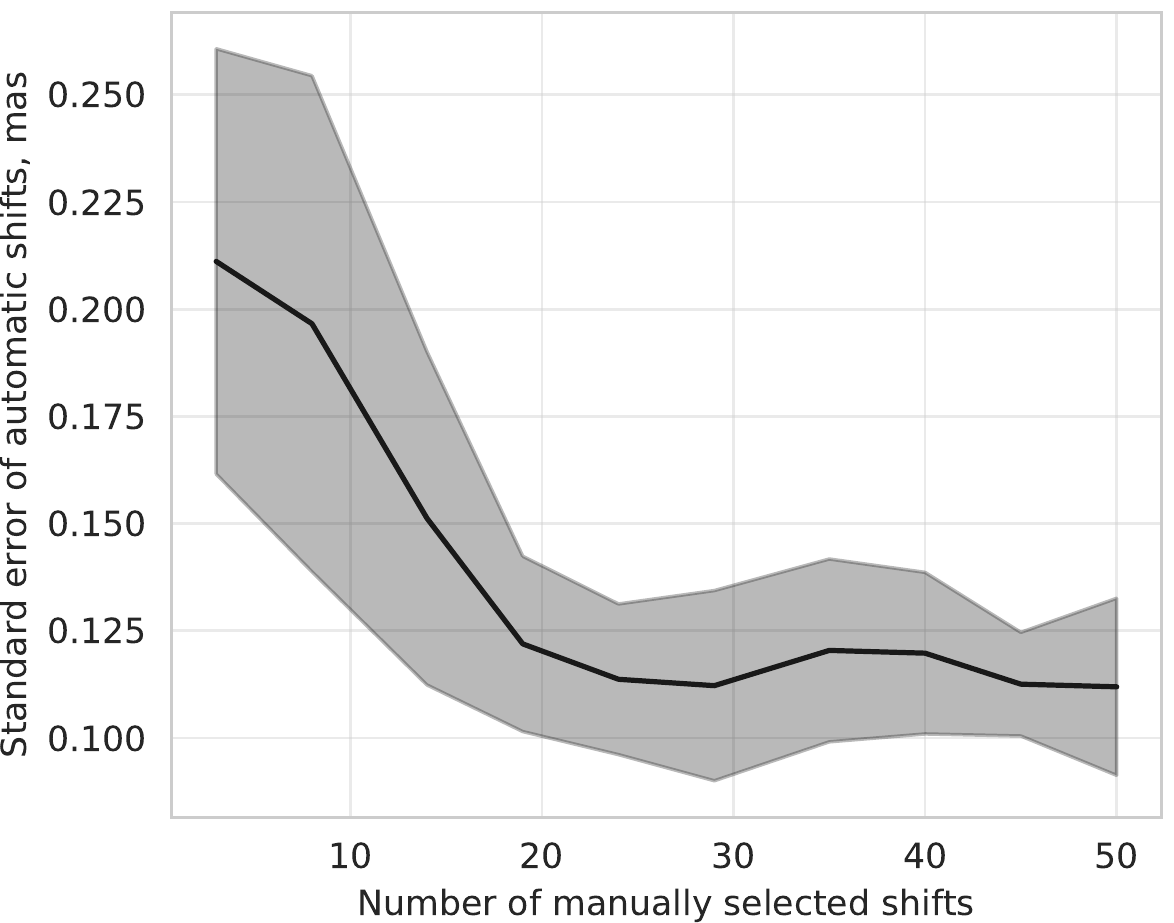}
    \caption{Learning curve of the automatic image alignment method. It shows the standard error of the shift achieved by the classifier with respect to number of manually selected samples. The shaded uncertainty region represents difference in random cross-validation splits.}
    \label{f:alignment_learningcurve}
\end{figure}

Using the classification features and ground truth answers described above, we trained a binary classifier using a tree boosting algorithm implementation in XGBoost \citep{Chen:2016:XST:2939672.2939785}. Such models can capture relationships between different features and are known to be robust to overfitting, which is important as we have relatively few training samples. Alignments marked as `correct' by the classifier are used for further analysis and combined as a weighted average when a single shift estimate is required. Major hyperparameters of the classifier algorithm were selected using cross-validation and they result in the median error of the estimated shift equal to 0.07~mas. The learning curve of the classifier for the chosen hyperparameters is shown in \autoref{f:alignment_learningcurve} confirming that using 50 manually-selected samples is sufficient to reach a $\sim 0.1$\,mas accuracy of the automated alignment procedure.
The curve remains stable with its variance slowly decreasing with more samples, and according to it one should not expect a significant improvement with more manually selected samples.

We assessed the resulting spectral index images visually and found out that generally they conform to our expectations (see above) and are consistent across observations of the same source at different epochs. \corr{The results of this method appear to be more robust compared to those discussed in \autoref{s:align-ccorr} and require less manual flagging and parameter selection, but otherwise are typically consistent with cross-correlation.}

\subsubsection{Combining results}

Having measured core positions at both frequencies ($\vec r_{2\C}$ at 2~GHz and $\vec r_{8\C}$ at 8~GHz) and the alignment shift between corresponding images $\vec r_{\text{img }2\to8}$ we calculate the core shift vector as $\Delta \vec r = \vec r_{2\C} - \vec r_{8\C} - \vec r_{\text{img }2\to8}$. The resulting core shift vectors are plotted in \autoref{f:cs_alignment} with respect to the jet position angle in respective objects. As noted before, our measurement methods give multiple values for each of these quantities. The spread of these values may be taken as representative of the uncertainty in their estimation. Here we combine them in all combinations, i.e.\ take both core positions $\vec r_{2\C}$ and $\vec r_{8\C}$ for all threshold values (see \autoref{s:method_corepos}) and all image alignment vectors $\vec r_{\text{img }2\to8}$ classified as `correct' (see \autoref{s:method_align}). This leads to multiple values of the resulting $\Delta \vec r$. To get a single estimate of the core shift magnitude, we compute a median of projection on the jet axis of all these values. Half-width of their 68\% interval is taken to be the $1\sigma$ error of this estimate, and this uncertainty is typically about 0.2~mas.

\section{Results}
\label{s:results}
\subsection{Evaluation of measurements}

\begin{figure}
    \centering
    \includegraphics[width=\columnwidth,trim=0cm 1.2cm 0cm 0.3cm]{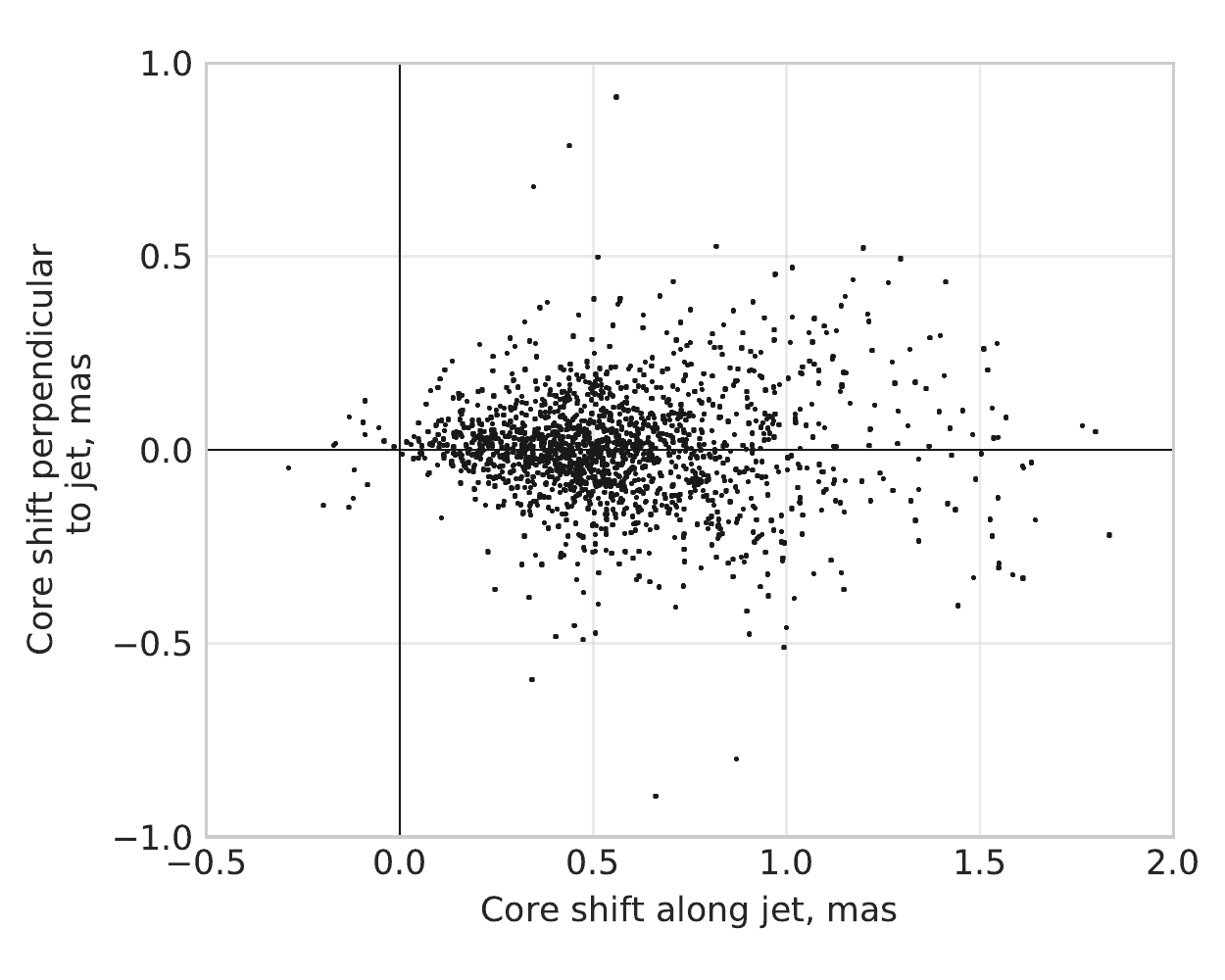}
    \caption{Core shift vectors relative to the jet direction. End points of the measured core shift vectors are plotted.}
    \label{f:cs_alignment}
\end{figure}

Previous studies of core shifts indicate that the core shift vector is generally well-aligned with the jet direction and its perpendicular component is likely to be noise-dominated \citep{2012A&A...545A.113P}. This conclusion is
confirmed by our results plotted in \autoref{f:cs_alignment}, which shows that about 80\% of the core shift vectors lie within $20^\circ$ of the jet direction. The assumption that the component perpendicular to the jet is noise-dominated is also directly confirmed: for about $2/3$ of observations the perpendicular component is within $1\sigma$ confidence interval from zero.

\begin{figure}
    \centering
    \begin{minipage}{0.5\columnwidth}
        \includegraphics[width=\columnwidth,trim=0cm 0.5cm 0cm 0.3cm]{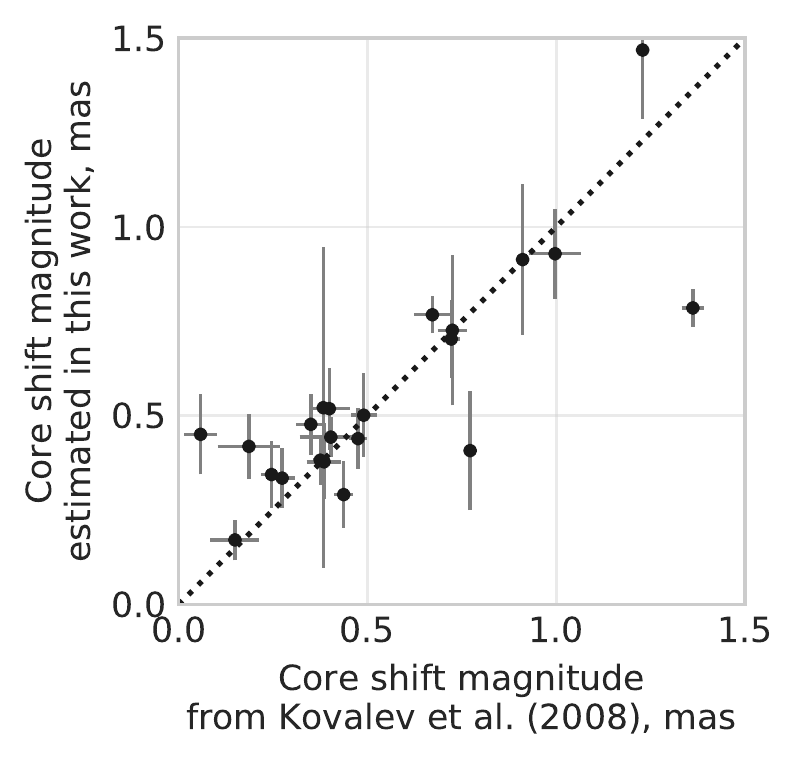}
    \end{minipage}%
    \begin{minipage}{0.5\columnwidth}
        \includegraphics[width=\columnwidth,trim=0cm 0.5cm 0cm 0.2cm]{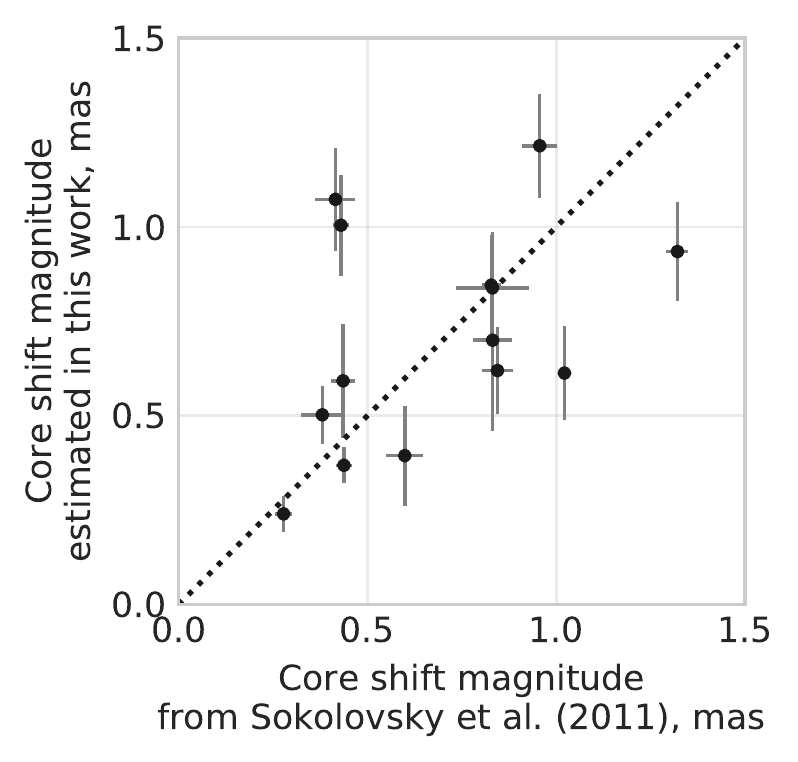}
    \end{minipage}
    \caption{Comparison between the core shift magnitudes obtained in this work with earlier measurements for the same objects made by \citet{2008A&A...483..759K} and \citet{2011A&A...532A..38S}.
    }
    \label{f:cs_agreement_kov}
\end{figure}

\begin{figure}
    \centering
    \includegraphics[width=\columnwidth,trim=0cm 0cm 0.5cm 0.3cm]{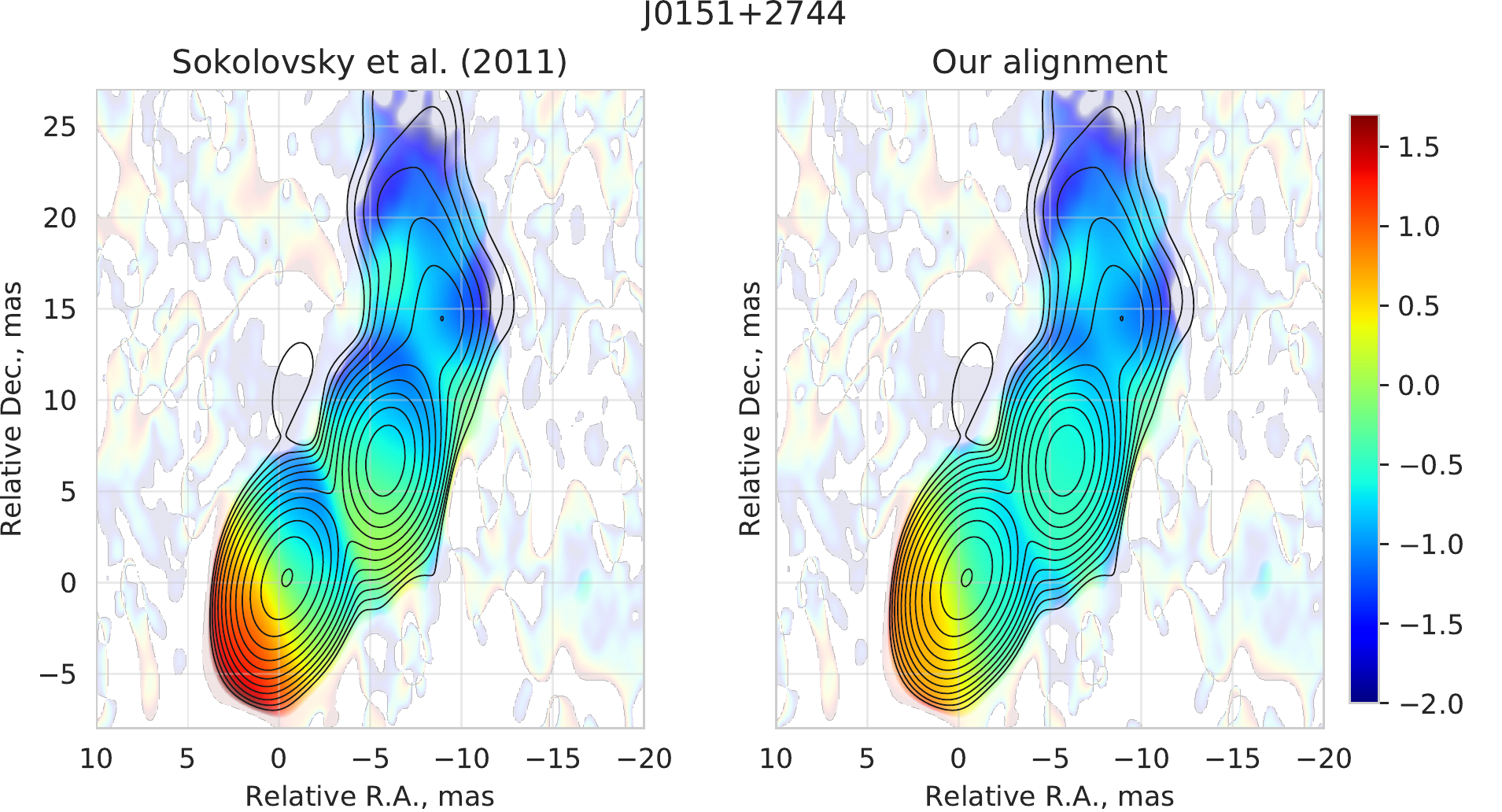}
    \includegraphics[width=\columnwidth,trim=0cm 1cm 0.5cm -0.1cm]{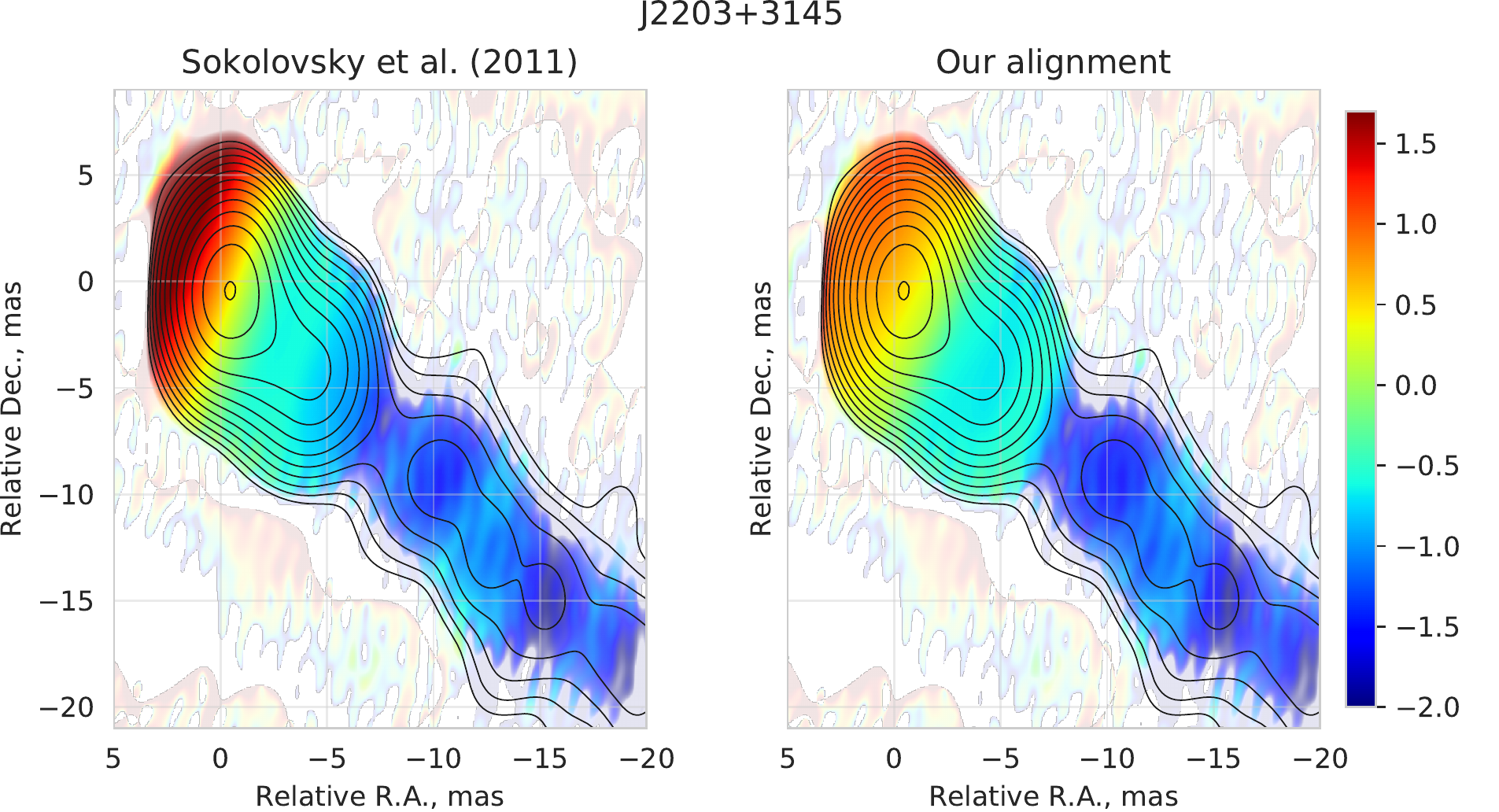}
    \caption{Comparison of spectral index images for J0151+2744 and J2203+3145 resulting from our image alignment and the alignment made in \citet{2011A&A...532A..38S}. These are sources with the largest difference of our core shift measurements from those made in \citet{2011A&A...532A..38S}. The spectral index value is shown in colour, and 2~GHz intensity contours are overlaid on top. Spectral index images for each source can be found in the electronic-only supplementary material.
    }
    \label{f:si_comparison}
\end{figure}

\begin{figure}
    \centering
    \includegraphics[width=\columnwidth,trim=0cm 1cm 0cm 0.3cm]{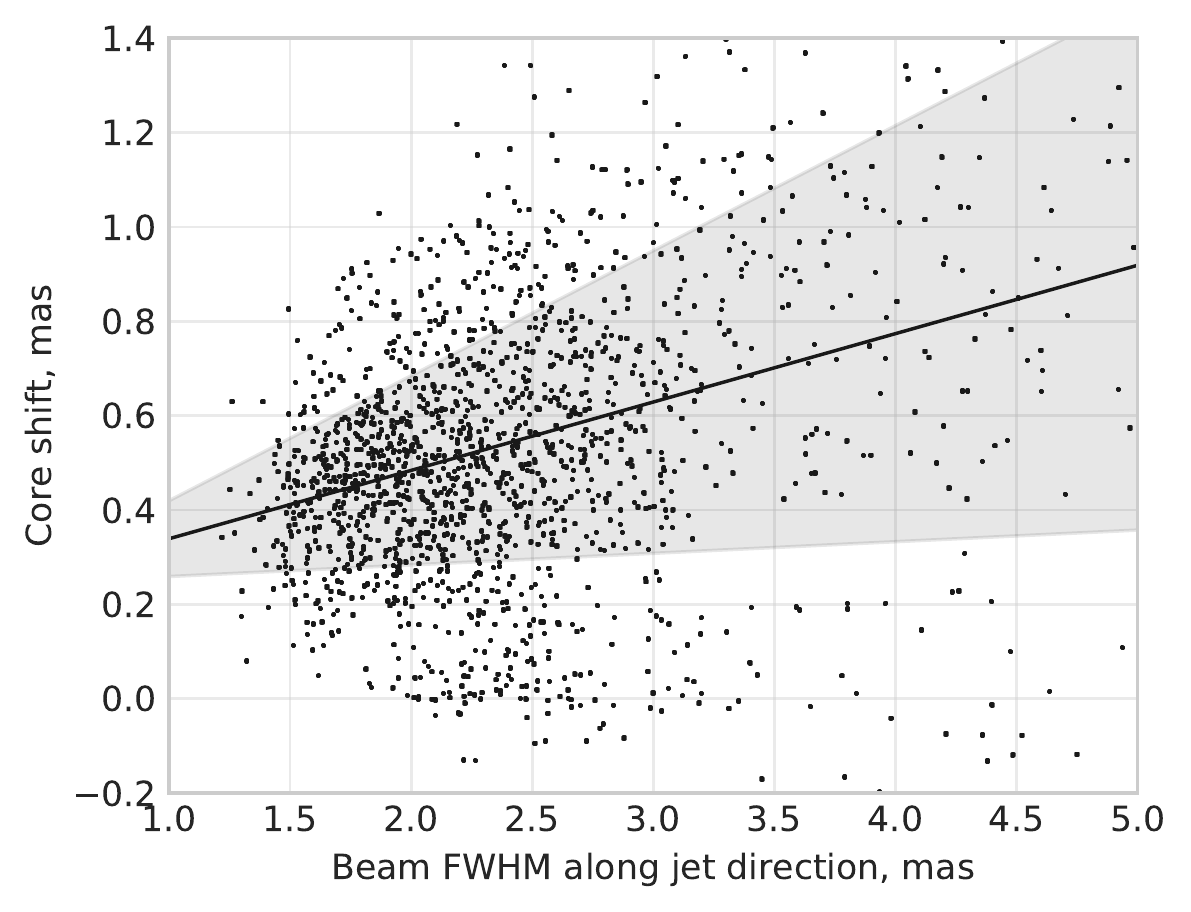}
    \caption{Relation between measured core shift magnitude and the synthesized beam size. Points represent individual measurements, solid line is the linear fit, and \corr{shaded} region is a 68\% uncertainty interval.}
    \label{f:beambias}
\end{figure}

As an additional test of the measurement method developed here, we compare our results in \autoref{f:cs_agreement_kov} with previous core shift measurements made \cite{2008A&A...483..759K} and \cite{2011A&A...532A..38S}, for the common sources and epochs. This comparison shows that the different estimates agree within the $1\sigma$ errors in 72\% and 61\% of sources, respectively. To understand possible reasons for larger discrepancies, we compare in \autoref{f:si_comparison} spectral index images for two sources with the largest difference between our measurements and those of \cite{2011A&A...532A..38S}. The present alignment results in smaller transverse gradients and a generally more homogeneous distribution of spectral index.

In a further test of the core shift results, we evaluate if the different ($u$,$v$)-coverages of the VLBI data for different sources and epochs have a systematic effect on our measurements. For this, we compare the estimated core shift magnitudes with the projected synthesized beam sizes along the jet direction. As described in \autoref{s:obsdata}, we use the average synthesized beam between 2 and 8~GHz. The relationship between the measured shifts and the projected beam sizes is shown in \autoref{f:beambias}. Assuming a linear dependence between the two parameters, the relation between them can be best described as
\begin{equation}
\label{eq:beambias}
r = 0.19\text{ mas} + 0.14 r_\text{beam} + \varepsilon \,,
\end{equation}
where \corr{$\varepsilon$ is a Gaussian random variable with standard deviation} $\sigma_\varepsilon = 0.12 r_\text{beam} - 0.04 \text{ mas}$. The resulting linear relation plotted in \autoref{f:beambias} shows \corr{that for larger beams both the average value of the core shift measurements and its spread increase, so measurements become less precise. However most of the measured core shifts and differences between them} are not dominated by changes of ($u$,$v$)-coverage from epoch to epoch \corr{as the points are spread out relative to the regression line}.

\corr{The sensitivity is different at S and X bands, and it changes from one observation to another. This combined with different spectral properties of jet regions can in principle affect our measurements. Using the same approach as with beam size, we compare the core shift measurements with image noise at 2 and 8 GHz, and with their ratio. In all cases the Spearman correlation is either insignificant or statistically significant with the coefficient of less than 0.06. We do not consider effects of sensitivity further as even if they exist they are much smaller than those of ($u$,$v$)-coverage.}

To study the \corr{relationship between flux density variability and core shift further in this paper}, we use the core flux density from the Gaussian component fit performed in \autoref{s:method_corepos} for identifying the core region. To verify the validity of using these results, we check and confirm that the core flux density is not lower than the unresolved flux density and not higher than the total flux density \citep[see their definitions in][]{2008AJ....136..580P}, where these flux densities are taken from the Radio Fundamental Catalogue.

\subsection{Core shift magnitudes}

\begin{figure}
    \centering
    \includegraphics[width=1\columnwidth,trim=0cm -0.2cm 0cm 0cm]{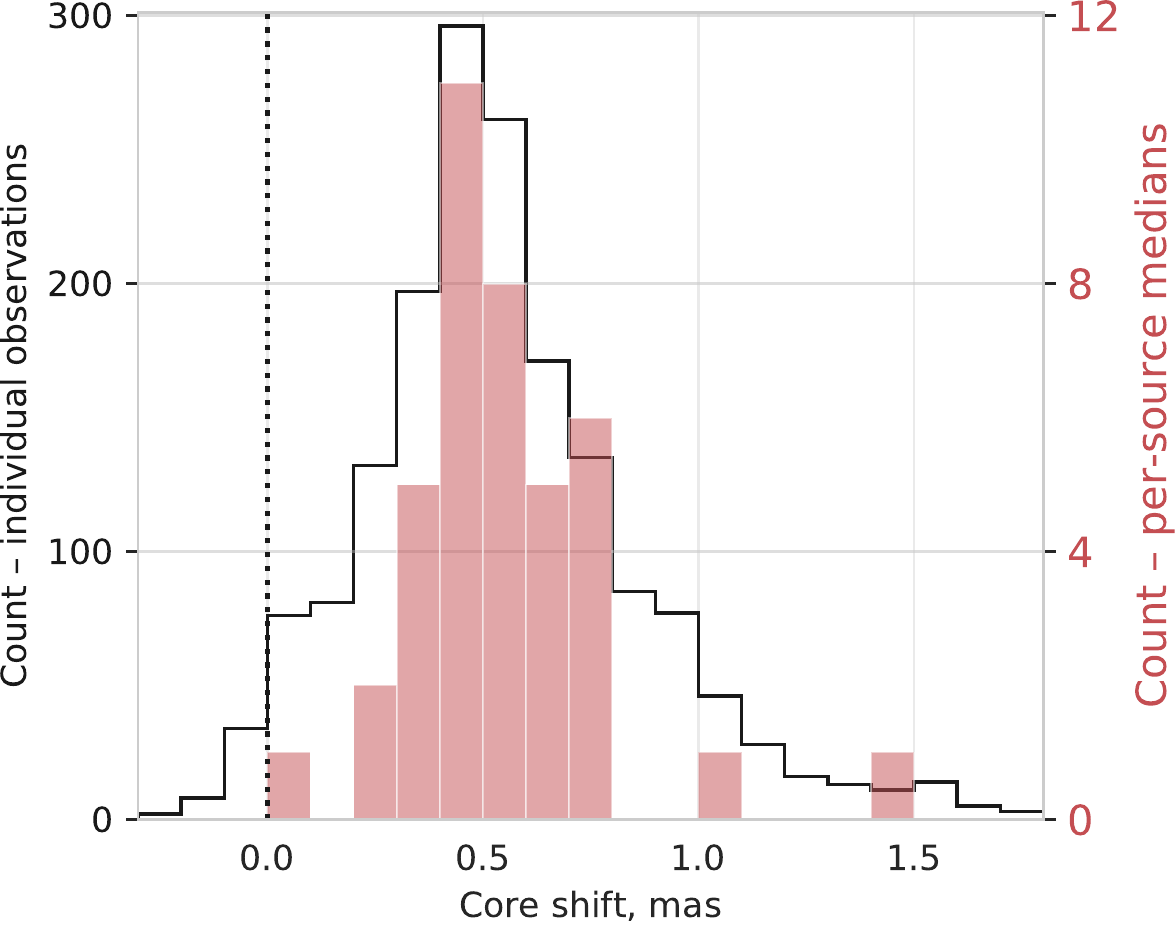}
    \includegraphics[width=1\columnwidth,trim=0cm 0.6cm 0cm 0cm]{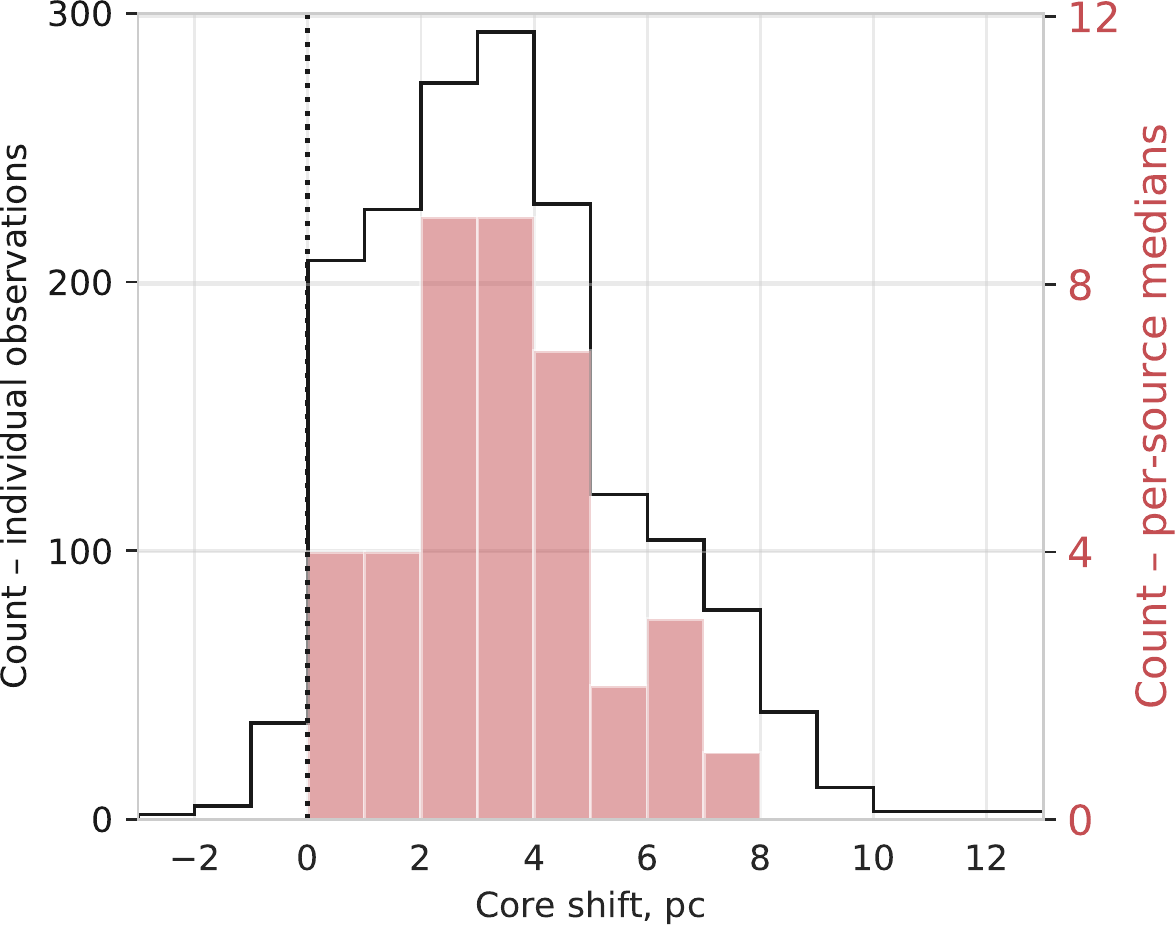}
    \caption{Distribution of 8-2 GHz core shift magnitudes measured in the plane of the sky in angular (top) and linear (bottom) units. The vertical dotted line denotes the zero value of the core shift.}
    \label{f:cs_dist}
\end{figure}

The overall distribution of the core shift magnitudes measured in the plane of the sky is shown in \autoref{f:cs_dist} for all sources, both in angular and linear units (derived using the angular size distances according to the standard $\Lambda$CDM cosmology). The median magnitudes of the core shifts are $0.53$~mas and $3.2$~pc. The core shift magnitudes are given in \autoref{t:measurements} for all sources and all epochs used in our analysis. \corr{About 2.5\% of all magnitudes are negative, and only 9\% of which differ from zero at $2\sigma$ level. Thus we do not find any significant evidence for cores at 8 GHz located downsteam from 2 GHz cores, and do not consider this case in all following discussions.}

Assuming inverse frequency dependence of core shift, $r_\mathrm{c}(\nu) \sim \nu^{-1}$, we obtain distance from the jet origin to the core at $\nu_2$:
\begin{equation}
r_\C(\nu_2) = \frac{\Delta r}{\frac{\nu_2}{\nu_1} - 1}.
\end{equation}
In our case, having $\nu_1 = 2.3$~GHz and $\nu_2 = 8.6$~GHz, the distances are $r_\mathrm{c}(\nu_2) \approx 0.35 \Delta r$ and $r_\mathrm{c}(\nu_1) \approx 1.35 \Delta r$. Their typical values can be estimated as $r_\mathrm{c}(\nu_2) \approx 1.1$~pc and $r_\mathrm{c}(\nu_1) \approx 4.3$~pc. Note that later in the paper we discuss conditions when this frequency dependence may or may not be applicable.

\begin{figure*}
    \centering
    \includegraphics[width=0.9\textwidth,trim=0cm 0.5cm 0cm 0cm]{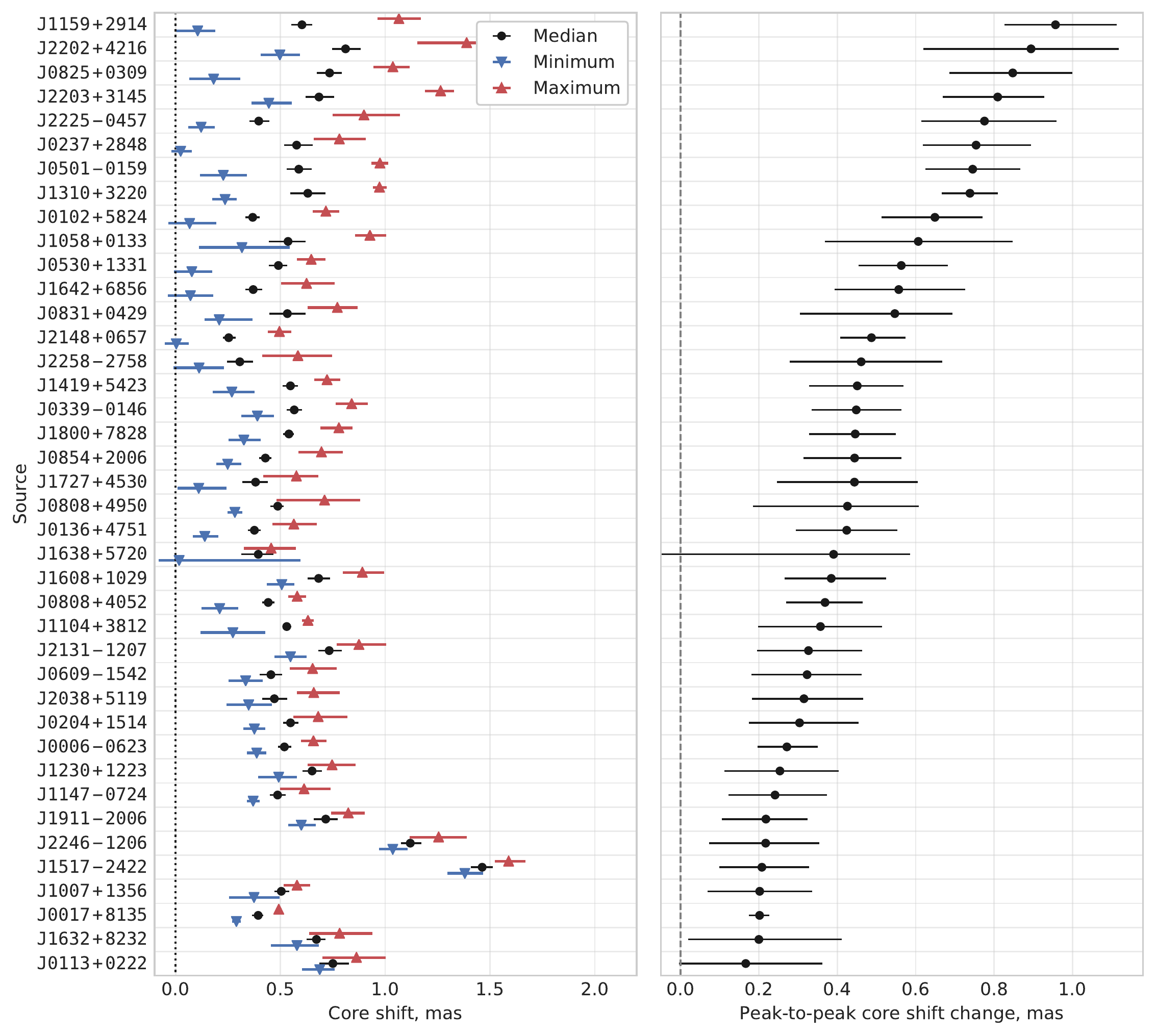}
    \caption{General properties of the 8-2~GHz core shift time series obtained for each source. The median, minimum and maximum values of the core shift are plotted in the left panel. Difference between the maximum and the minimum measured shifts is shown in the right panel as a proxy for the amplitude of core shift variations in a given object. Error bars indicate 68\% credible intervals. Sources are ordered by decreasing of the core shift variations.}
    \label{f:cs_stats}
\end{figure*}

\begin{table}
\caption{Estimated parameter values. Columns are as follows:
(1) J2000 name;
(2) epoch of observation;
(3) 8-2~GHz core shift magnitude in angular units;
(4) core shift magnitude in linear units as projected on the sky.
This table is available in its entirety in the online journal.
A portion is shown here for guidance regarding its form and content.
}
\label{t:measurements}
\centering
\begin{tabular}{cccc}
\hline
J2000 name & Epoch & $r_\C$, mas & $r_\C$, pc \\
(1) & (2) & (3) & (4) \\
\hline
J0006$-$0623 & 1995-10-12 & 0.83 & 4.13 \\
J0006$-$0623 & 1997-01-30 & 0.43 & 2.13 \\
J0006$-$0623 & 1997-03-31 & 0.56 & 2.76 \\
J0006$-$0623 & 1997-05-07 & 0.50 & 2.51 \\
J0006$-$0623 & 1997-05-19 & 0.66 & 3.29 \\
\hline
\end{tabular}
\end{table}

\subsection{Variability of core shifts}

To analyze how core shift magnitude changes with time, first we plot time series of the 8 and 2~GHz core flux density and 8-2~GHz core shift measurements for each source. These time series are shown in \autoref{f:timeseries} as individual points and are approximated with smooth curves, representing Gaussian process fits (see \autoref{s:fit_gp} for fitting details).

From the Gaussian process fits, we estimate the variability amplitude of the core shift. We take realizations of a Gaussian process and compute the minimum, maximum and median values for each object. The results are summarized in Fig.~\ref{f:cs_stats}. For 33 out of 40 sources, the 95\% intervals for $\max \Delta r_\C -\min \Delta r_\C$ difference do not include zero, implying significant detections of core shift variability.
We estimate that the median variability amplitude of $\Delta r_\C$ is 0.4~mas and the maximum amplitude reaches almost 1~mas.

\begin{figure*}
\centering
\begin{minipage}{\textwidth}
\includegraphics[width=0.23\linewidth,trim=0cm 0cm 0cm 0.3cm]{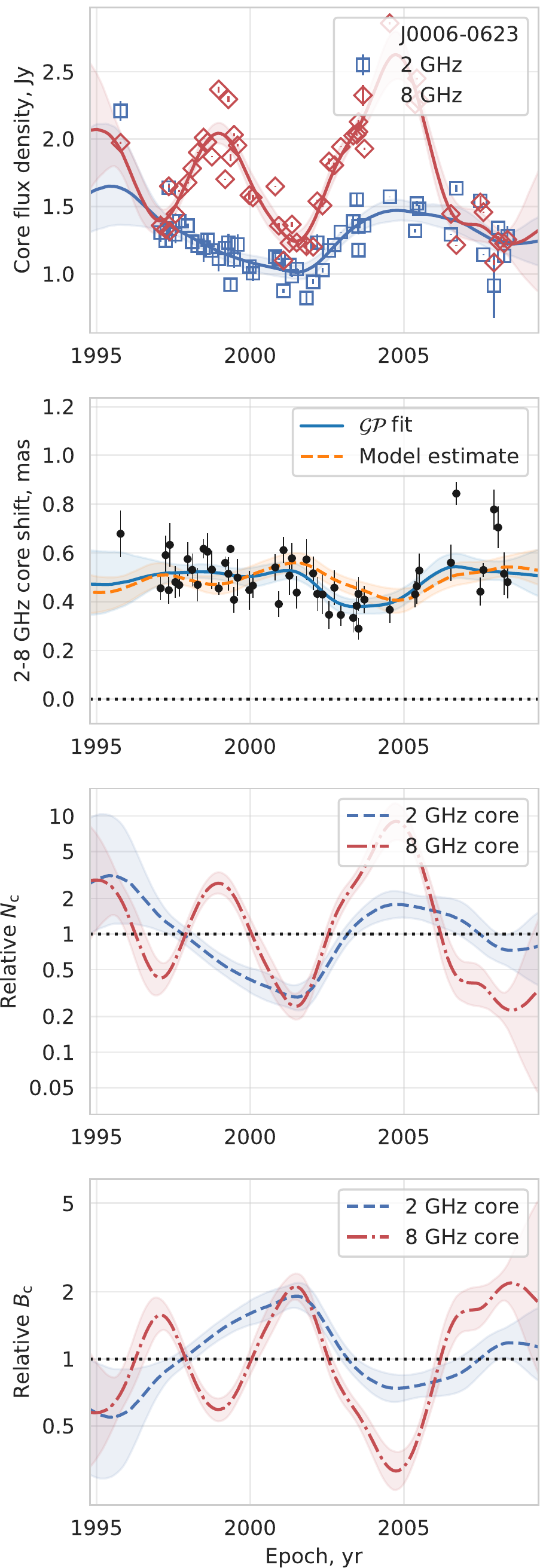}
\includegraphics[width=0.23\linewidth,trim=0cm 0cm 0cm 0.3cm]{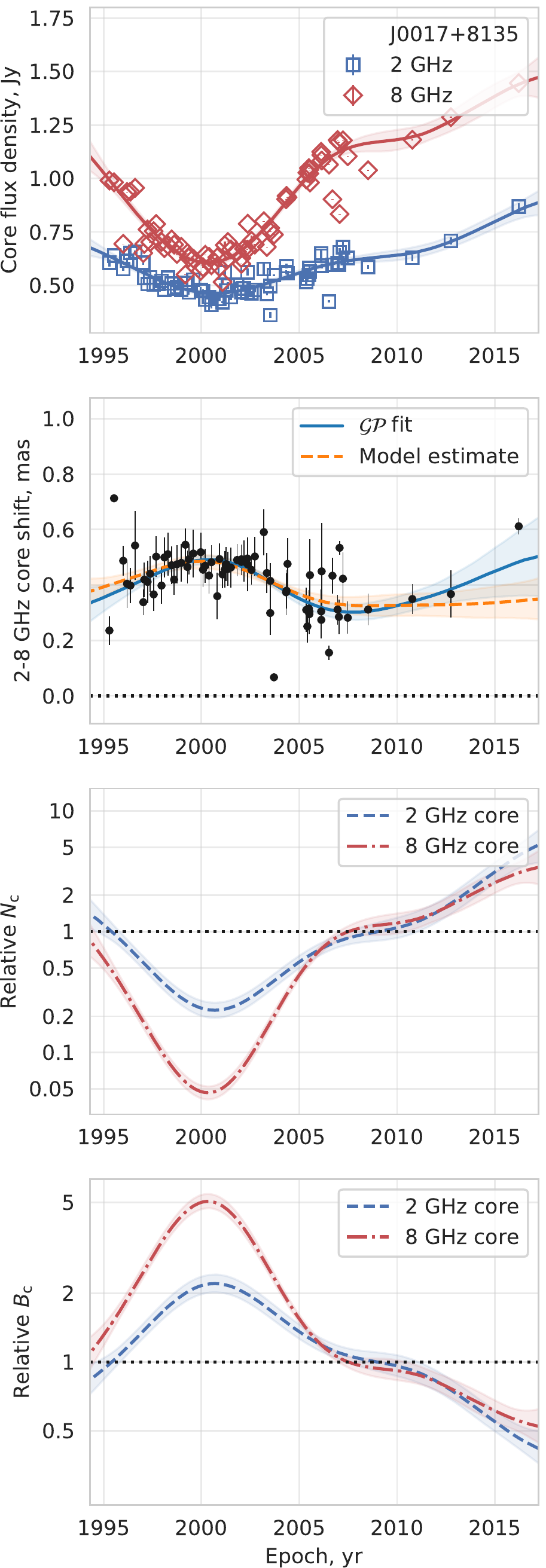}
\includegraphics[width=0.23\linewidth,trim=0cm 0cm 0cm 0.3cm]{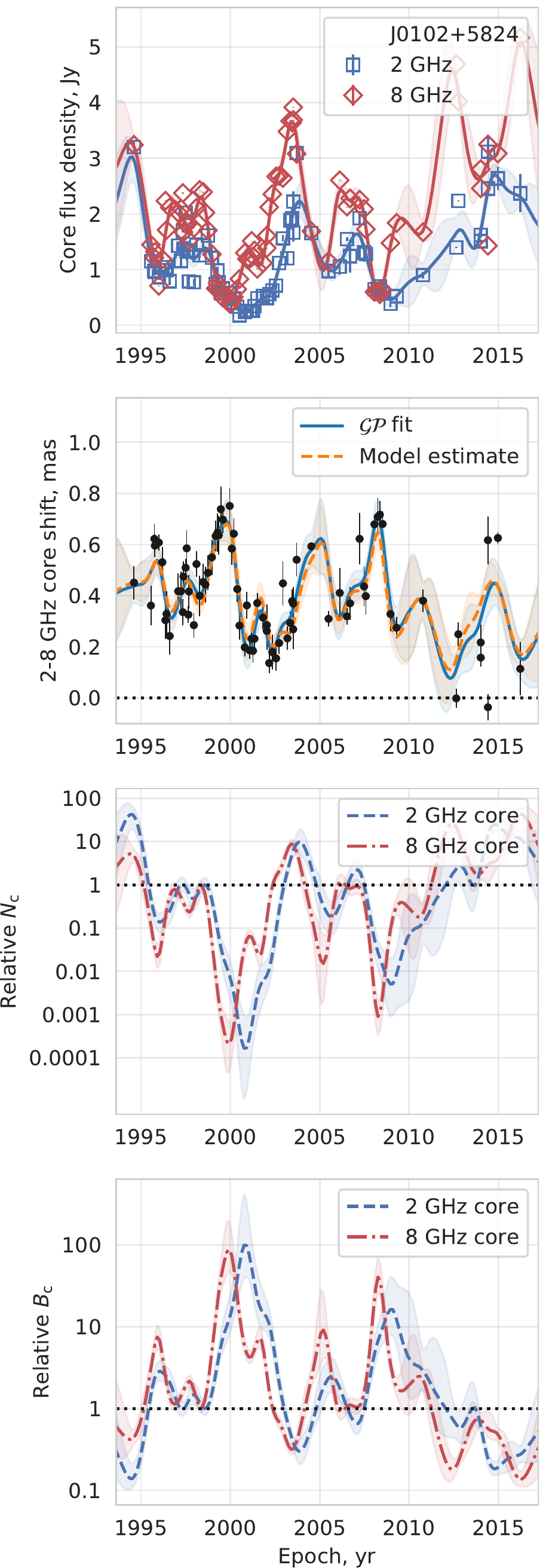}
\includegraphics[width=0.23\linewidth,trim=0cm 0cm 0cm 0.3cm]{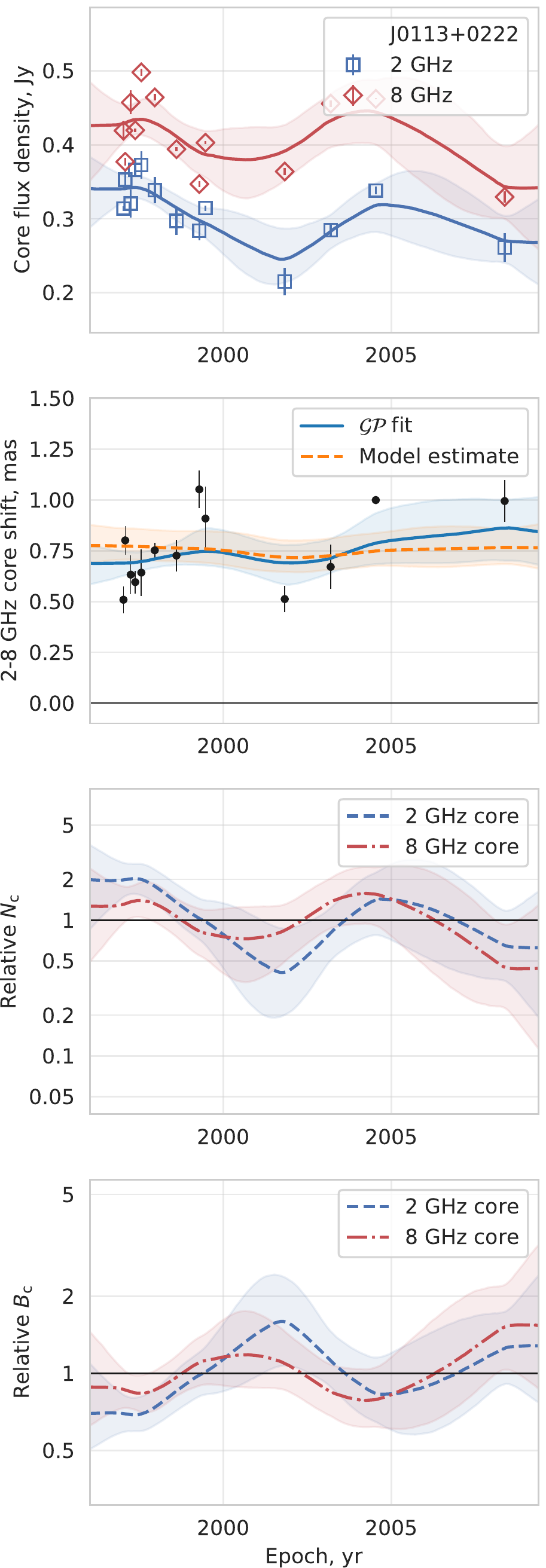}
\end{minipage}
\begin{minipage}{\textwidth}
\includegraphics[width=0.23\linewidth,trim=0cm 0.5cm 0cm 0cm]{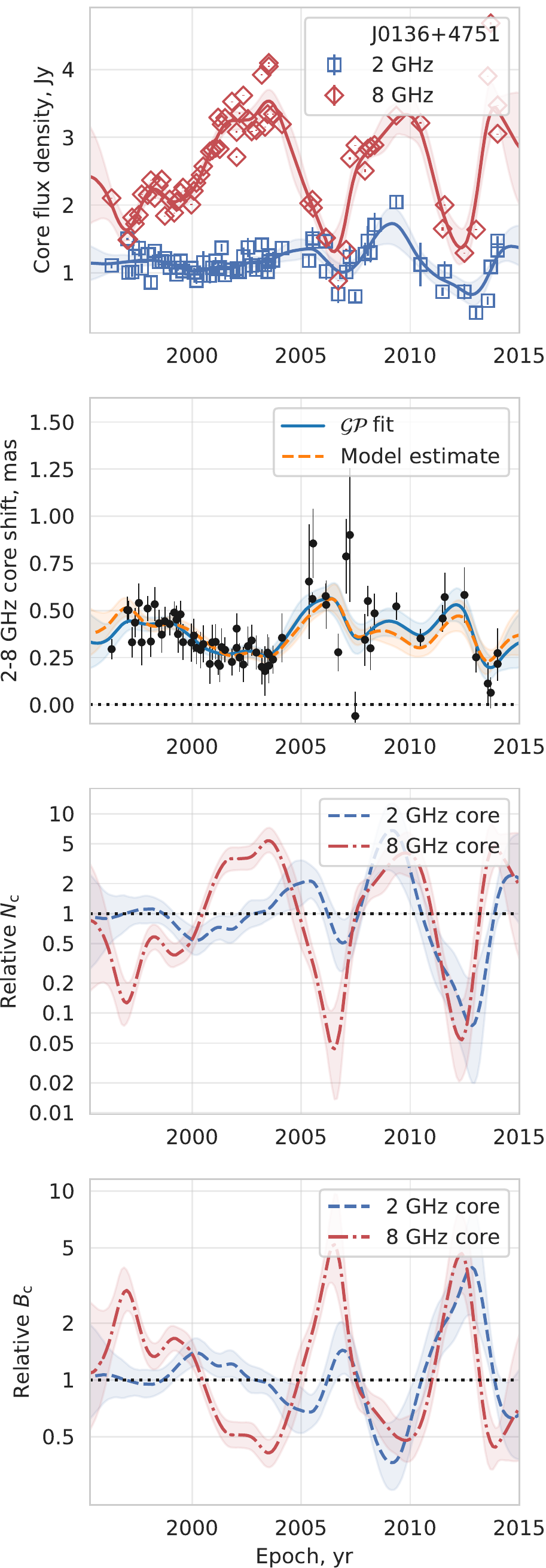}
\includegraphics[width=0.23\linewidth,trim=0cm 0.5cm 0cm 0cm]{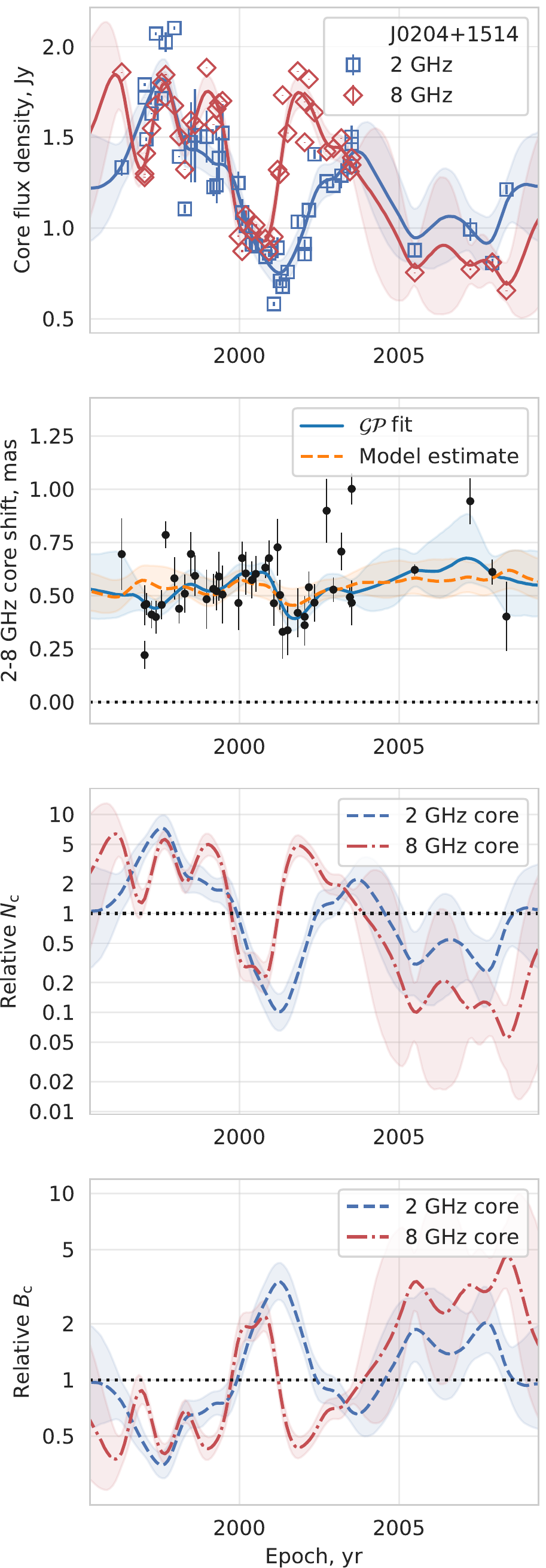}
\includegraphics[width=0.23\linewidth,trim=0cm 0.5cm 0cm 0cm]{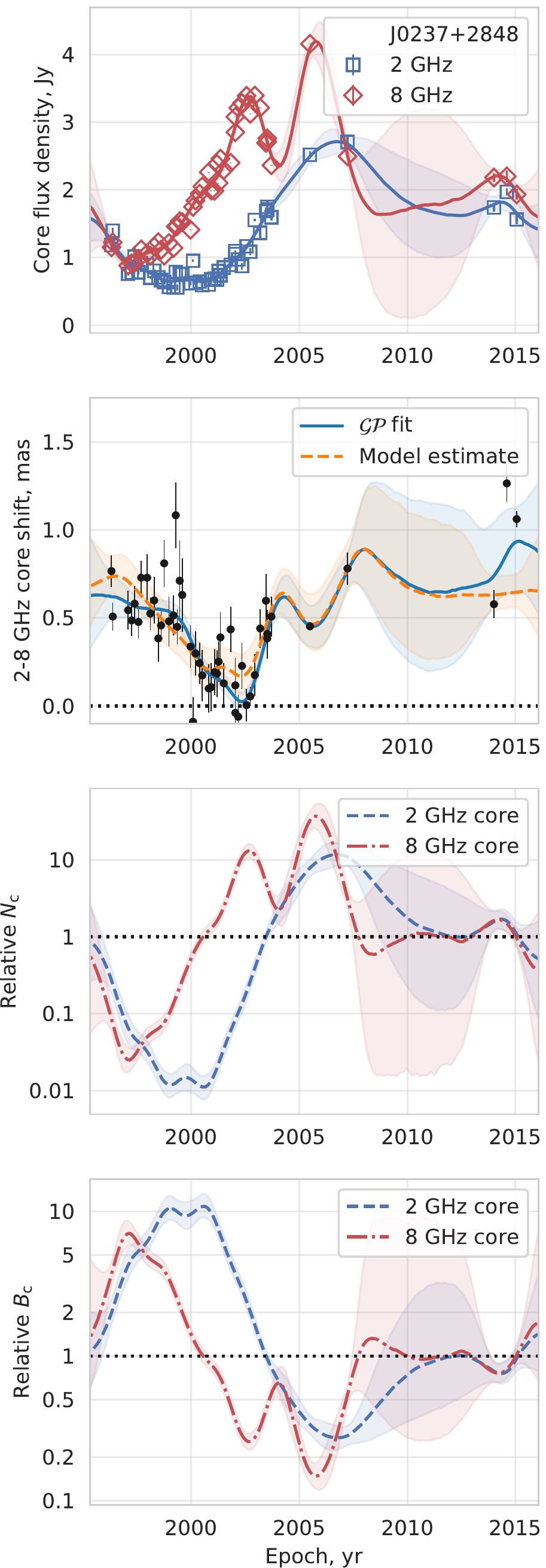}
\includegraphics[width=0.23\linewidth,trim=0cm 0.5cm 0cm 0cm]{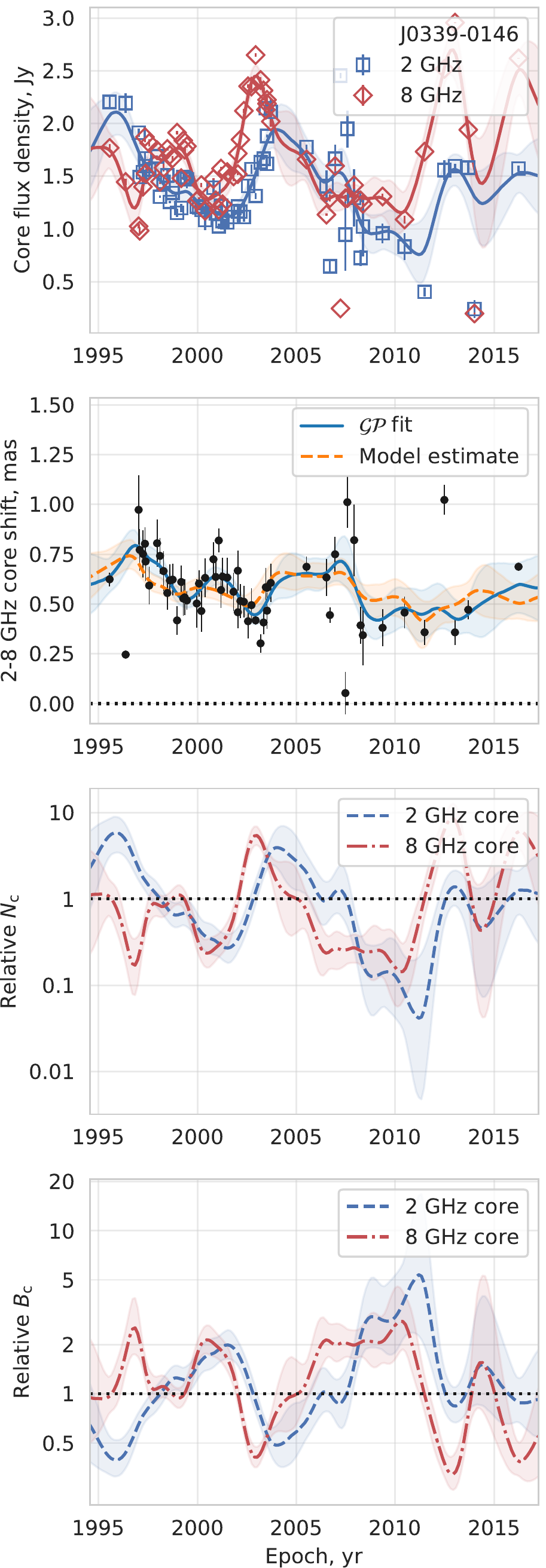}
\end{minipage}
\caption{Time series of the measured 2 and 8~GHz core flux density and 8-2~GHz core shift, as well as derived estimates of relativistic particles density and magnetic field variability at the position of apparent core at 2 and 8~GHz separately. Error bars represent $1\sigma$ errors of individual measurements, and filled areas correspond to 68\% credible intervals of the fit. See \autoref{s:fit_gp} for model fitting details.}
\label{f:timeseries}
\end{figure*}
\begin{figure*}
\centering
\begin{minipage}{\textwidth}
\includegraphics[width=0.23\linewidth]{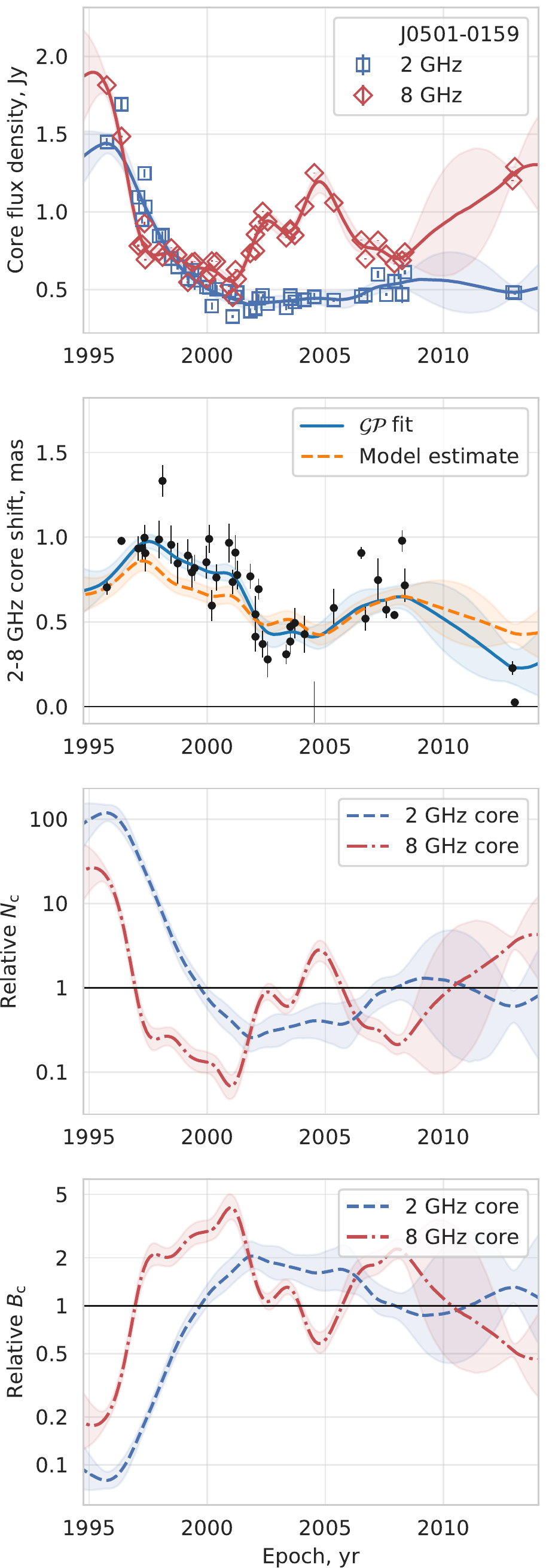}
\includegraphics[width=0.23\linewidth]{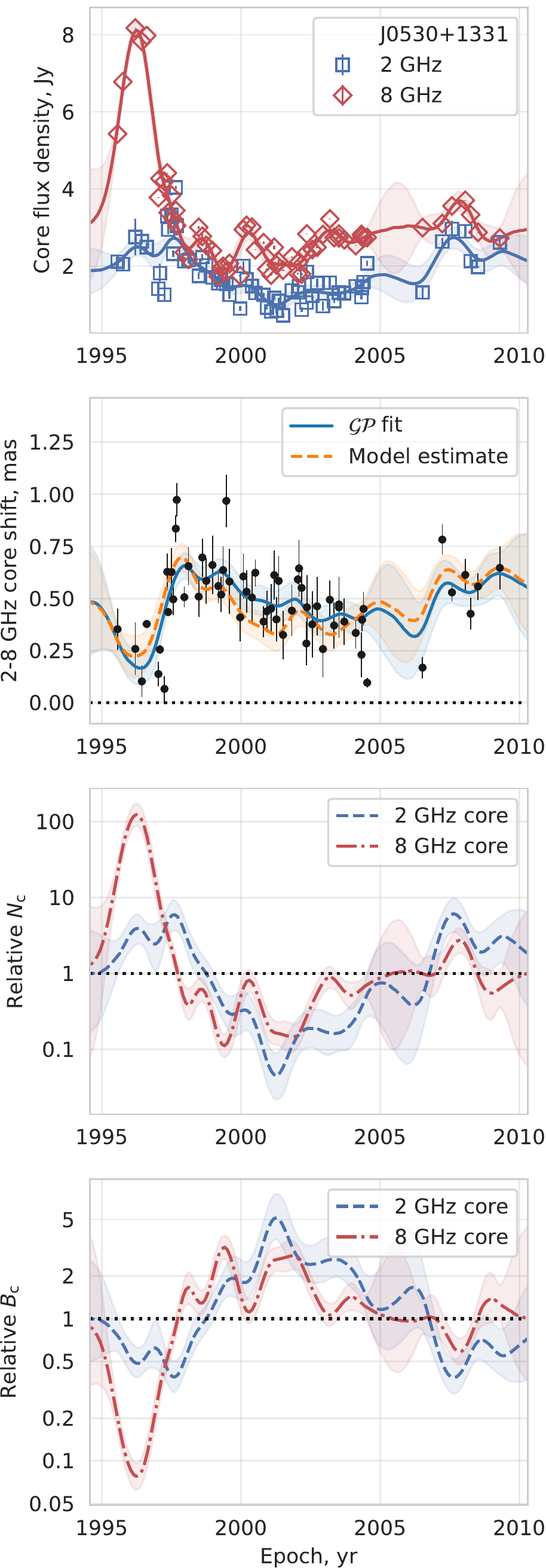}
\includegraphics[width=0.23\linewidth]{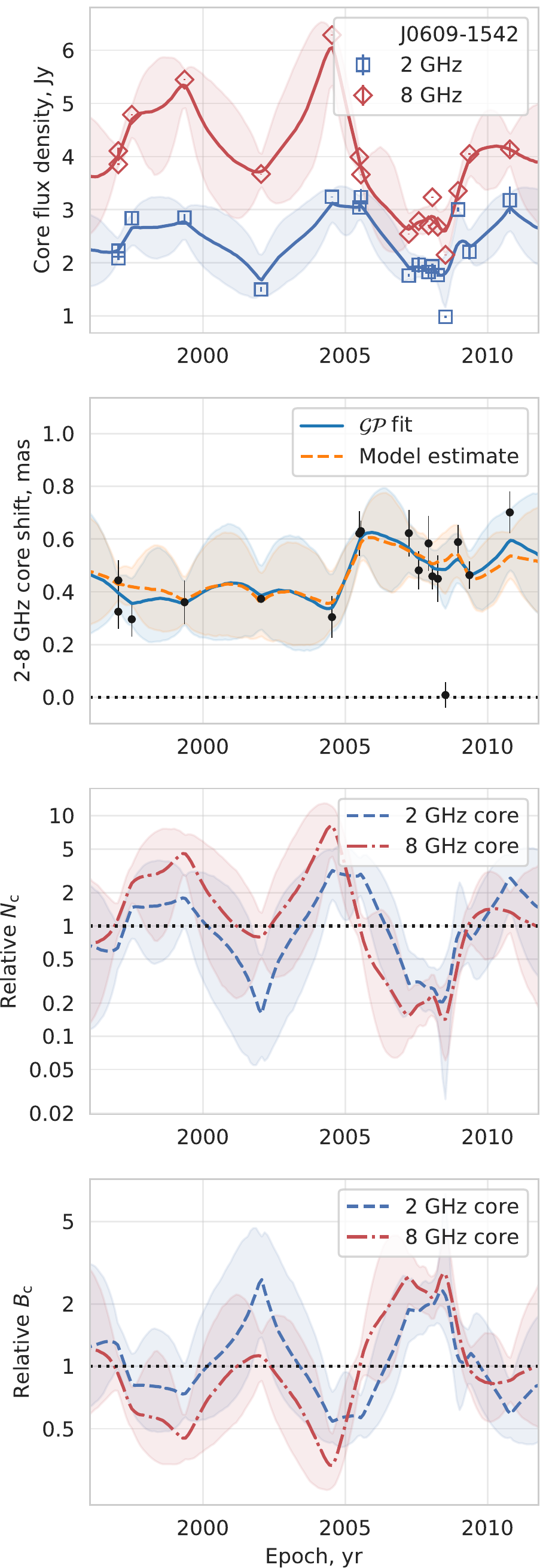}
\includegraphics[width=0.23\linewidth]{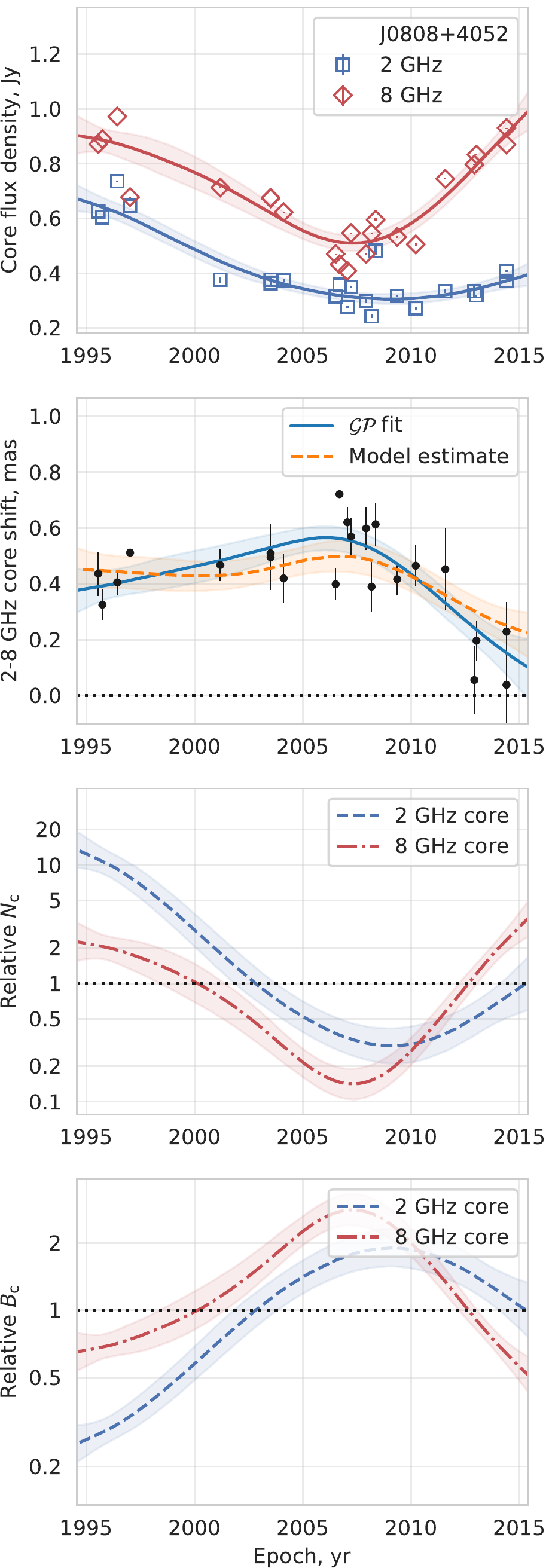}
\end{minipage}
\begin{minipage}{\textwidth}
\includegraphics[width=0.23\linewidth]{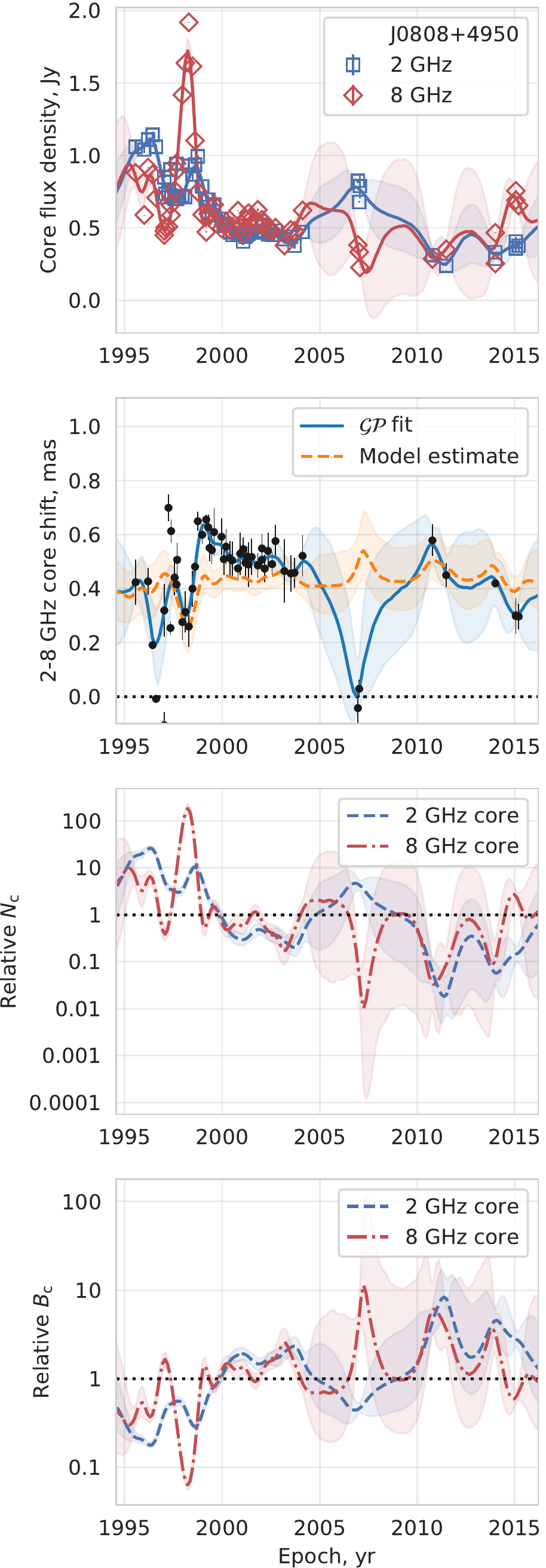}
\includegraphics[width=0.23\linewidth]{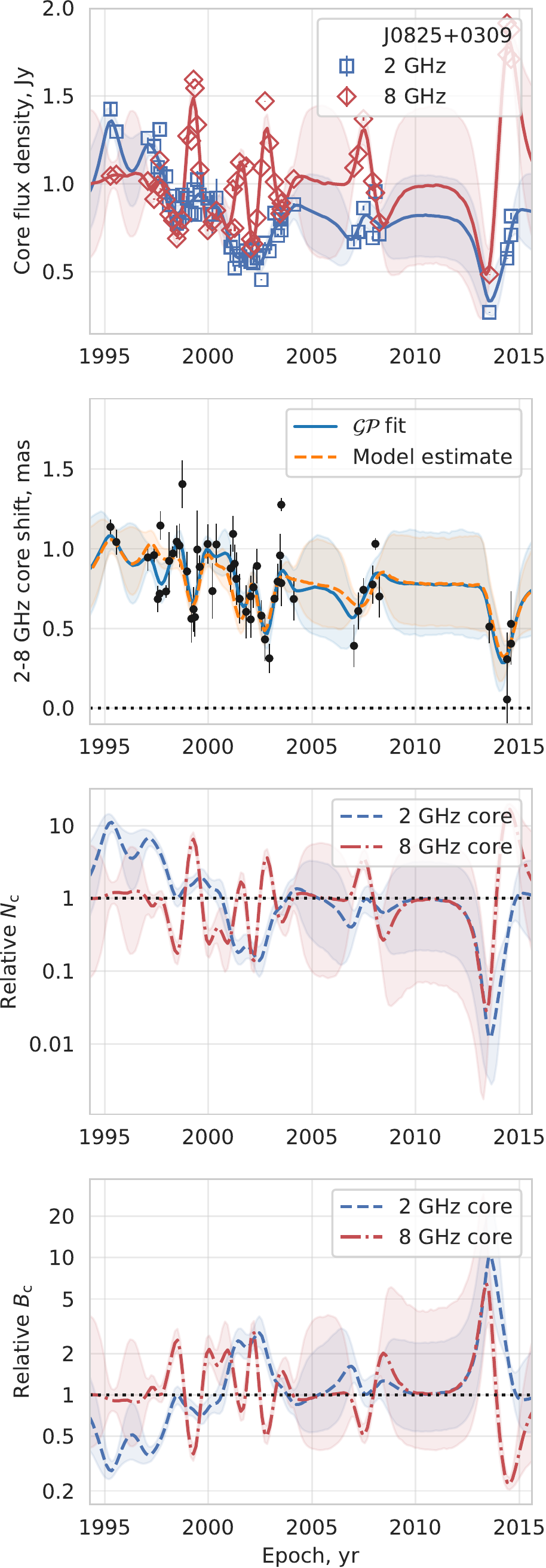}
\includegraphics[width=0.23\linewidth]{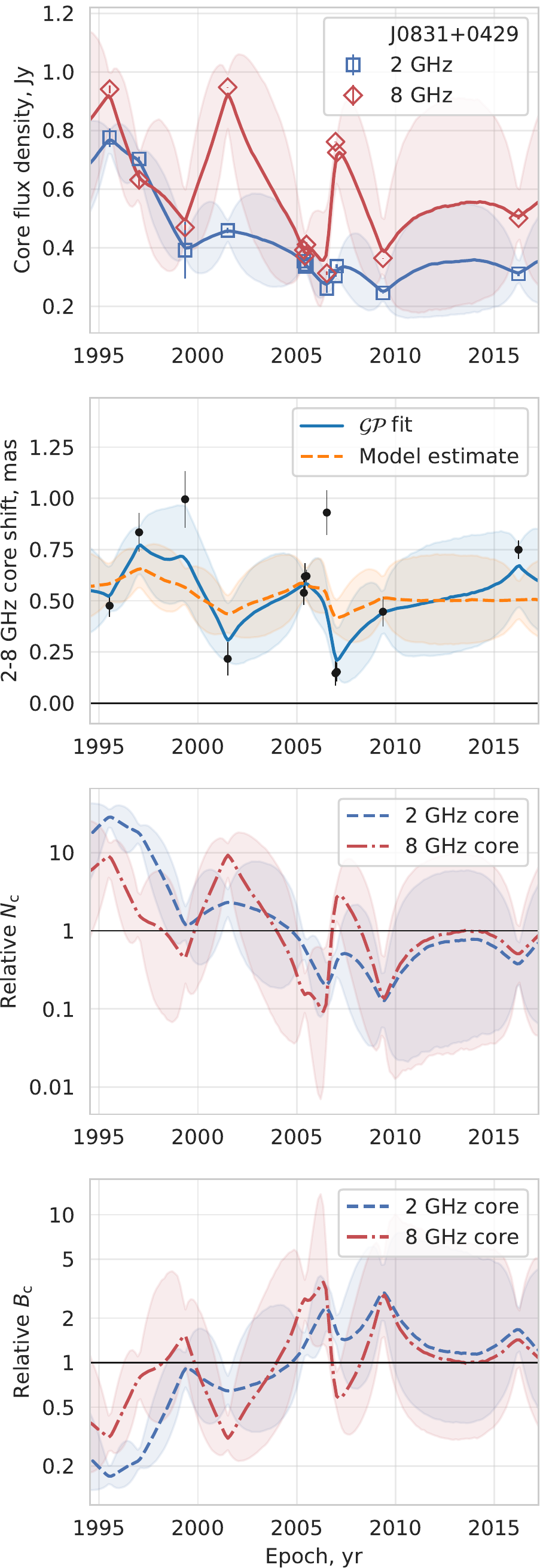}
\includegraphics[width=0.23\linewidth]{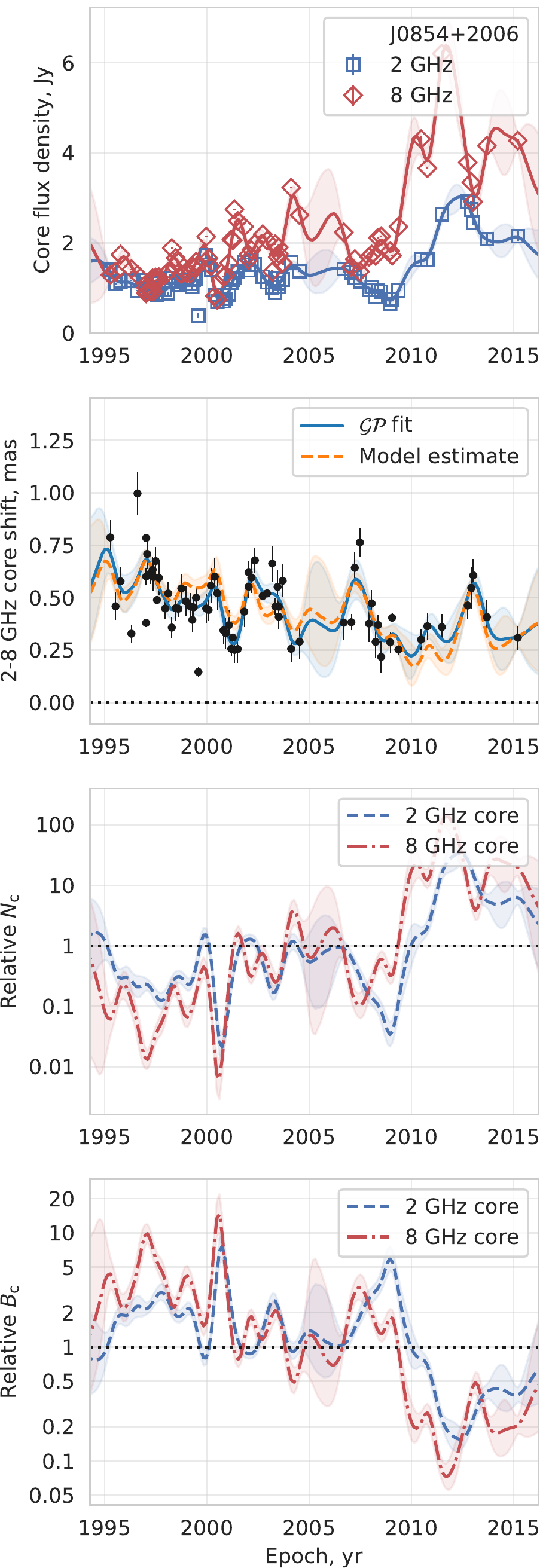}
\end{minipage}
\contcaption{}
\end{figure*}
\begin{figure*}
\centering
\begin{minipage}{\textwidth}
\includegraphics[width=0.23\linewidth]{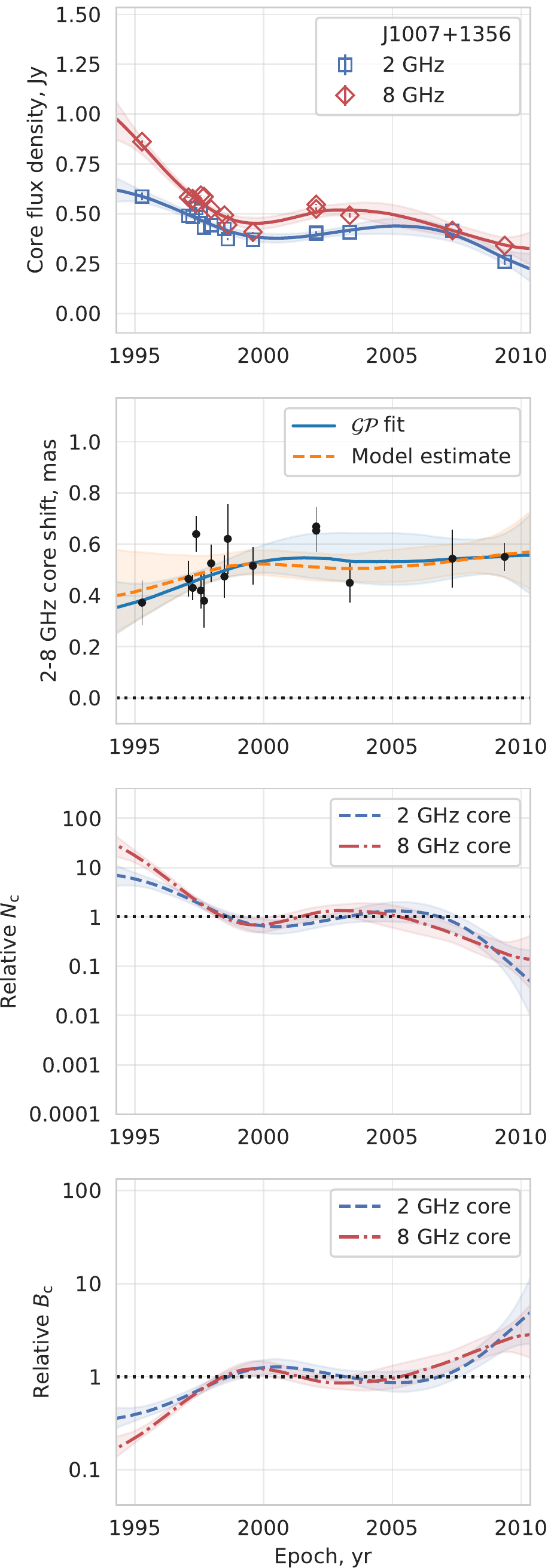}
\includegraphics[width=0.23\linewidth]{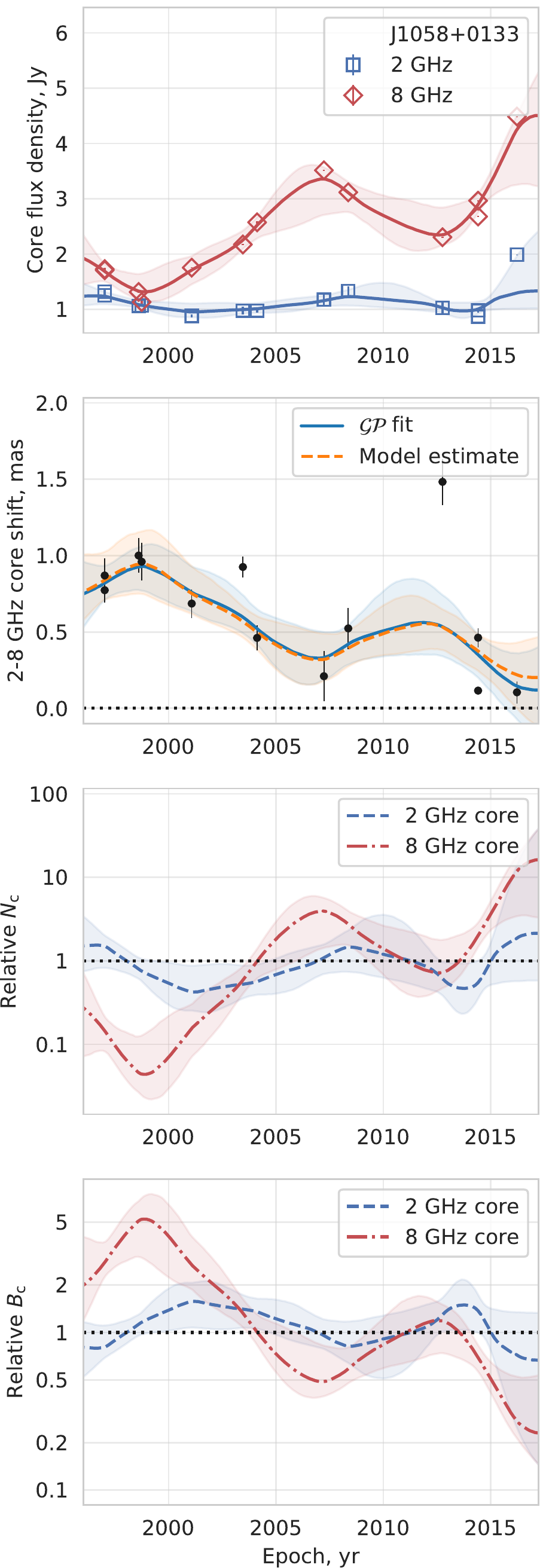}
\includegraphics[width=0.23\linewidth]{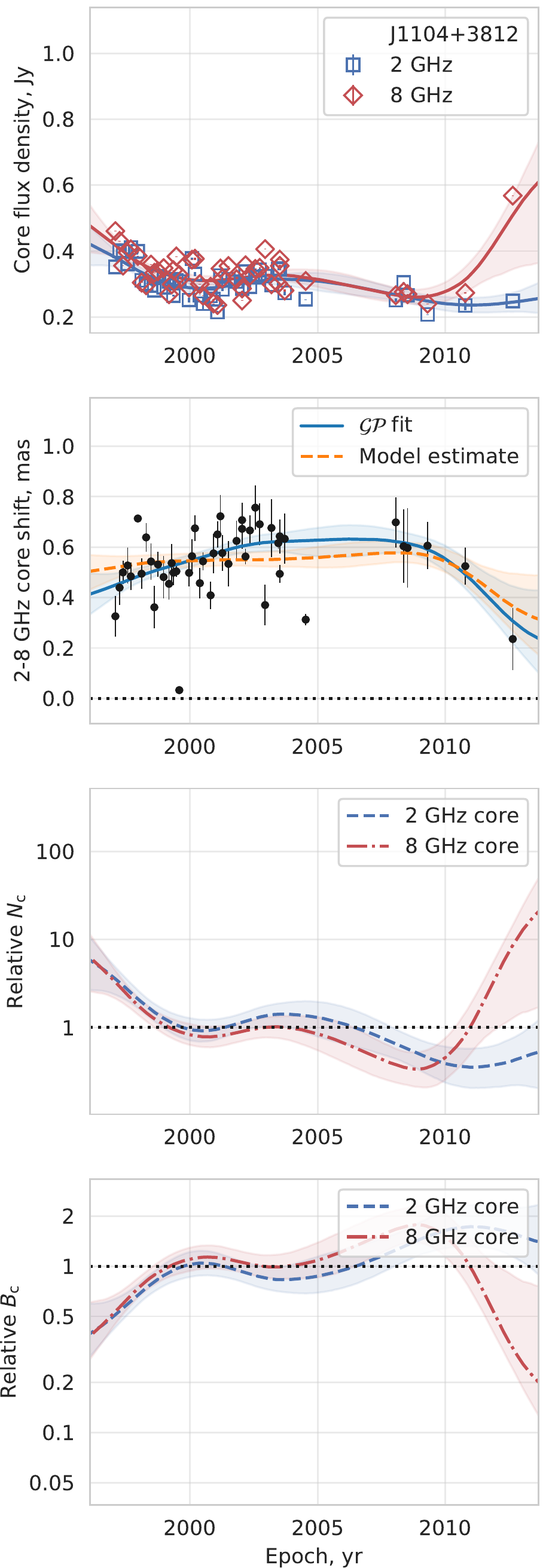}
\includegraphics[width=0.23\linewidth]{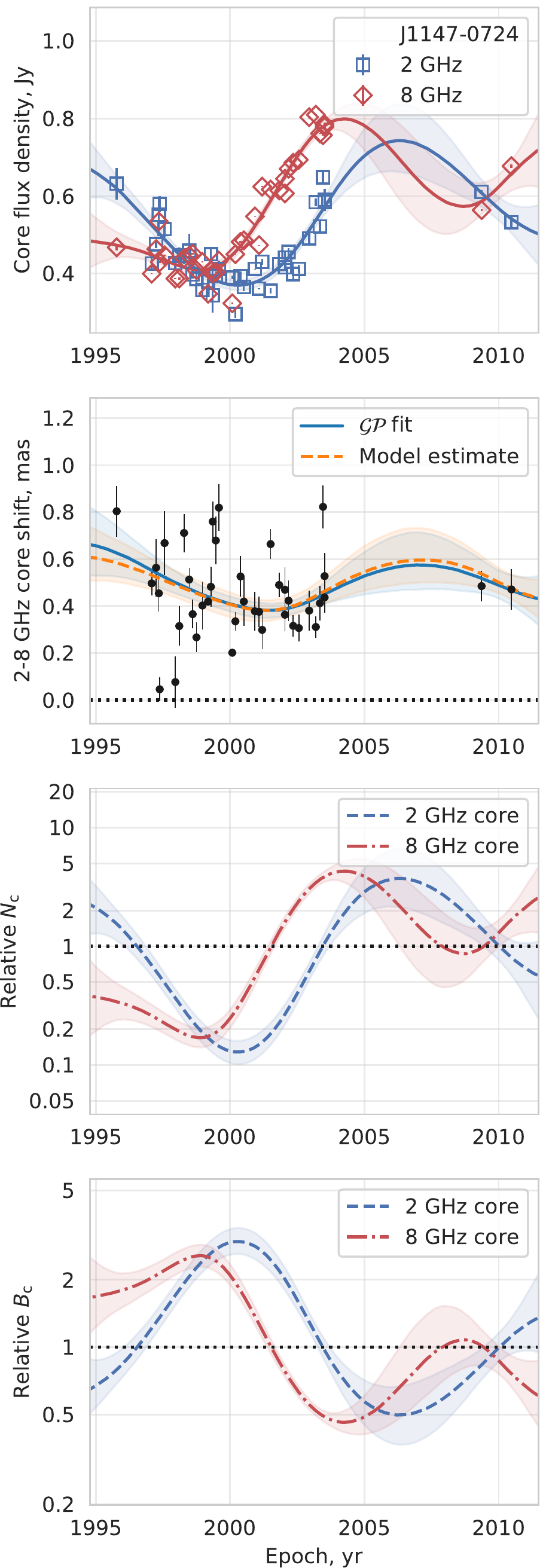}
\end{minipage}
\begin{minipage}{\textwidth}
\includegraphics[width=0.23\linewidth]{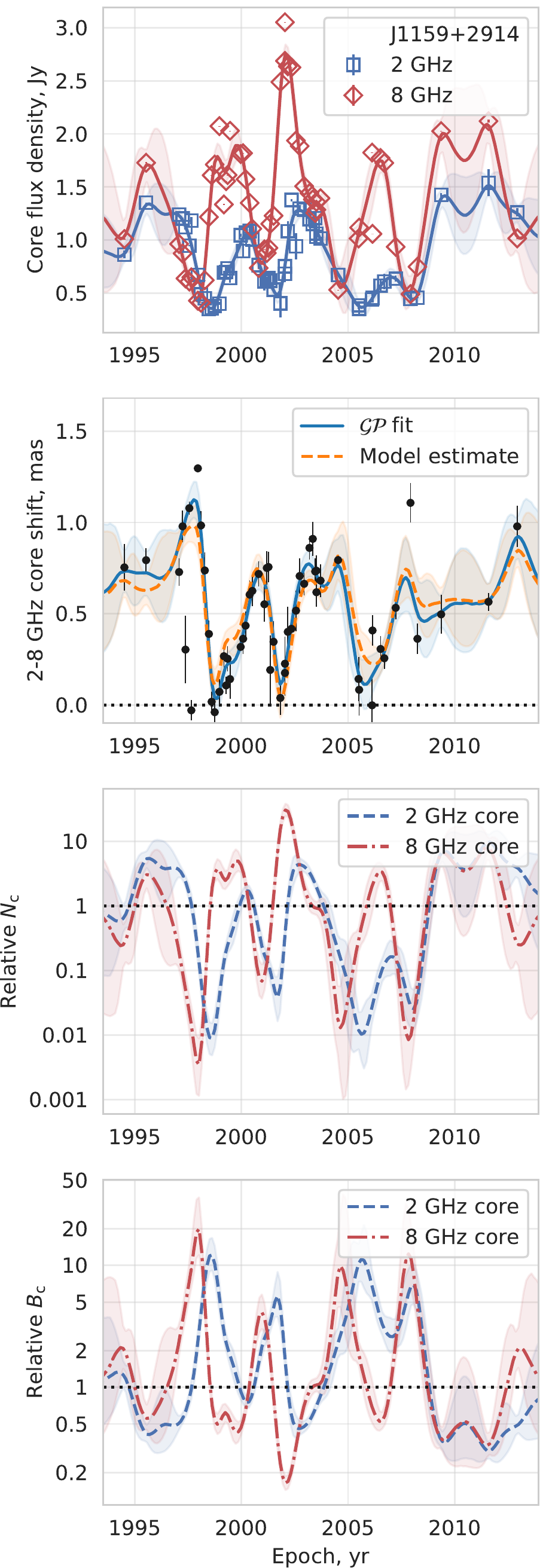}
\includegraphics[width=0.23\linewidth]{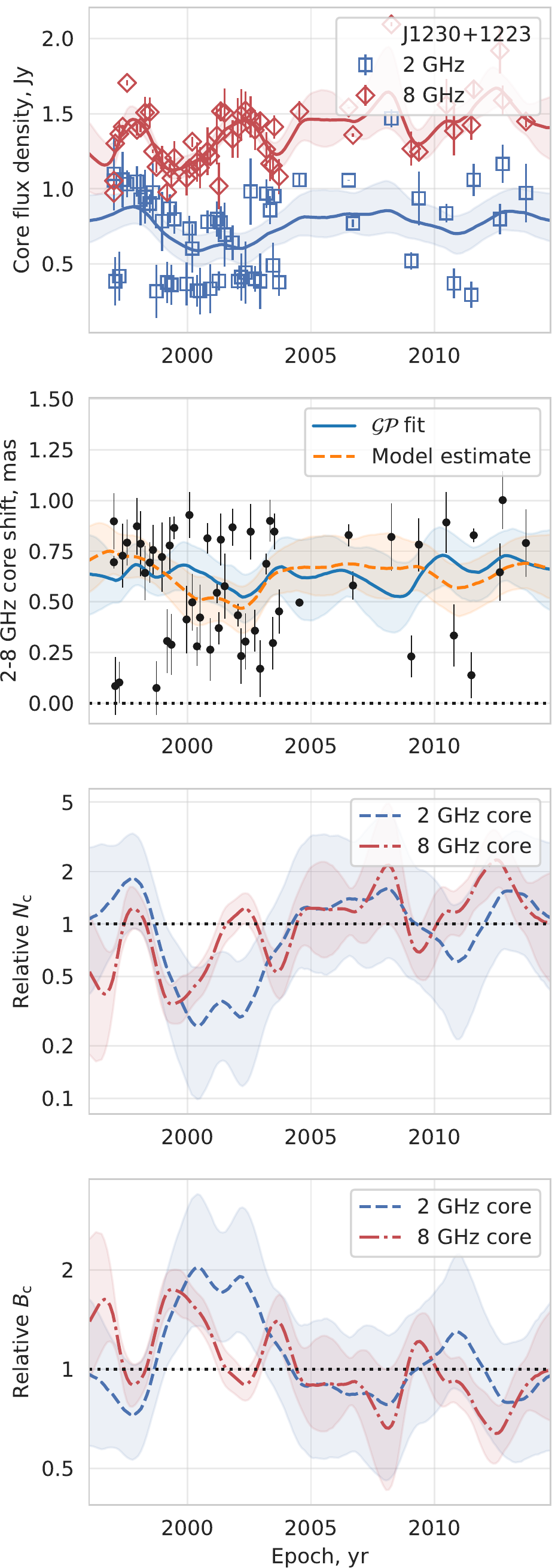}
\includegraphics[width=0.23\linewidth]{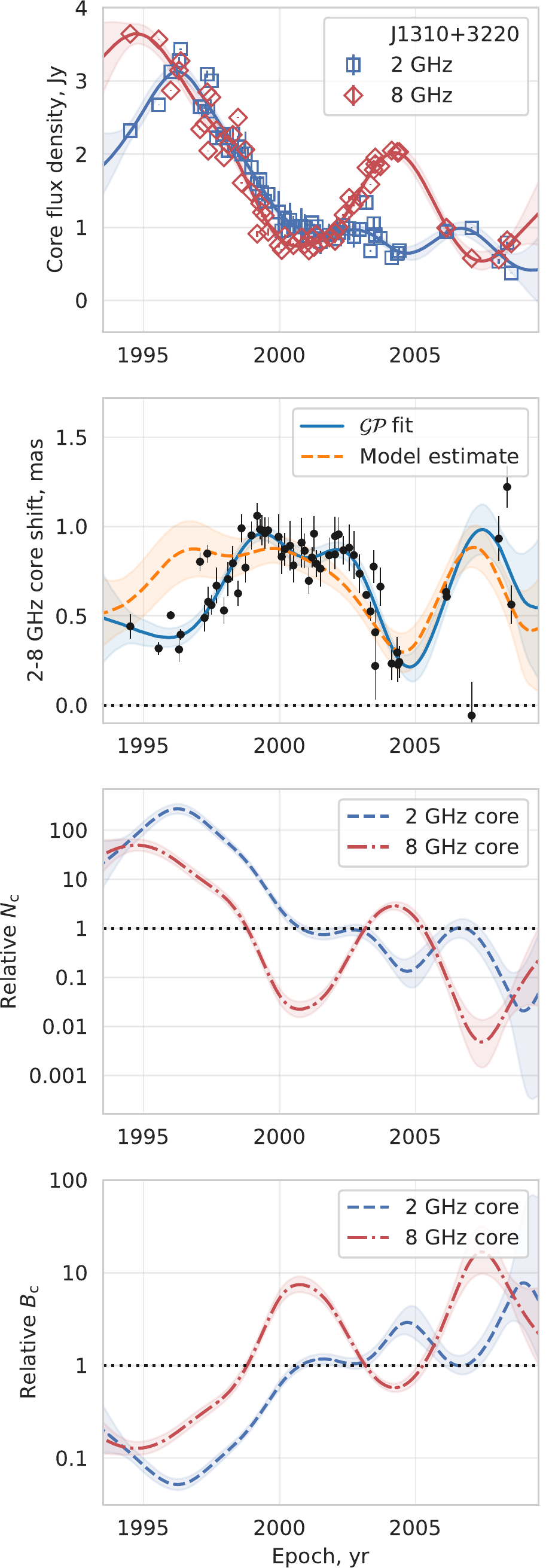}
\includegraphics[width=0.23\linewidth]{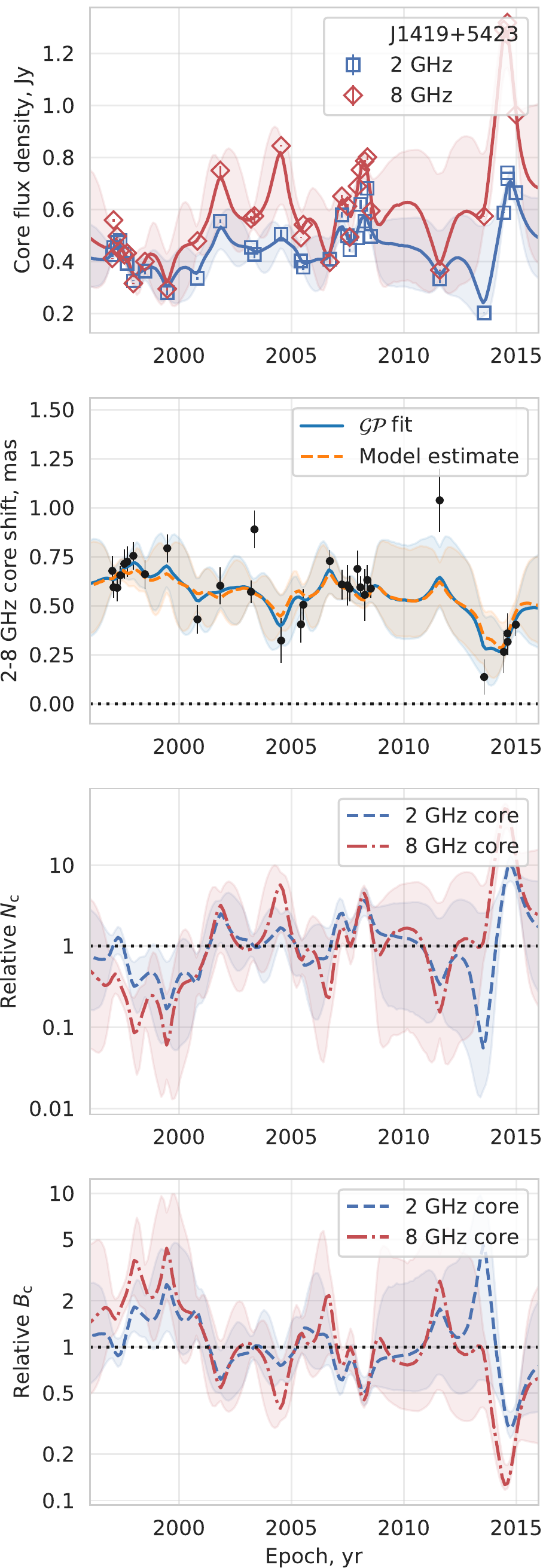}
\end{minipage}
\contcaption{}
\end{figure*}
\begin{figure*}
\centering
\begin{minipage}{\textwidth}
\includegraphics[width=0.23\linewidth]{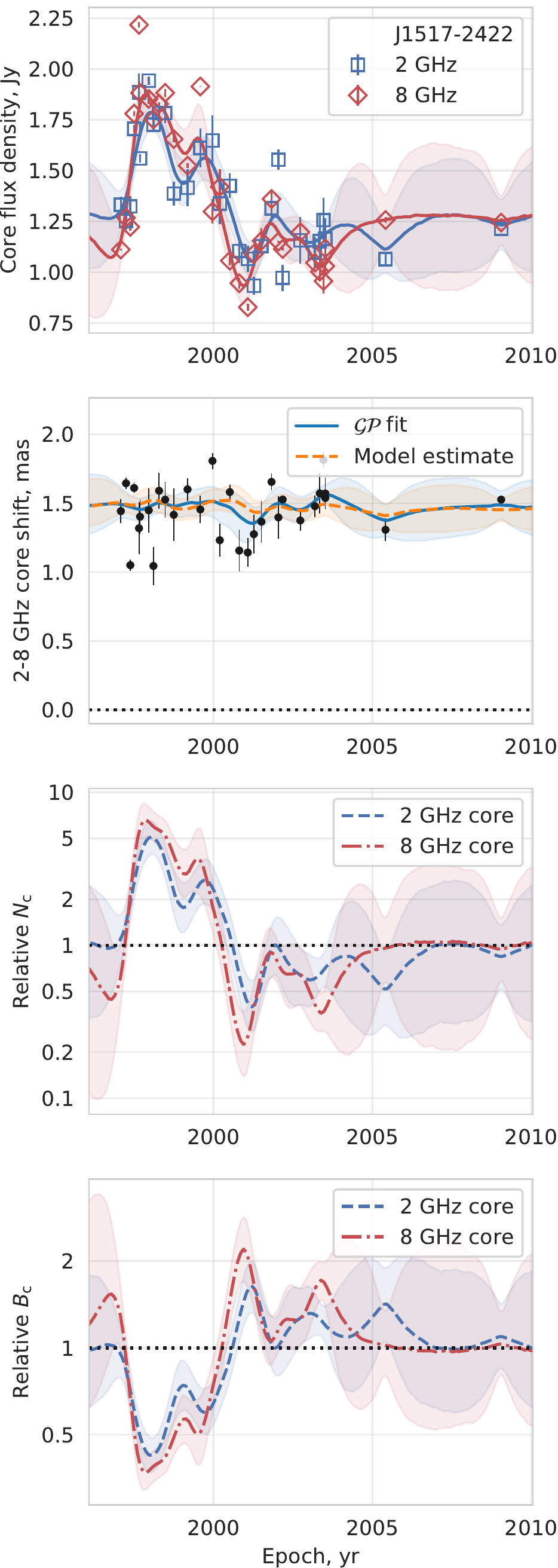}
\includegraphics[width=0.23\linewidth]{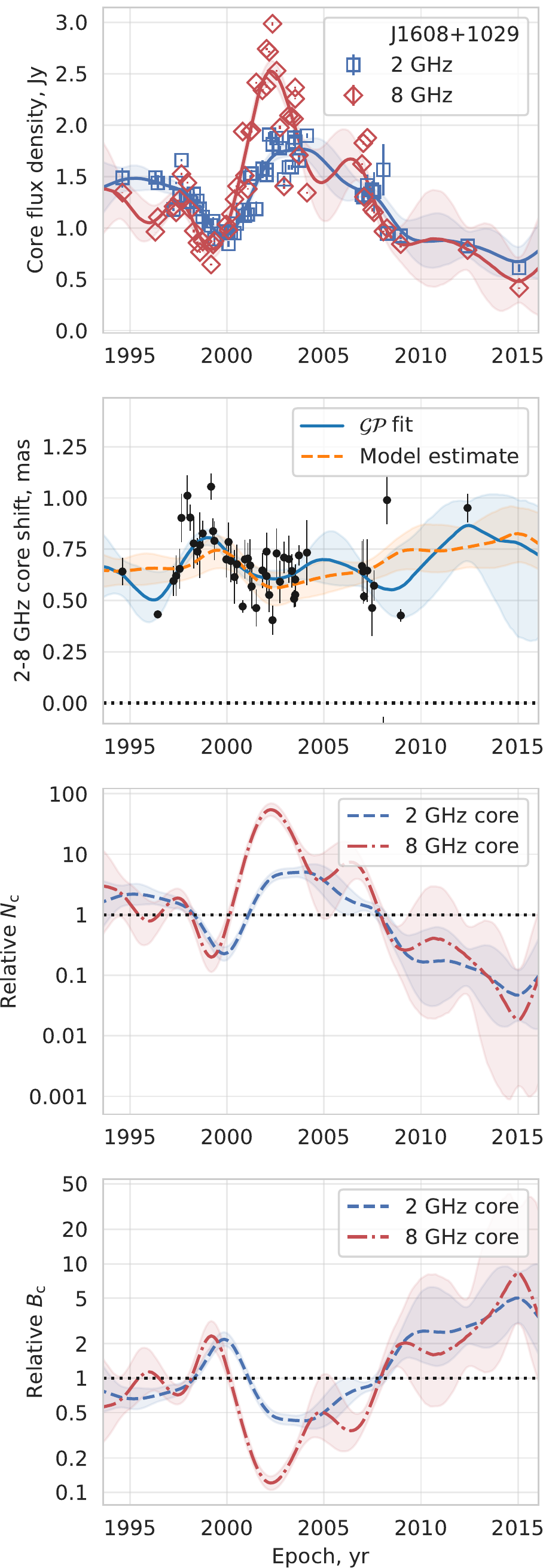}
\includegraphics[width=0.23\linewidth]{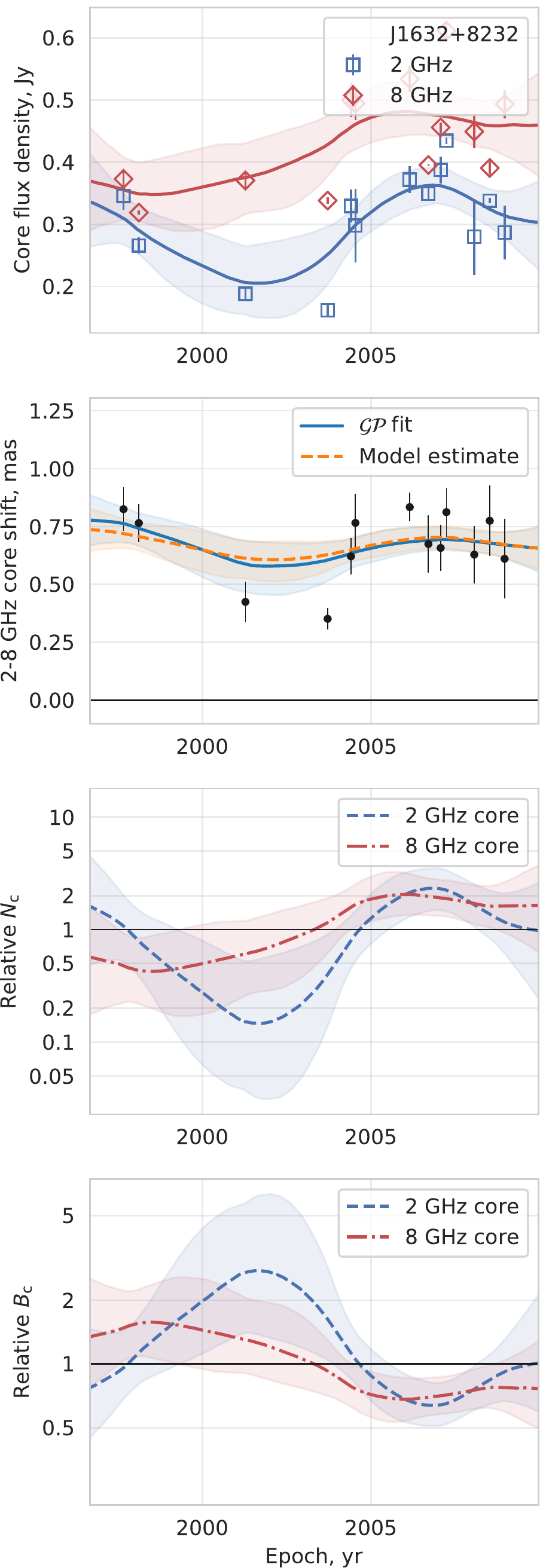}
\includegraphics[width=0.23\linewidth]{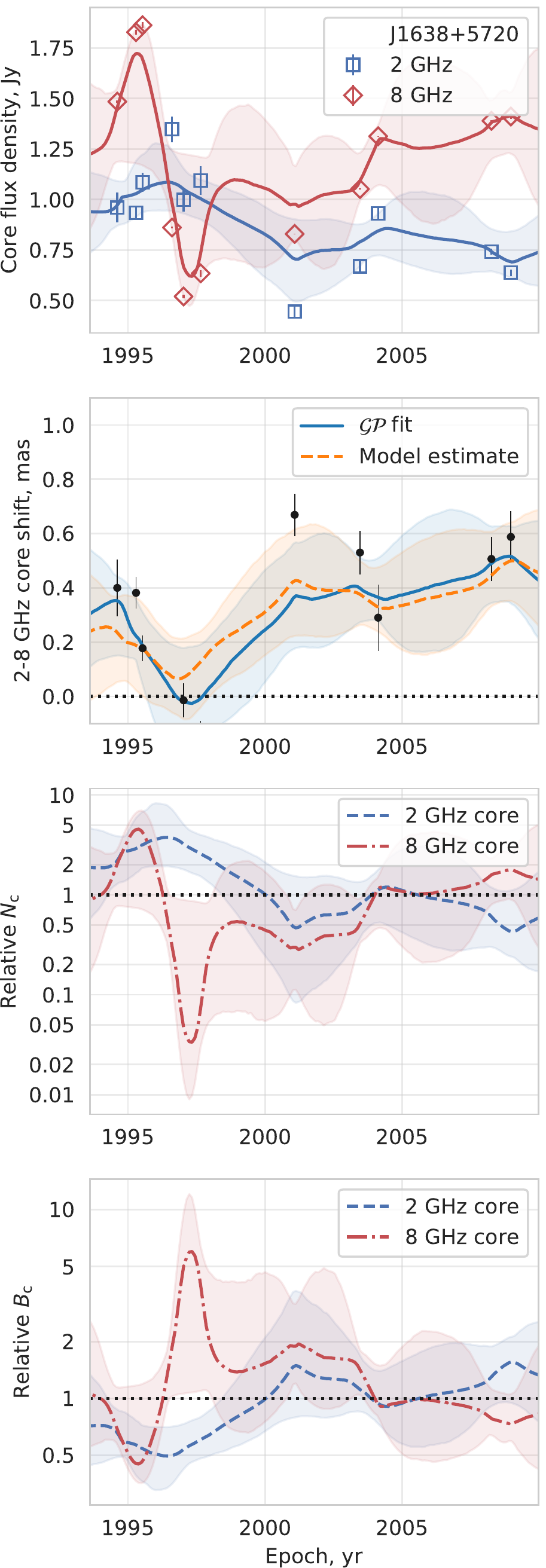}
\end{minipage}
\begin{minipage}{\textwidth}
\includegraphics[width=0.23\linewidth]{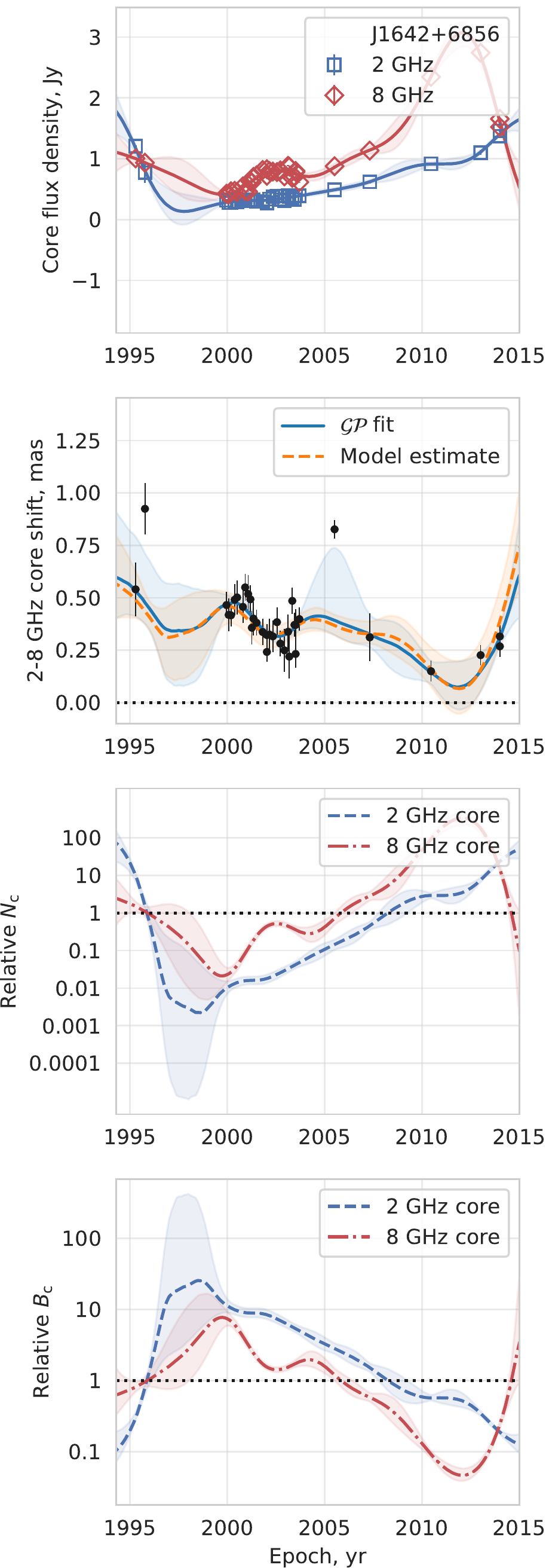}
\includegraphics[width=0.23\linewidth]{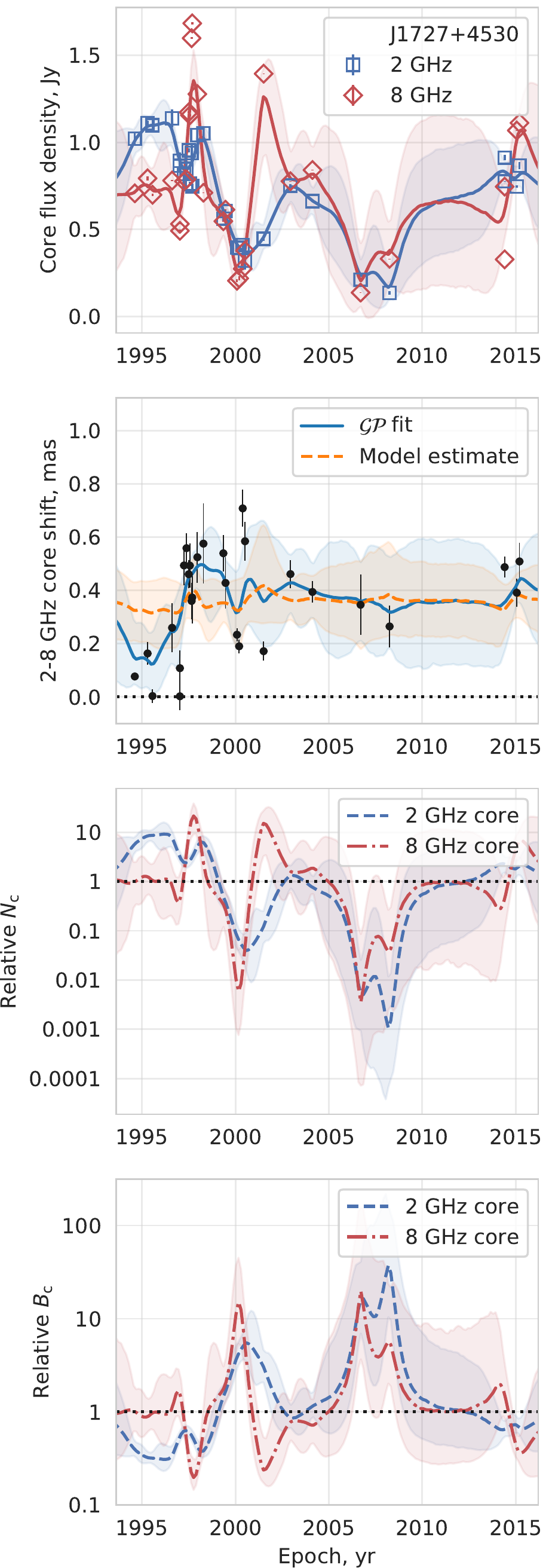}
\includegraphics[width=0.23\linewidth]{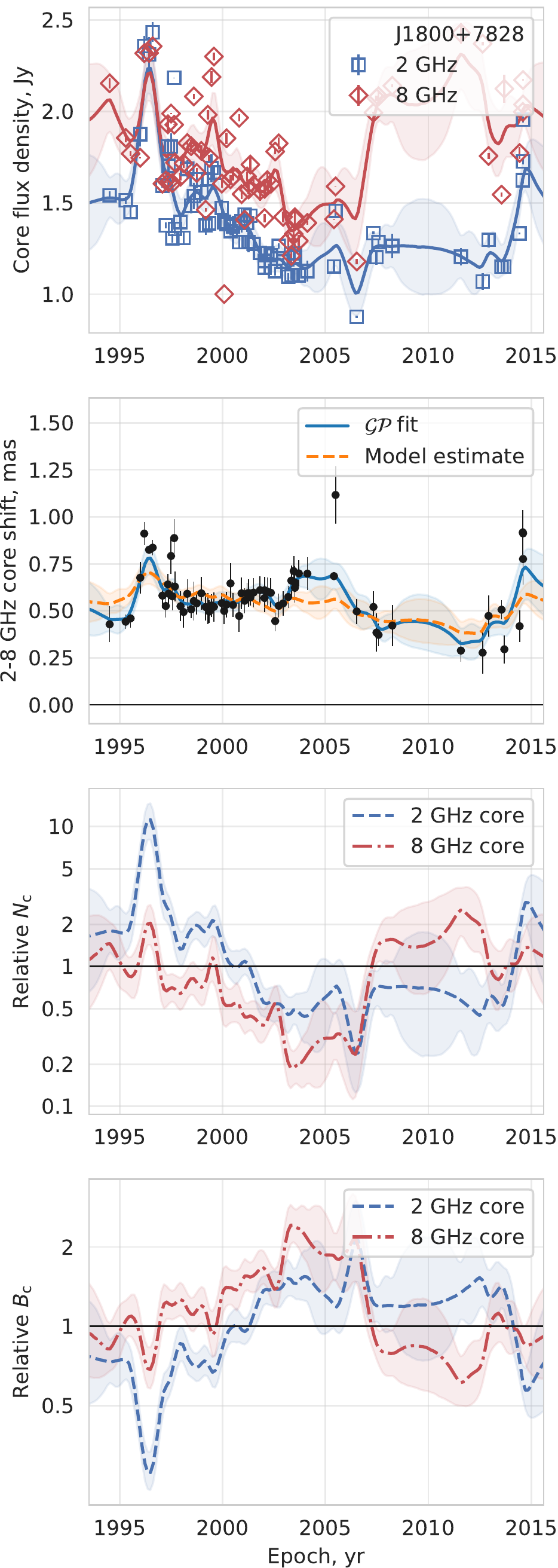}
\includegraphics[width=0.23\linewidth]{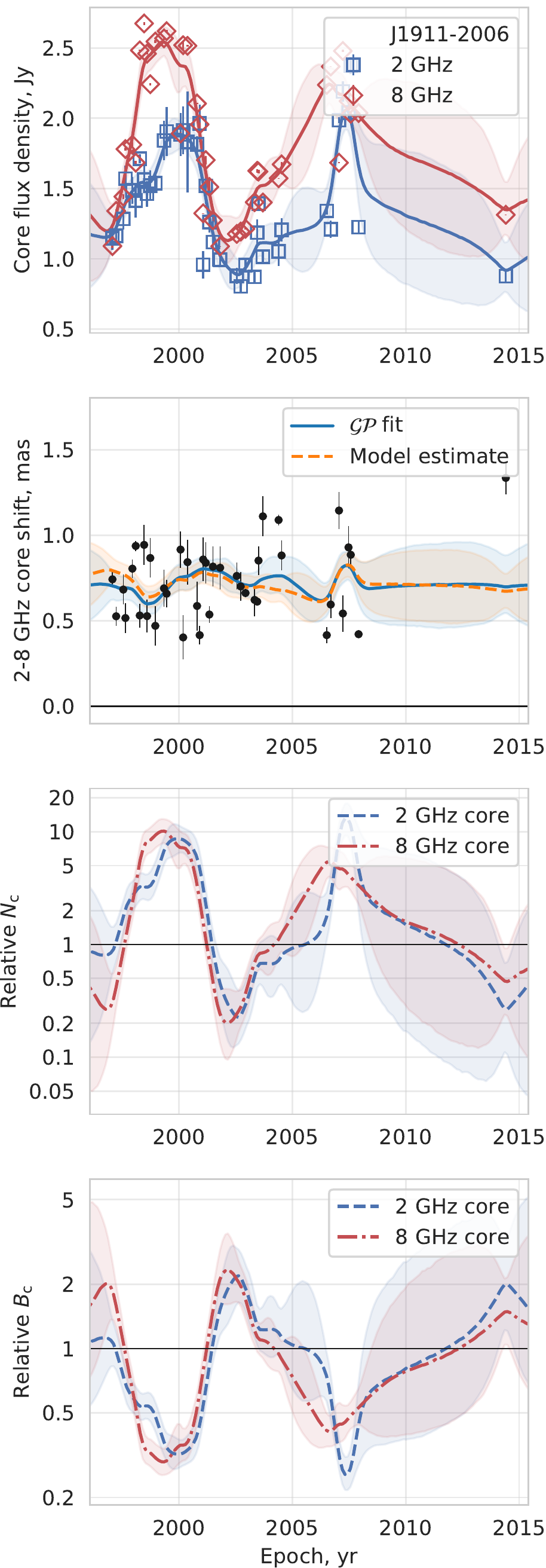}
\end{minipage}
\contcaption{}
\end{figure*}
\begin{figure*}
\centering
\begin{minipage}{\textwidth}
\includegraphics[width=0.23\linewidth]{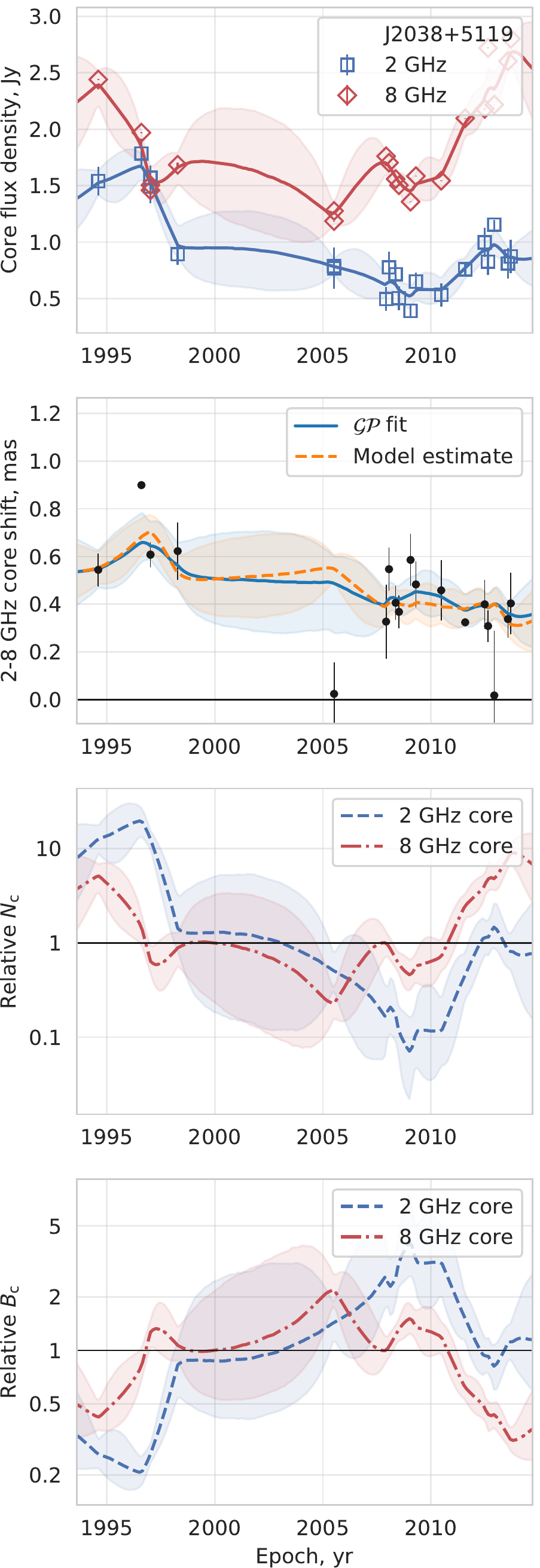}
\includegraphics[width=0.23\linewidth]{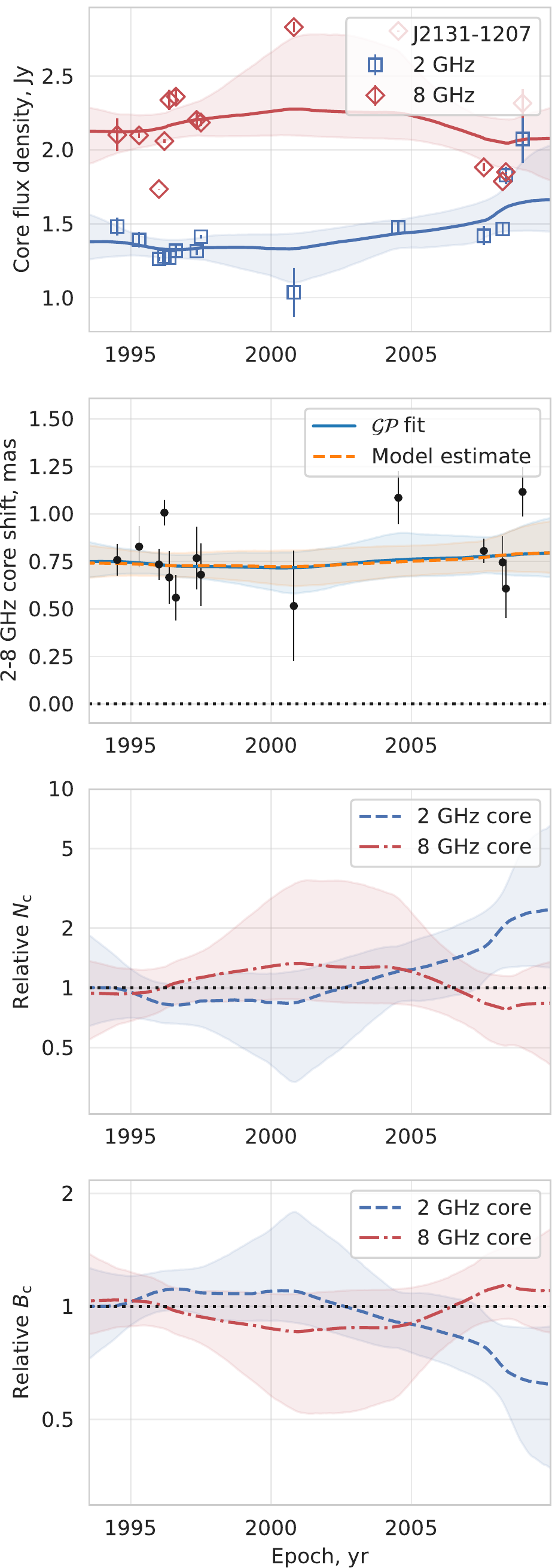}
\includegraphics[width=0.23\linewidth]{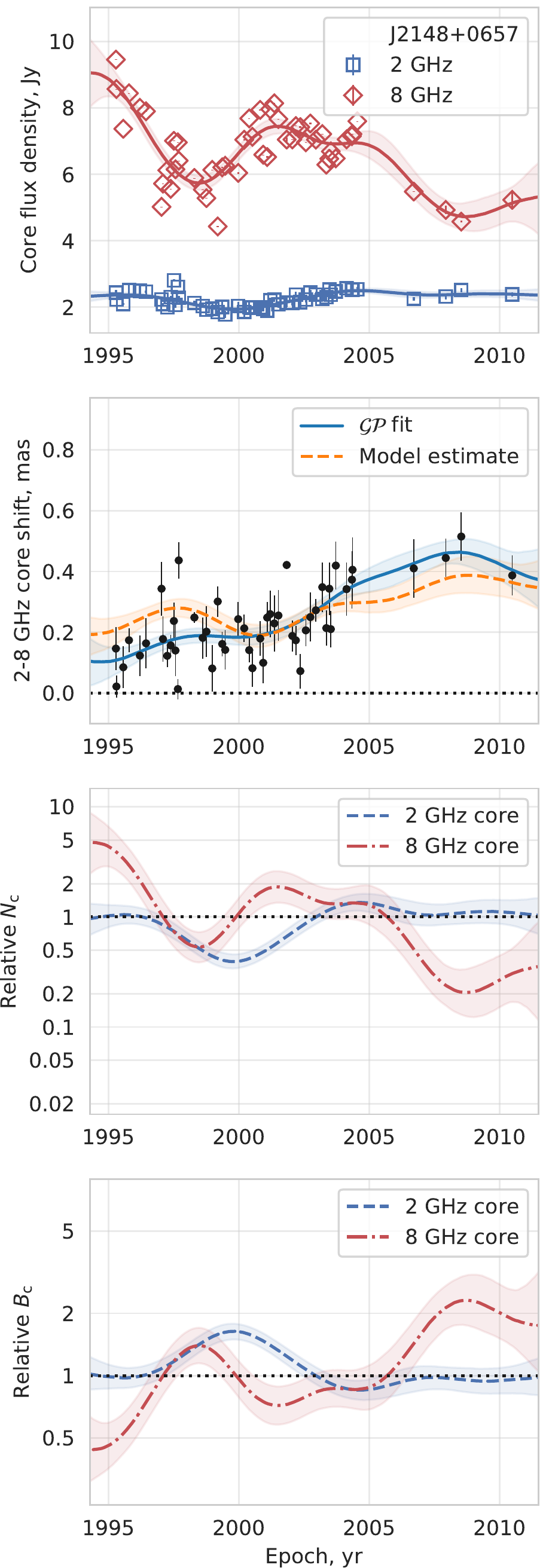}
\includegraphics[width=0.23\linewidth]{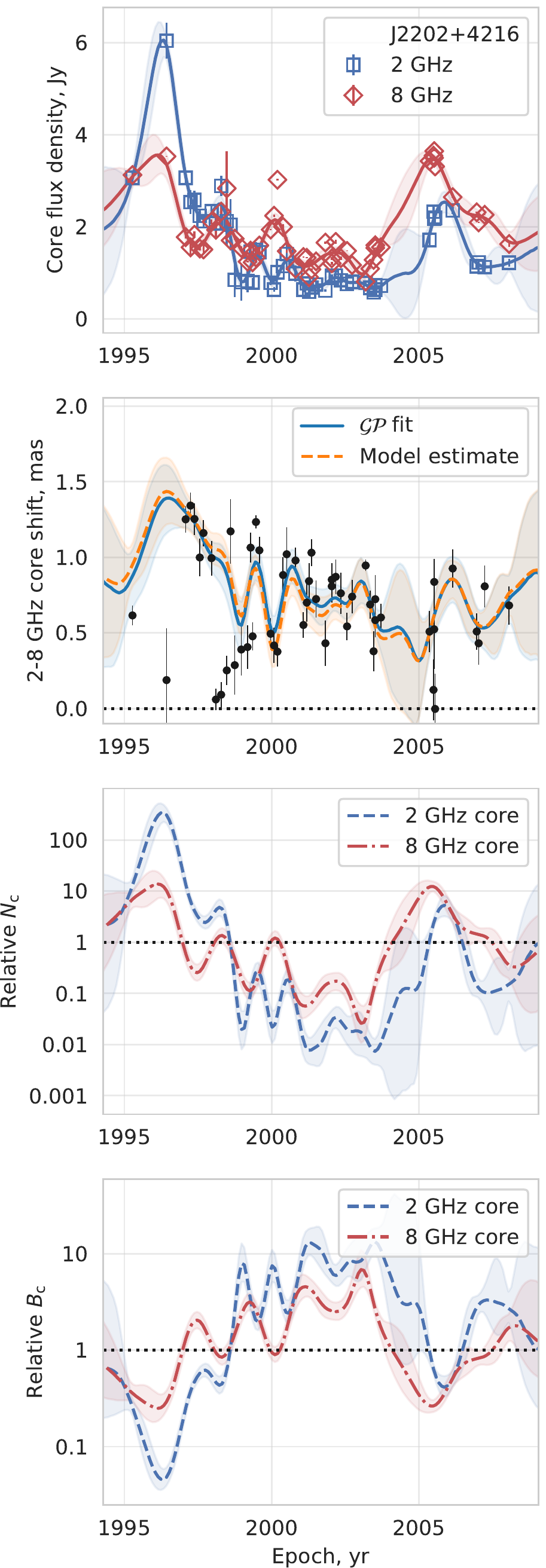}
\end{minipage}
\begin{minipage}{\textwidth}
\includegraphics[width=0.23\linewidth]{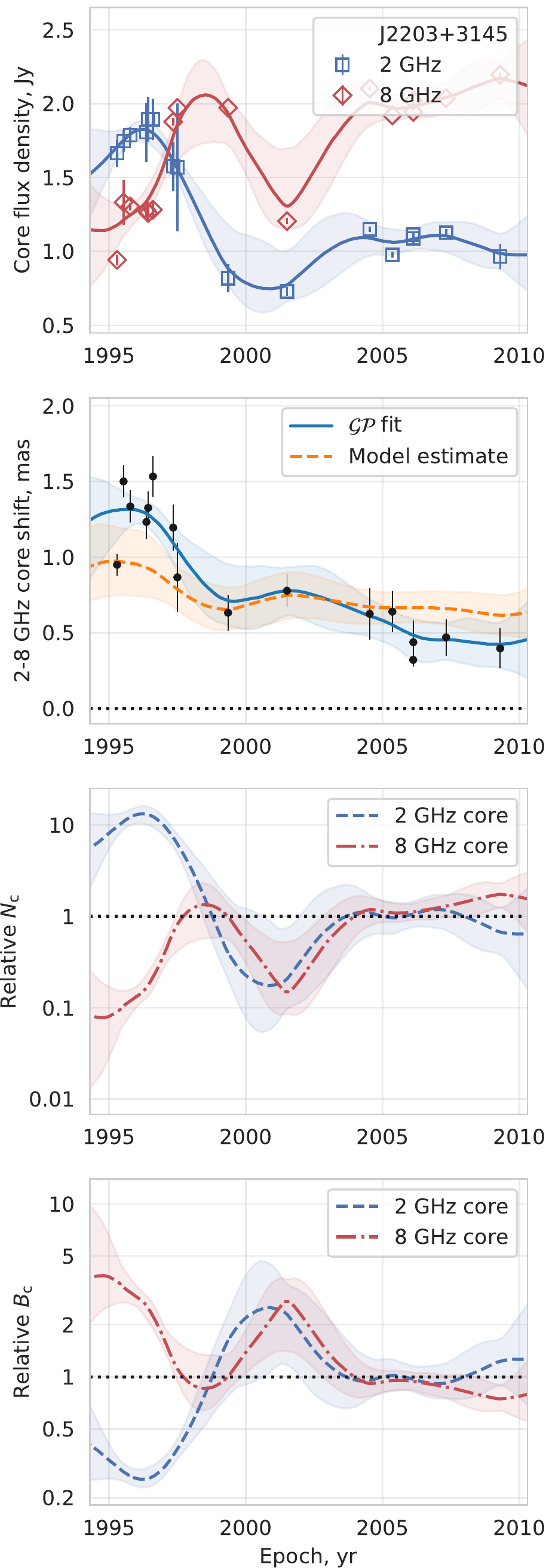}
\includegraphics[width=0.23\linewidth]{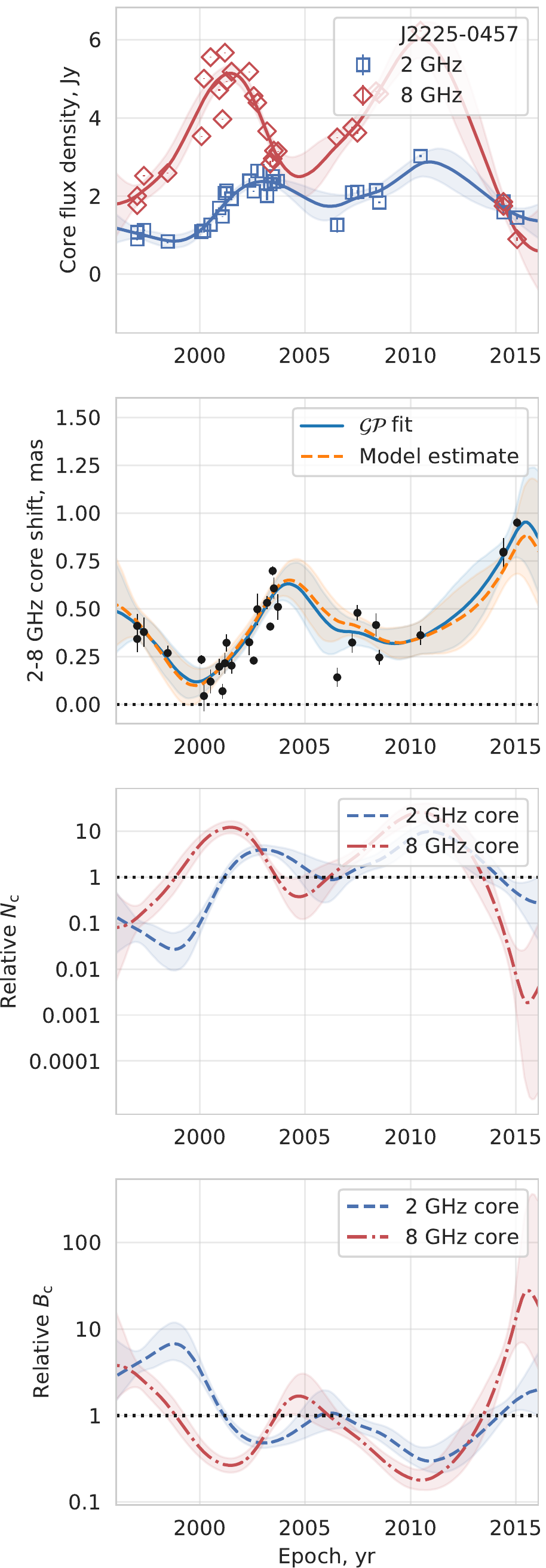}
\includegraphics[width=0.23\linewidth]{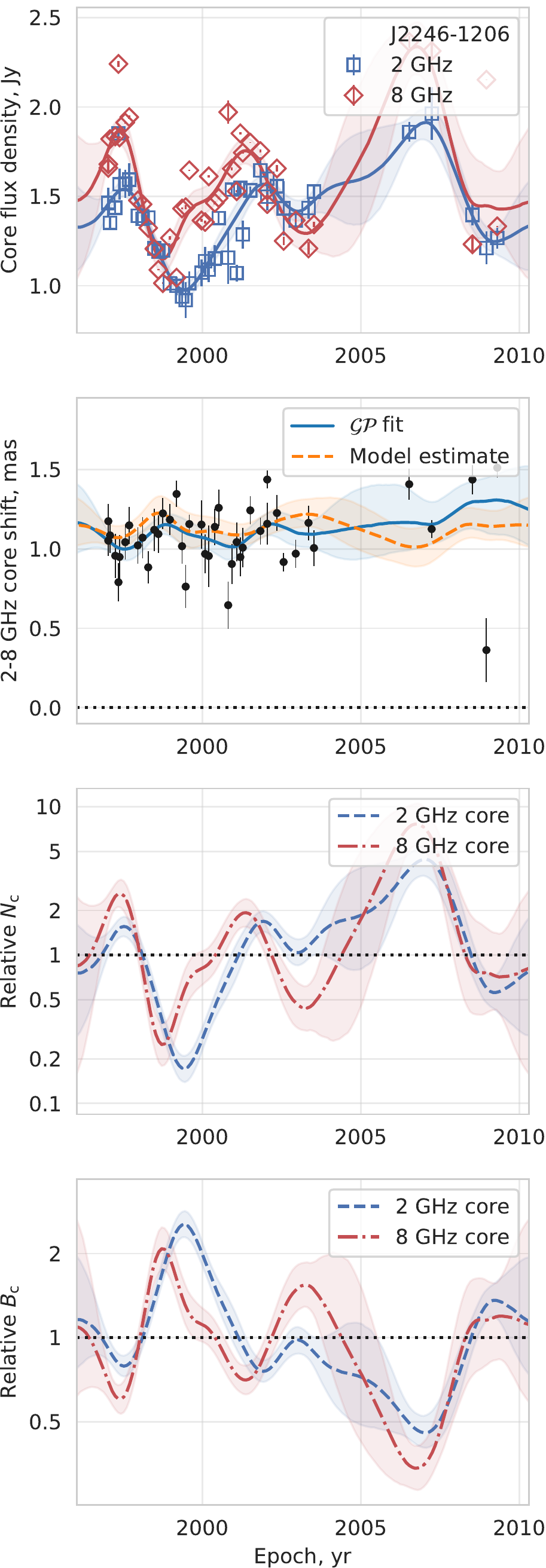}
\includegraphics[width=0.23\linewidth]{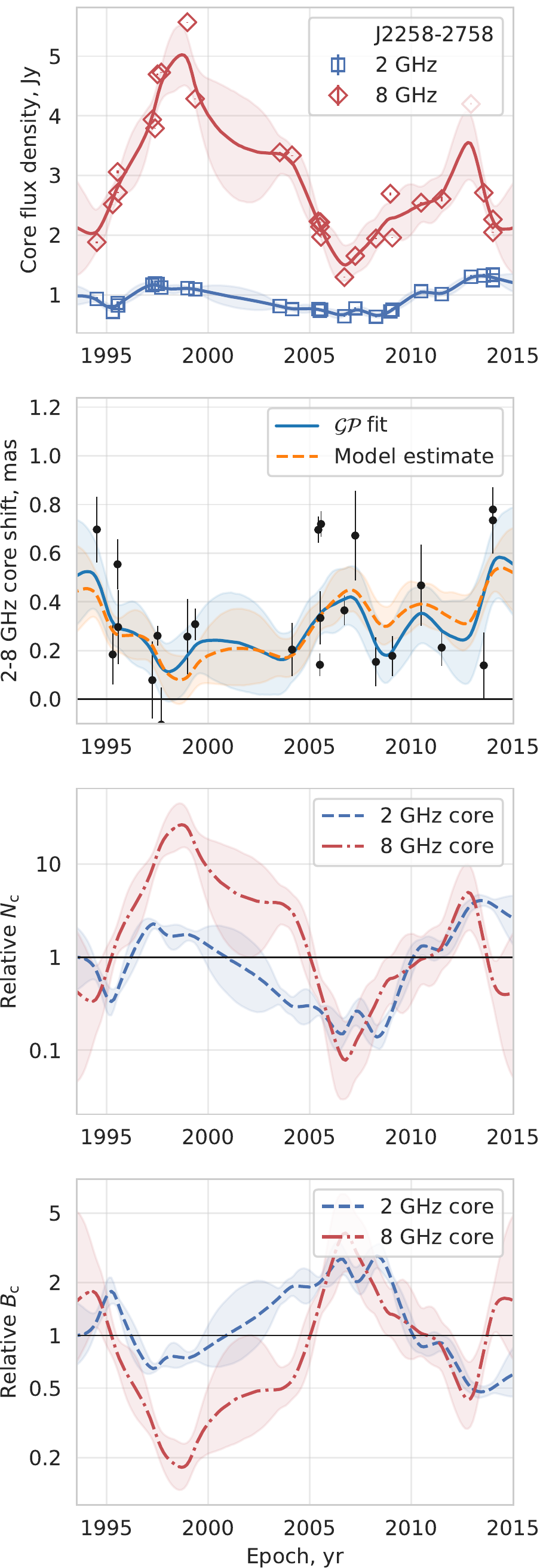}
\end{minipage}
\contcaption{}
\end{figure*}

\section{Discussion}
\label{s:discussion}

The detected significant variations of the measured core shifts have a number of implications for the physics of compact jets, the use of core shifts for estimating basic physical parameters of the flow, and for the use of compact radio sources as astrometric references. To understand these implication in a better detail, we first attempt to relate the observed variability of the core shift to nuclear flares in compact radio sources.

\subsection{Core shift and flux density changes in a flaring jet}
\label{s:jet_theory}

To model the effect of flaring activity in a jet on observed core shifts, we assume that the jet has a conical shape \citep{2017MNRAS.468.4992P} with a constant opening angle, where both magnetic field and particle density evolve as power laws:
\begin{equation}
\label{eq:BNr}
    B(r)\propto r^{-m}\quad \text{and} \quad N(r)\propto r^{-n},
\end{equation}
with the distance $r$ measured along the flow direction. We denote values at the fixed distance of $r=1$~pc from the jet base by $B_1$ and $N_1$, and values at the apparent core position at a specific frequency by $B_\C$ and $N_\C$.

During a flare the
jet opacity is affected by the transient plasma component responsible
for the flaring emission. This can cause variations in the apparent core
position $r_\C$ measured at different epochs at the same frequency and,
respectively, variations of the core shifts $\Delta\,r_\C$
measured between two given frequencies,
$\nu_1$ and $\nu_2$
($\nu_1<\nu_2$).

As the flaring material propagates
downstream, it first affects the core position observed at the
higher frequency and then acts, with a delay due to synchrotron opacity, on the core position
observed at the lower frequency. As a result, a non-trivial relation
can arise between the measured core shift and the variations of the
flux density in the core region. This relation can be modeled, to the
first order, by assuming that:
\begin{enumerate}
\item
The flare originates upstream from the opaque core observed at $\nu_2$ and it does not disrupt the jet.
\item
The downstream evolution of the magnetic field and particle density of the flaring component of the plasma is the same as the quiescent plasma of the jet. This implies that $N_\C$ and $B_\C$ of the flaring component also follow Eq.~\eqref{eq:BNr} when it passes along the jet up to the location of the core at the lower frequency $\nu_1$.
\item
Emission of the apparent core at each observing frequency $\nu_\text{obs}$ is well represented by a synchrotron spectrum with an
optically thin spectral index $\alpha$ and
the peak frequency $\nu_\text{max} \approx \nu_\text{obs}$. This is
similar to the condition required for the core in the conical jet
model \citep{1981ApJ...243..700K}.
\end{enumerate}

With these assumptions, variability of the core flux density would be
primarily caused by temporal variations of one of three basic
properties of the relativistic plasma flowing through the core region:
particle density, $N_\C$, magnetic field $B_\C$, or the bulk Doppler factor, $\delta_\C$. The resulting variable core flux density $S_\C \propto N_\C\,B_\C^{1-\alpha}\, \delta_\C^{3-\alpha}$ \citep{1987slrs.work..280M,1999ApJ...521..509L}.

To simplify further derivations, the power indices can be expressed in terms of
two new variables: $\epsilon = 3/2-\alpha$ and $\kappa = 1/(\epsilon\,m+n-1)$.
With these designations,
\begin{equation}
\label{eq:s_c}
S_\C \propto N_\C\,B_\C^{\epsilon-1/2}\, \delta_\C^{\epsilon+3/2}\,.
\end{equation}
Correspondingly, the observed location of the core, $r_\C$
\citep{1998A&A...330...79L}, will vary as
\begin{equation}
\label{eq:r_c}
r_\C \propto N_1^{\kappa}\, B_1^{\epsilon\,\kappa}\, \delta_\C^{\epsilon\,\kappa}\,.
\end{equation}
Here $N_1$ and $B_1$ are the flare particle density and magnetic field strength at a 1~pc deprojected distance from the jet origin. Relationships between $N_1, B_1$ and $N_\C, B_\C$ are provided by Eq.~\eqref{eq:BNr}.

Combining Eq.~\eqref{eq:s_c} and \eqref{eq:r_c}, one can consider various possible scenarios for the
flares, depending on the actual evolution of $N_\C, B_\C$, and $\delta_\C$. To assess different possible physical settings, four basic combinations of the jet parameters  can be considered:
\begin{enumerate}
    \item Flaring emission is in equipartition: $N_\C\propto B_\C^2$, $\delta_\C \approx const$.
    \item Flux density variability is dominated by changes in the particle density, with $N_\C=N_\C(t)$, $B_\C\approx const$, $\delta_\C \approx const$.
    \item Flux density variability is dominated by variations of the magnetic field, with $N_\C \approx const$, $B_\C=B_\C(t)$, $\delta_\C \approx const$.
    \item Flux density variability is caused by changes of the bulk speed of the plasma, with $N_\C \approx const$, $B_\C\approx const$, $\delta_\C = \delta_\C(t)$.
\end{enumerate}
Arguments have been made that variations of the particle density in the jet may offer the most plausible physical mechanism for the nuclear flares \citep{1999ApJ...521..509L,2008A&A...483..759K}.

To explore further the potential relations between these jet parameters, we consider a more general case with an arbitrary power-law relationship $B_\C(t) \sim N_\C(t)^{K_{\mathrm{BN}}}$ while $\delta_\C$ is held constant. This trivially includes cases (i) with $K_{\mathrm{BN}} = 1/2$, (ii) with $K_{\mathrm{BN}} = 0$, and (iii) with $K_{\mathrm{BN}} \to \infty$. We substitute this into Eq.~\eqref{eq:s_c}, \eqref{eq:r_c} and have
\begin{equation}
\label{eq:srn1}
S_\C \sim N_\C^{1 + (\epsilon - 1/2) K_{\mathrm{BN}}},\quad r_\C \sim N_\C^{-1-\epsilon K_{\mathrm{BN}}},
\end{equation}
which implies the following relationship between core flux density and its position:
\begin{equation}
\label{eq:rs_c1}
r_\C \sim S_\C^{-\frac{1 + \epsilon K_{\mathrm{BN}}}{1 + (\epsilon - 1/2)K_{\mathrm{BN}}}} \equiv S_\C^{K_{\mathrm{rS}}}.
\end{equation}
To derive specific relations between the magnitude of core shift and flux density variations, we assume that the exponents are $m=1$, $n=2$ and the optically thin spectral index takes its typical value $\alpha=-0.5$ \citep[e.g.][]{2014A&A...566A..59A}, so that $\epsilon = 2$ and  $\kappa = 1/3$.

Under these assumptions, the core flux density and its position can be related as $r_\C = b S_\C^{K_{\mathrm{rS}}}$, with some constant, $b$. Then, for core shift measured between two frequencies, $\nu_1$ and $\nu_2$, it follows that
\begin{equation}
\Delta r_\C \equiv r_{\C1} - r_{\C2} = b_1 S_{\C1}^{K_{\mathrm{rS}}} - b_2 S_{\C2}^{K_{\mathrm{rS}}}\,,
\end{equation}
where the subscripts "1" and "2" refer to measurements made at the respective frequency. If we fit this model to our measurements, hence finding the values of $b_1$, $b_2$, and $K_{\mathrm{rS}}$, we can effectively determine how the positions of cores vary at both frequencies, and how the corresponding jet parameters change.

Note that $r_\C = b S_\C^{K_{\mathrm{rS}}}$ can be considered a phenomenological model independent of the physical assumptions made above. This means our estimates of individual core positions and their variability do not require the particular assumptions about the nature of jet flares.

\subsubsection{General relation between core shift and flux density}

In order to robustly determine the $\Delta r_\C(S_\C)$ dependence the $\Delta r_\C = b_1 S_{\C1}^{K_{\mathrm{rS}}} - b_2 S_{\C2}^{K_{\mathrm{rS}}}$ needs to be modified so as to account for possible bias in $\Delta r_\C$ measurements. We achieve this by adding a constant offset $a$ and including a linear term, $c\,r_\text{beam}$, which describes the effect of variable beam size in different observations (see Eq.~\eqref{eq:beambias}):
\begin{equation}
\label{eq:individual_model}
\Delta r_\C = a + b_1 S_{\C1}^{K_{\mathrm{rS}}} - b_2 S_{\C2}^{K_{\mathrm{rS}}} + c r_\text{beam}.
\end{equation}
The addition of the linear term also accounts for possible effects of different ($u$,$v$)-coverage at different epochs. With these modifications, we still can identify the variability of individual core positions, while making predictions about their absolute locations with respect to the jet origin would require further model assumptions.

We fit this model to all of the data, using a single value of exponent $K_{\mathrm{rS}}$ for all sources, thus implicitly assuming that the underlying physical process governing the core shift variations acts in a similar way in all objects. The remaining model parameters $a$, $b_1$ and $b_2$ are fit individually for each source. The fitting procedure yields $K_{\mathrm{rS}} = 0.23 \pm 0.03$ for the entire sample. We then verify that excluding the $c\,r_\text{beam}$ term from the fitting does not affect estimate of $K_{\mathrm{rS}}$ significantly. This indicates that while the core shift measurements are biased upwards higher values for larger beam sizes, the core flux and position relationship stays the same and hence the observed core shift variability is a real effect. When the $c\,r_\text{beam}$ term is included, the estimated value for $c$ is consistent with $0.12$~mas we give in Eq.~\eqref{eq:beambias}.

\subsubsection{Time dependent model of $r_\C(S_\C)$ relation}
\label{s:fit_gp}

We consider here a more general model, which accounts for time dependent deviations from the expected relationship $r_\C \sim S_\C^{K_{\mathrm{rS}}}$.
To address time dependence of the core shifts, we model the difference between the model-based core shift values from Eq.~\eqref{eq:individual_model} and their measurement representation by Gaussian processes \citep{rasmussen2005gaussian}. We model flux densities at both bands using Gaussian processes as well, which also allows us to calculate temporal derivatives of all of the modeled parameters.

This results in the following description of flux density and core shift variations in a single source:
\begin{equation}
\label{eq:timeseries_model}
\begin{split}
S_{\C1}(t) &\sim \mathcal{GP}(\mu_1, \beta_1 \text{RatQuad}(T, \alpha)),\\
S_{\C2}(t) &\sim \mathcal{GP}(\mu_2, \beta_2 \text{RatQuad}(T, \alpha)),\\
\Delta r_\C(t) &\sim a + b_1 S_{\C1}(t)^{K_{\mathrm{rS}}} - b_2 S_{\C2}(t)^{K_{\mathrm{rS}}} + c r_\text{beam} +\\&+ \mathcal{GP}(0, \beta_3 \text{RatQuad}(T, \alpha)),
\end{split}
\end{equation}
where $\mathcal{GP}(\mu, \text{cov})$ is a Gaussian process with a mean  $\mu$ and a covariance function $\text{cov}(t_1, t_2)$. The term $\text{RatQuad}(T, \alpha)$ represents a rational quadratic covariance function with a timescale $T$, i.e.\ $\text{cov}(t_1, t_2) = (1 + (t_1 - t_2)^2 / (2 \alpha T^2))^{-\alpha}$. The terms $\beta_1$ and $\beta_2$ provide magnitude scaling for respective covariance functions and are treated as free parameters.

We assume that individual measurement errors, i.e.\ difference between our $i$-th measurement $S_{\C1}^i, S_{\C2}^i, \Delta r_{\C}^i$ and the corresponding modeled value at time $t^i$ follow the Student $t$ distribution so that
\begin{equation}
\label{eq:tsmodel_err}
\begin{split}
S_{\C1}^i - S_{\C1}(t^i) \sim t(0, \sigma_1, \nu_S),\\
S_{\C2}^i - S_{\C2}(t^i) \sim t(0, \sigma_2, \nu_S),\\
\Delta r_{\C}^i - \Delta r_{\C}(t^i) \sim t(0, \sigma_3, \nu_r),
\end{split}
\end{equation}
where $t(\mu, \sigma, \nu)$ is the $t$-distribution with a mean $\mu$, scale $\sigma$, and $\nu$ degrees of freedom. This description is also used to account for possible outlier measurements.

In the combined model applied to the entire data, different values of $\mu_{1,2}, \beta_{1,2,3}, T, a, b_{1,2}, \sigma_{1,2,3}$ are allowed for each source, and a single value of the parameters $\alpha, K_{\mathrm{rS}}, c, \nu_S, \nu_r$ is used for the entire sample.

\begin{figure}
    \centering
    \includegraphics[width=\columnwidth,trim=0cm 1cm 0cm 0.3cm]{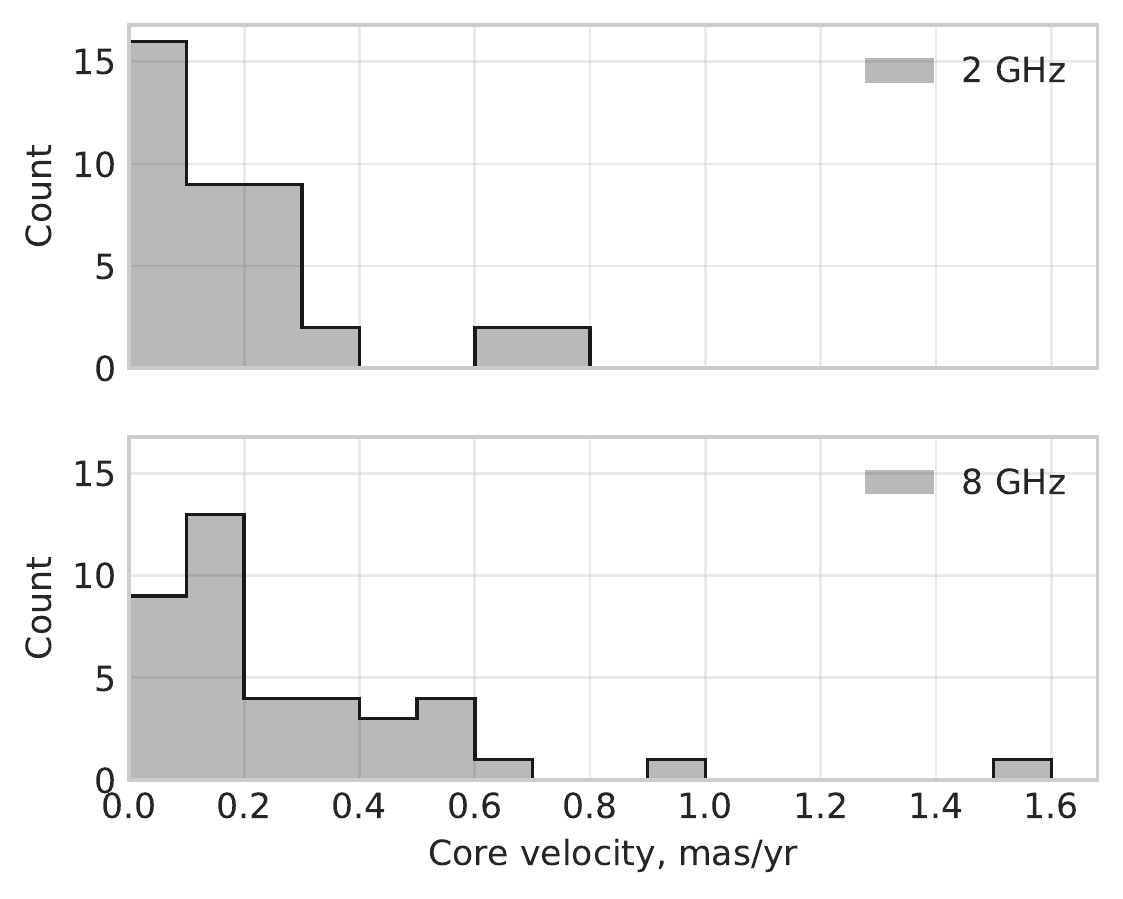}
    \caption{Histogram of maximal projected core velocity per source and band in angular units estimated from the core shift variability modeling.}
    \label{f:core_velocity_dist}
\end{figure}

\begin{figure}
    \centering
    \includegraphics[width=\columnwidth,trim=0cm 0cm 0cm 0.2cm]{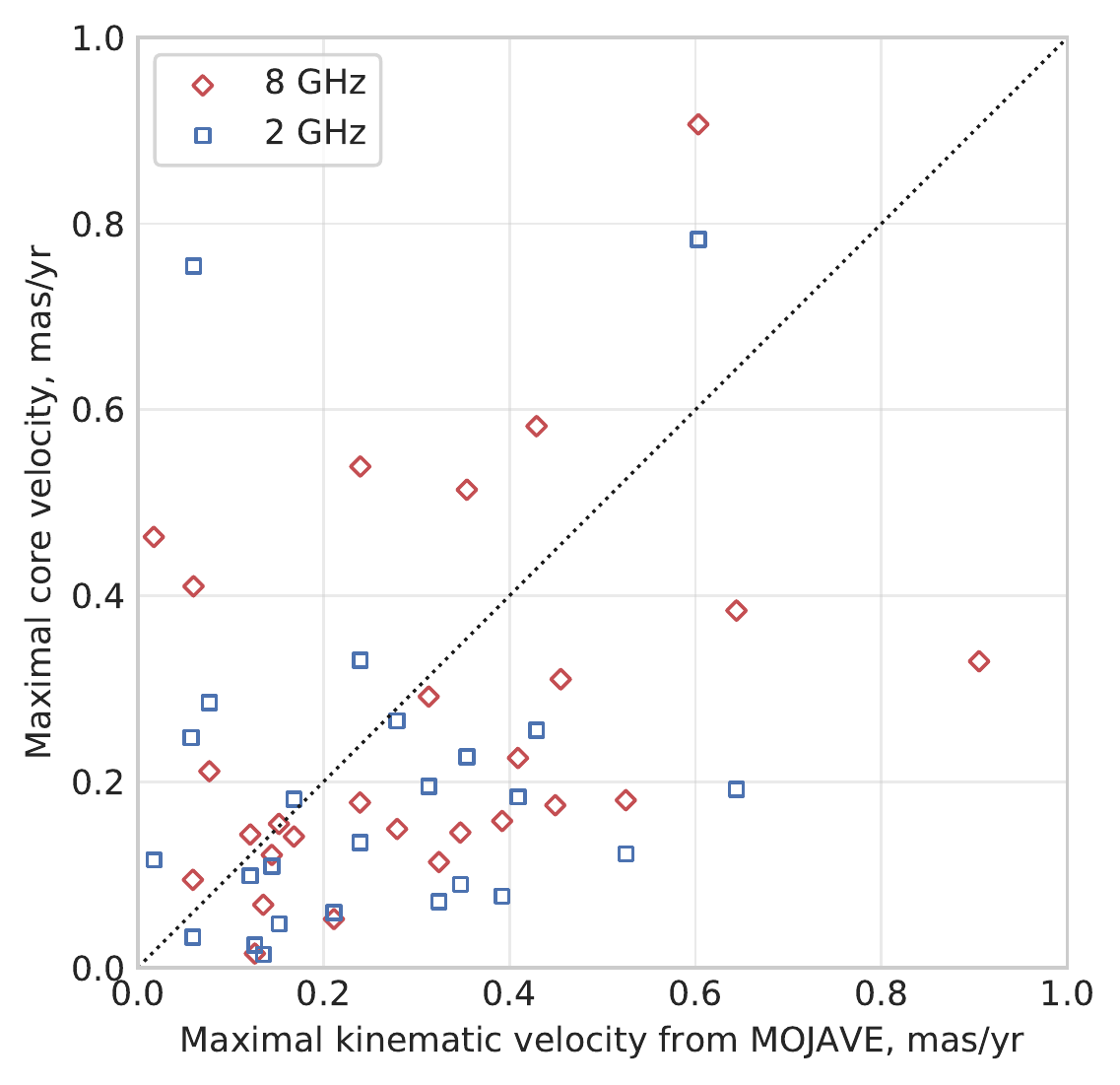}
    \includegraphics[width=\columnwidth,trim=0cm 1cm 0cm 0cm]{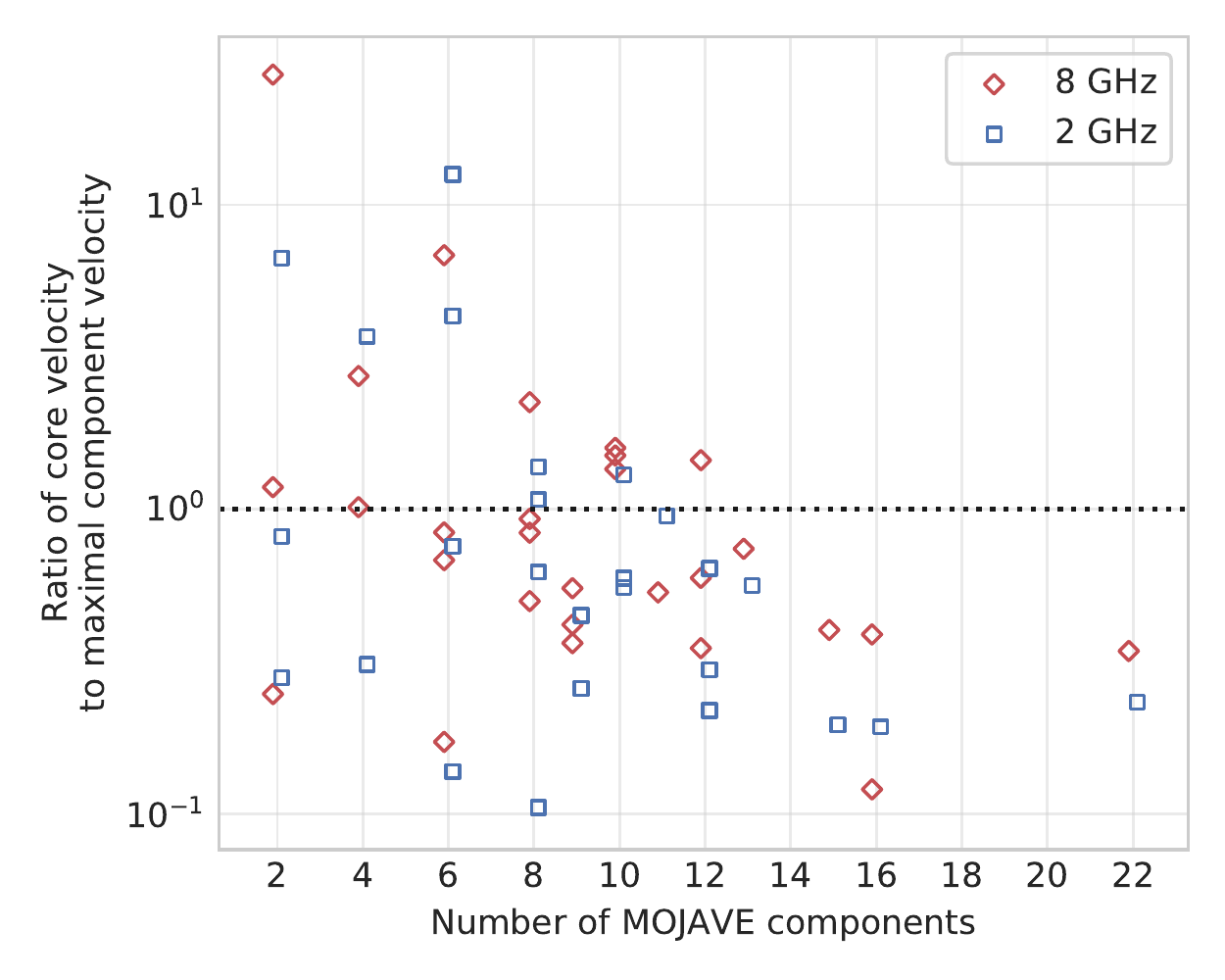}
    \caption{Comparison of per-source maximal projected core velocity with MOJAVE kinematic measurements \citep{2013AJ....146..120L,2016AJ....152...12L}.}
    \label{f:core_velocity_mojave}
\end{figure}

\begin{figure}
    \centering
    \includegraphics[width=\columnwidth,trim=0cm 1cm 0cm 0.2cm]{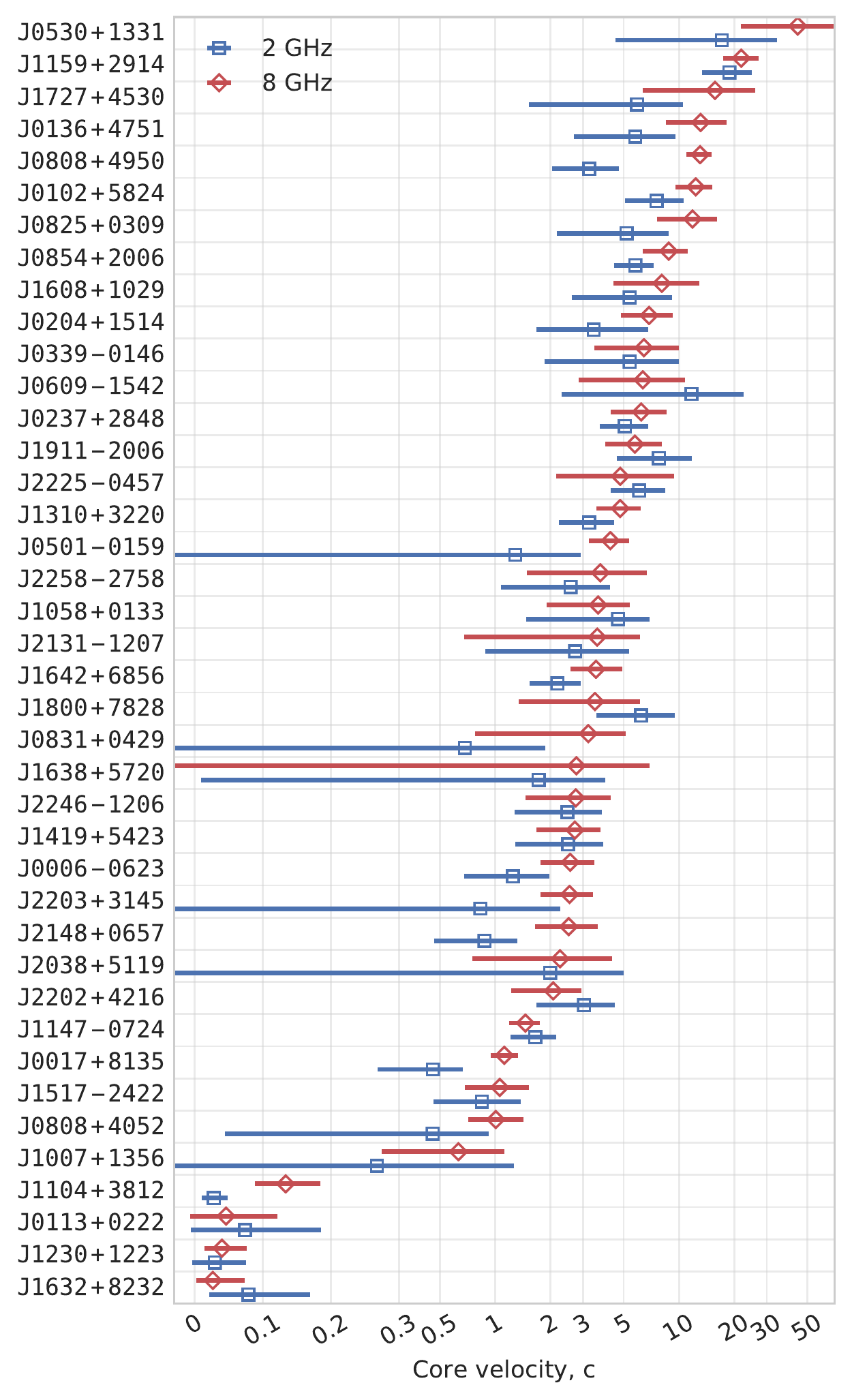}
    \caption{Maximal projected apparent core velocity for each object and band separately in units of speed of light. Medians and 68\% credible intervals are shown, sources are sorted by their core velocity at 8~GHz.}
    \label{f:core_velocity}
\end{figure}

We fit this model using the NUTS Markov Chain Monte Carlo sampler as implemented in \textsc{PyMC3} \citep{10.7717/peerj-cs.55}, which automatically accounts for uncertainties of all the parameters in further inferences. As a result of this fitting, we obtain $K_{\mathrm{rS}} = 0.28 \pm 0.05$, which is consistent with the model described in the previous subsection. The estimates of $b_{1,2}, c$ coefficients for each  individual source are also consistent between the two models. Typical values of $b_{1,2}$ are clustered around $1 \text{ mas/Jy}^{K_{\mathrm{rS}}}$. In 33 of 40 sources, we find at least one of the $b_{1,2}$ coefficients to be significantly greater than zero. For 26 of 40 objects, both $b_1$ and $b_2$ are positive. The only source in which the derived $b_{1,2}$ are negative is J1638+5720. We do not find anything special about this object, and it happens to have the lowest number of measurements in our sample, so additional checks would be required there to draw a definite conclusion.


\subsection{Opacity changes in the core and jet flow velocity}
\label{s:velocity}

\begin{figure*}
    \centering
    \includegraphics[width=1.5\columnwidth,trim=0cm 0cm 0cm 0cm]{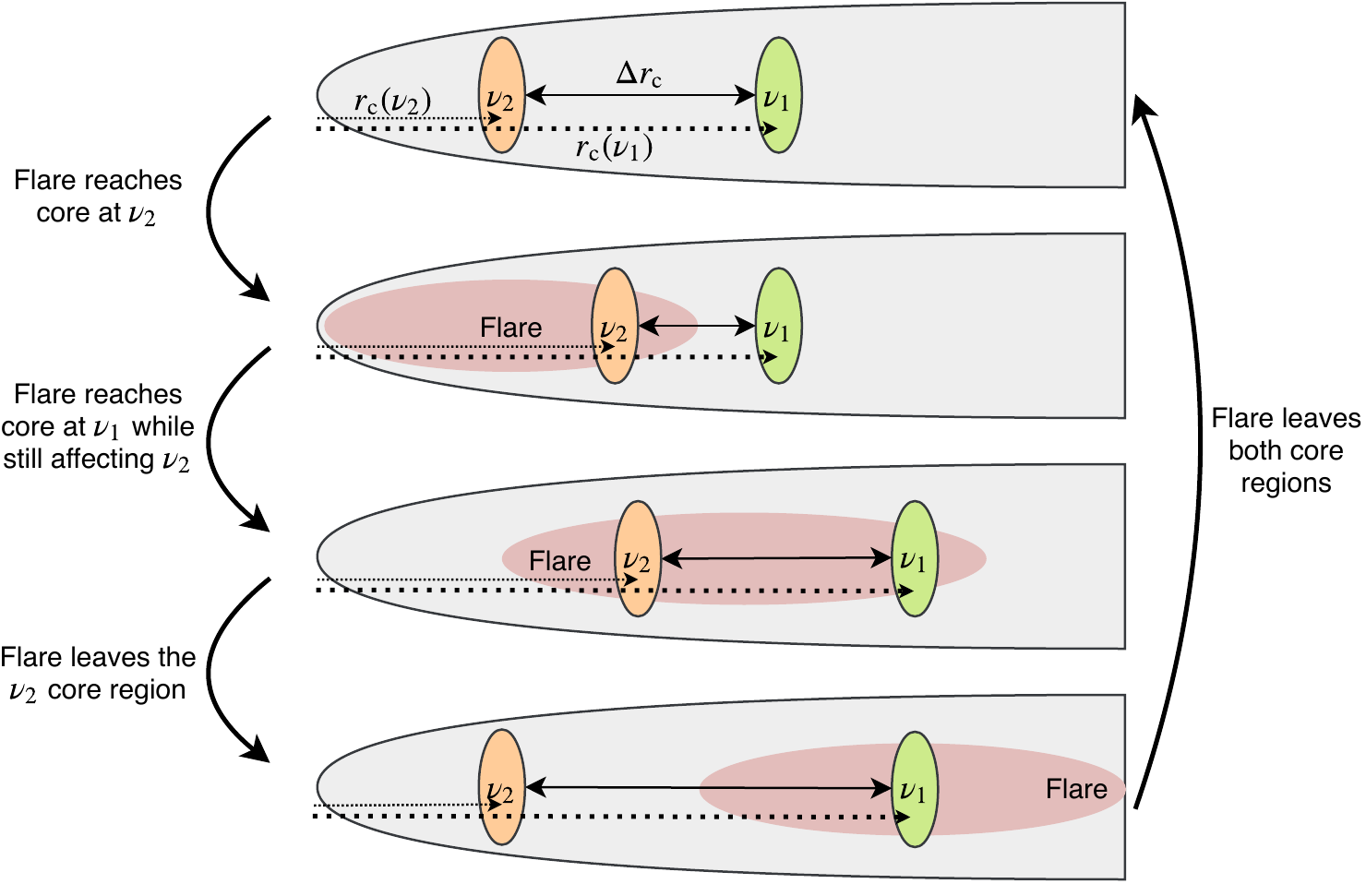}
    \caption{Propagation of a flare along the jet, shown schematically.}
    \label{f:flare_diagram}
\end{figure*}

Recently, core shift measurements have been applied to estimating the velocity of jet plasma flow \citep[see e.g.][]{2014MNRAS.437.3396K,2017MNRAS.468.4478L,2018arXiv180905536K}. These works take the ratio of the core shift magnitude and the time delay between flares at the same pair of frequencies as the flow velocity. This method assumes implicitly that the distance between the cores observed at different frequencies does not vary in time. In the following discussion, we relax this assumption and introduce a conceptually different approach to constrain jet bulk motion using multi-epoch measurements of the core shift.

Fitting the model from \autoref{s:fit_gp} gives us the ability to study temporal evolution of $S_\C$ and $r_\C$ at both frequencies individually. In addition to the smoothing and interpolation as illustrated in \autoref{f:timeseries}, we can use the Gaussian process models to estimate the rate of change (a time derivative) of each of the model parameters. Of special interest here is the time derivative of $r_\C$ which reflects an effective velocity of an individual core induced by variable opacity in the jet. The distribution of per-source maximal angular proper motions estimated this way is given in \autoref{f:core_velocity_dist}, and it has the median value of 0.2 mas/yr. This means that the position of the apparent core in AGN may move at such an angular speed at least during flares. It would be particularly interesting to see if this speed can be related to the true velocity of the jet plasma flow.

The apparent velocity of the plasma in AGN jets is typically estimated using apparent motion of individual components \citep[e.g.][]{2013AJ....146..120L}. However, it is not clear whether these estimates represent the bulk plasma flow or result from some pattern motion, for instance, of a shock wave or a plasma instability propagating in the jet. In the latter case, the true flow velocity may be higher than the one estimates from the observed apparent motions. If flow velocities are underestimated, it may explain the extreme brightness temperatures observed e.g.\ by \textit{RadioAstron} \citep{2016ApJ...820L...9K,2016ApJ...817...96G,2017SoSyR..51..535K,2018MNRAS.474.3523P,2018MNRAS.475.4994K} and significantly alter various other estimates of jet parameters. Variability of the core position reported above can provide an independent way to estimate the plasma speed in the jet. If the core location is affected by propagation of a plasma condensation, the maximal velocity obtained from the variable core position can physically be expected to be close to the true plasma flow speed reduced by the opacity gradient in the flaring region. This implies that we can use our measurements of the core positional variability to give lower bounds on the bulk motion of the jet plasma.

We compare the core velocities estimated using our model from \autoref{s:fit_gp} to jet kinematics measurements from the MOJAVE program \citep{2013AJ....146..120L,2016AJ....152...12L} for the same sources. This comparison is shown in \autoref{f:core_velocity_mojave}, indicating that for the majority of sources (27 at 2~GHz and 26 at 8~GHz), the core velocity does not exceed the maximal velocity of jet components. This is the expected outcome if jet components represent the true underlying motion and the core position moves slower due to opacity gradient. We can argue that the remaining cases with larger core velocities do not indicate any systematic effect, but occured by chance. The bottom plot in \autoref{f:core_velocity_mojave} shows ratio of maximal core velocity to maximal component velocity for sources having different number of components in the MOJAVE dataset. Indeed, strong excess occurs only for objects with 6 or fewer components, while for larger numbers the kinematic velocity is significantly higher. We do not find any significant dependence of the estimated core velocity on the number of core shift measurements we have.
This comparison justifies the usage of variability of the core position to put a lower bound on the jet plasma flow velocity. The respective estimates of projected linear speeds are shown in \autoref{f:core_velocity} for each source.

The dependence of velocity ratio on the number of VLBA observing epochs for a given target is much less prominent, than on the number of components with measured kinematics shown in \autoref{f:core_velocity_mojave}. This is consistent with the MOJAVE conclusion that not all observed features move at the characteristic velocity of the bulk flow \citep{2016AJ....152...12L}. Our result implies that for reliable plasma flow velocity estimation one typically needs kinematic measurements for about 10 or more separate components.

\subsection{Nature of jet flares}
\label{s:jet_evolution}

\begin{figure}
    \centering
    \includegraphics[width=\columnwidth,trim=0cm 0.5cm 0cm 0cm]{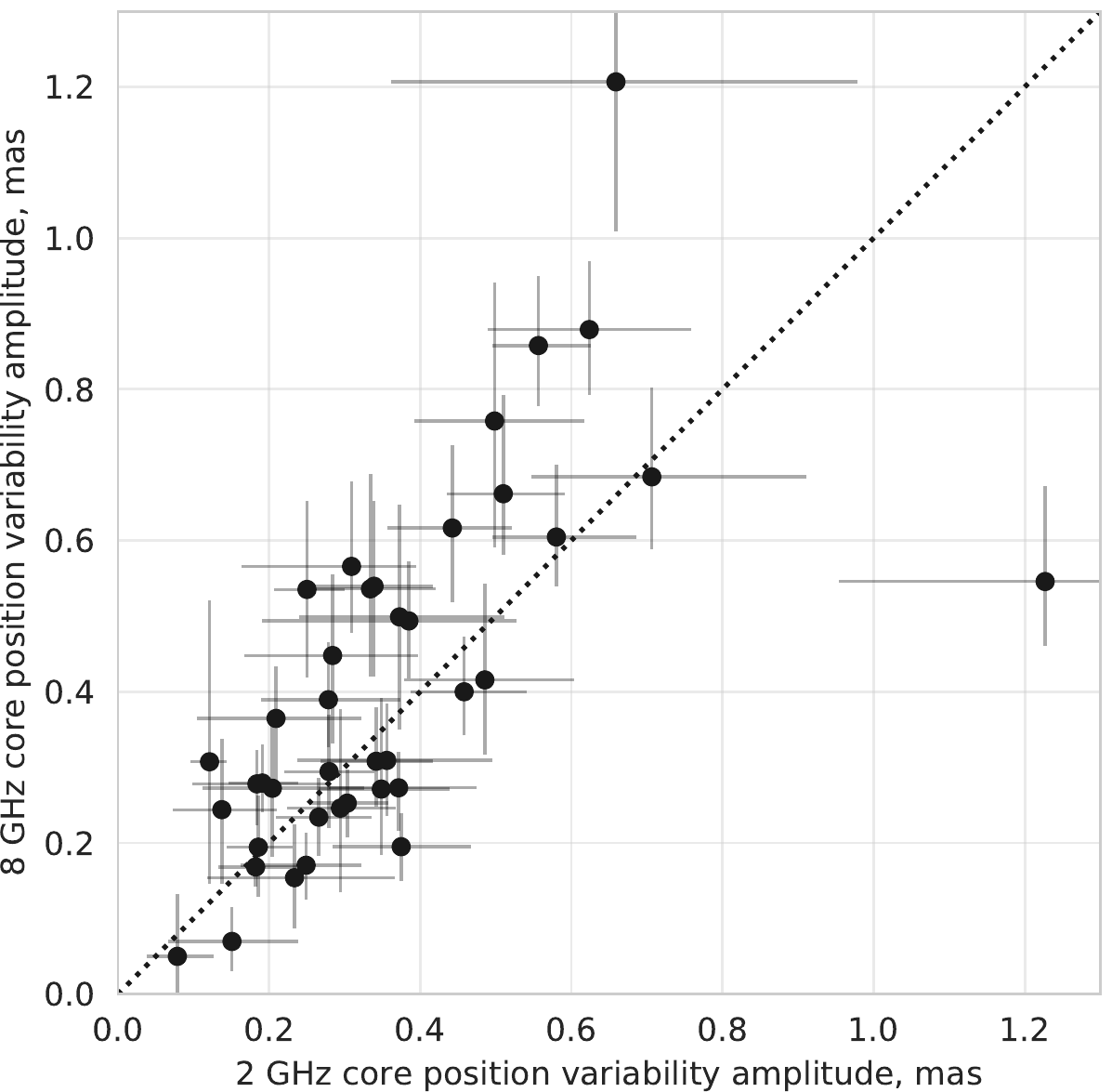}
    \caption{Variability amplitudes (difference between maximum and minimum values) of core positions at different frequencies. Error bars represent 68\% credible intervals.}
    \label{f:core_2_8_comparison}
\end{figure}

The individual light curves presented in \autoref{f:timeseries} exhibit complicated behavior, but often contain flares visible at both 2 and 8~GHz. Such flares show substantial time overlaps between these two bands, which implies that the size of a typical flaring region should be at least comparable to the spatial separation between the cores at both frequencies. \corr{The overlapping flares are unlikely to be caused by illumination of an optically thin component close to the core, as a bright component would lead to a steep apparent core spectrum. The ratio of 8 GHz to 2 GHz core flux densities and spectral index images (see \autoref{f:spectral_index_sample}) both indicate that this is not the case. Thus the flares do often affect cores at both frequencies at the same time.} As the core position is shown to vary together with its flux density (see \autoref{s:fit_gp}), we propose a qualitative flare propagation model shown as a diagram in \autoref{f:flare_diagram}.

We estimate the variability amplitude of each core position and compare them between the two frequencies. This comparison is shown in \autoref{f:core_2_8_comparison}, indicating that the magnitudes of the core position changes are similar at both frequencies. In the framework of model \autoref{f:flare_diagram}, this means that the core shift variability magnitude should be smaller than the variability of each individual core position when the flare region is large compared with the separation between cores at the two frequencies.

Using Eq.~\eqref{eq:srn1}, \eqref{eq:rs_c1} and the value of $K_{\mathrm{rS}} = 0.28 \pm 0.05$ determined in \autoref{s:fit_gp}, we obtain $K_{\mathrm{BN}} = -0.529\pm0.005$ for the magnetic field changes described by $B_\C(t) \sim N_\C(t)^{K_{\mathrm{BN}}}$. For the variations of the particle density described by $S_\C \sim N_\C^{K_{\mathrm{SN}}}$, this procedure yields $K_{\mathrm{SN}} = 0.207\pm0.007$. We use these estimates to calculate the relative variability of the magnetic field and emitting particle density in all of the sources studied. These changes are plotted in the two lower panels in \autoref{f:timeseries}, for each of the sources. These plots illustrate that major variations of these parameters, especially the particle density, are required to explain the observed flux density variability and the variations of the core position.

\subsection{Core shift frequency dependence}
\label{s:freqdep}

Previous studies of the core shift effect \citep[e.g.][]{2011A&A...532A..38S} show that its magnitude is close to inverse frequency dependence $r(\nu)\propto \nu^{-1}$ which was predicted in the case of equipartition in the jet \citep{1979ApJ...232...34B}. But clearly if cores at different frequencies move at different moments in time due to flares, as we infer from our analysis, this dependency cannot always hold. This apparent disagreement cannot be resolved by arguing that previous studies had measurements during `quiet' periods without flares: there was no such selection performed and it is highly unlikely that sources were observed at a quiet state just by chance.

To verify that there is no disagreement indeed, we took measurements from \cite{2011A&A...532A..38S} and fit inverse frequency dependence to core shift magnitude for each source. Then we calculated average deviations of their measurements at 2 and 8 GHz from the $r(\nu) \sim \nu^{-1}$ curve: the median sum of these two deviations exceeds $0.3$~mas and is comparable to the variability we detect in \autoref{s:results}. This means that core positions routinely deviating from the inverse frequency dependence does not contradict that paper conclusions. Results of \cite{2011A&A...532A..38S} are averages over multiple frequencies in a wide range, so they are less sensitive to core movements. In addition, the previous section shows how the separation between cores, which is the quantity measured in that paper, may vary less than individual core positions.

\subsection{Astrophysical implications}

Core shift measurements are commonly used for making estimates of various geometrical and physical parameters of AGN jets \citep[see, e.g.][]{1998A&A...330...79L,2009MNRAS.400...26O,2008A&A...483..759K,2011Natur.477..185H,2012A&A...545A.113P,2014MNRAS.437.3396K,2018MNRAS.475.4994K,2016MNRAS.462.2747K,2018arXiv180806138P}. Our analysis shows that core shift measured between two particular bands can vary substantially over time and that these variations are well connected to flares observed in compact jets. Hence  estimates of the jet parameters based on core shift measurements made at two frequencies and a random epoch may be affected by an ongoing flare in a source. One remedy to that would be to resort to making core shift measurements during periods of quiescence in a given source. We stress that making measurements at multiple frequencies does not always eliminate the variability effect because a flare may influence cores at a range of frequencies at the same time.

Our simplified variability model given in \autoref{s:jet_theory} describes relationships between physical jet parameters and observable values as power laws. Using the estimated exponents from \autoref{s:jet_evolution} one can derive magnetic field and emitting particle density variability from core flux density monitoring, as we show in \autoref{f:timeseries}. Note that it does not require core shift measurements and can be done even for observations at a single frequency. In addition, as we show in \autoref{s:velocity}, core shift variability measurements and the model we employ can be used to estimate the plasma flow velocity in the jet \citep[see also on this subject][]{2018arXiv180905536K}.

\subsection{Astrometric implications}

The relationships inferred in this paper imply that for high-accuracy astrometric measurements of source positions (e.g.\ for ICRF3\footnote{\url{http://hpiers.obspm.fr/icrs-pc/newwww/icrf/index.php}} or RFC\footnote{\url{http://astrogeo.org/rfc/}} VLBI catalogs) one has to take the core shift into account, in particular if the source is strongly variable. Coordinates measured using single-band phase delays are directly affected by the variable position of the core, which is usually the brightest and most compact part of the source. Even astrometric measurements based on group delays between two different frequencies, which is not sensitive to positional offset following $r \sim \nu^{-1}$ \citep{2009A&A...505L...1P}, can be affected by the variability of core positions. Indeed, our results imply that strong flares disrupt any regular frequency dependence, hence also affecting astrometric inferences made during such a flare.

As mentioned by \cite{2009A&A...505L...1P}, taking the extended source structure into account when doing astrometric measurements does not help in getting rid of the core shift effect. These two are completely different effects, both affecting the astrometric positions of AGNs, and they should separately be taken care of. The core shift effect should also be present for point-like objects with no extended jet structure visible. We are unable to measure it in the same way for such cases, so different methods should be used. See discussion of phase-referencing core shift measurements \citep[e.g.][]{voitsik2018} earlier in this paper.

A study by \cite{2017A&A...598L...1K} and \cite{r:gaia5} comparing VLBI and optical \citep[Gaia,][]{Gaia_DR1_CRF,Gaia_DR2_CRF} coordinates suggests that the core shift may have an effect on the measured positions.
The core shift \corr{can at least partly explain the detected} offsets of up to 2~mas between the radio and optical positions of AGNs for cases when radio is shifted downstream the jet.
These offsets are understood as the distance of the apparent VLBI core from the nucleus which dominates in the optical band.
This would imply that it does not follow the inverse frequency dependence \citep{2009A&A...505L...1P}, as can easily be explained by the variability discussed here.
Moreover, \cite{r:gaia4} \corr{showed that statistically} significant AGN proper motions \corr{derived} with VLBI occur predominantly along the jet direction. This can be at least partly explained by the core shift variability due to radio flares.

\section{Summary}
\label{s:summary}

In this paper, we have developed and applied an automatic method of measuring the core shift effect in compact extragalactic jets using multi-frequency VLBI observations. We apply this method to multi-epoch \it{S}/\it{X}-band (2 and 8~GHz) observations of 40 compact jets and show that it yields more robust and consistent measurements of the core shift compared with results from `manual' methods employed in earlier works.

The dataset we use contains 1691~source-epoch pairs: 40~sources with 11 to 78 observations each. This database allows us for the first time to study temporal variability of the core shift in compact jets. Significant variability is detected for 33 out of 40 sources. The typical magnitude of \corr{8-2 GHz} core shift changes is 0.4~mas or 2~pc, with the strongest variations reaching up to 1~mas. \corr{For comparison, the median core shift magnitude is 0.5~mas or 3.2~pc.} As a byproduct, spectral index maps are generated, based on the image alignment required for the core shift measurements. These maps are presented as a supplementary material.

The core shift measurements combined with the core flux densities at the same frequencies allow us to model positional changes of the core at each individual frequency. We estimate that the core position at both 2 and 8~GHz vary by about 0.35~mas or 2~pc on average. These variations are interpreted as the effect of nuclear flares, with flaring plasma component propagating down the jet and affecting the individual apparent cores at different times for different frequencies. Our analysis shows that the observed flares are caused by significant increase in emitting particle density in the core region. The magnetic field is found to decrease at the same time by a smaller factor. We introduce a way to estimate the plasma particle density and magnetic field variability using only core flux density measurements, based on the jet model we use and evaluate in this paper.

We show that apparent velocities of individual cores can serve as lower bounds of the bulk plasma flow velocity in the jet. This constitutes a novel method of constraining the bulk motion velocity, and for a few sources it gives stronger bounds than kinematics of jet components.
Our results imply that one needs about 10 or more separate components to reliably estimate the plasma flow velocity using VLBI kinematic measurements, as not all components move at the characteristic plasma flow velocity.

As the core shift magnitude changes with time, its variations should be taken into account when estimating physical jet parameters. High-accuracy AGN astrometry is also affected by this variability, as any fixed frequency dependence such as $r \sim \nu^{-1}$ is disrupted during flares, even if true during the quiescent state.

\section*{Acknowledgements}

This research was supported by Russian Science Foundation (project 16-12-10481).
We cordially thank the teams referred to in \autoref{s:obsdata} for making their fully calibrated VLBI FITS data publicly available as well as Richard Porcas, Leonid Petrov and anonymous referee for discussions and paper comments.
This research has made use of NASA’s Astrophysics Data System.



\bibliographystyle{mnras}
\bibliography{cs_var}



\appendix

\section{Electronic only supplementary materials}

\begin{figure*}
\centering
\includegraphics[width=0.24\linewidth]{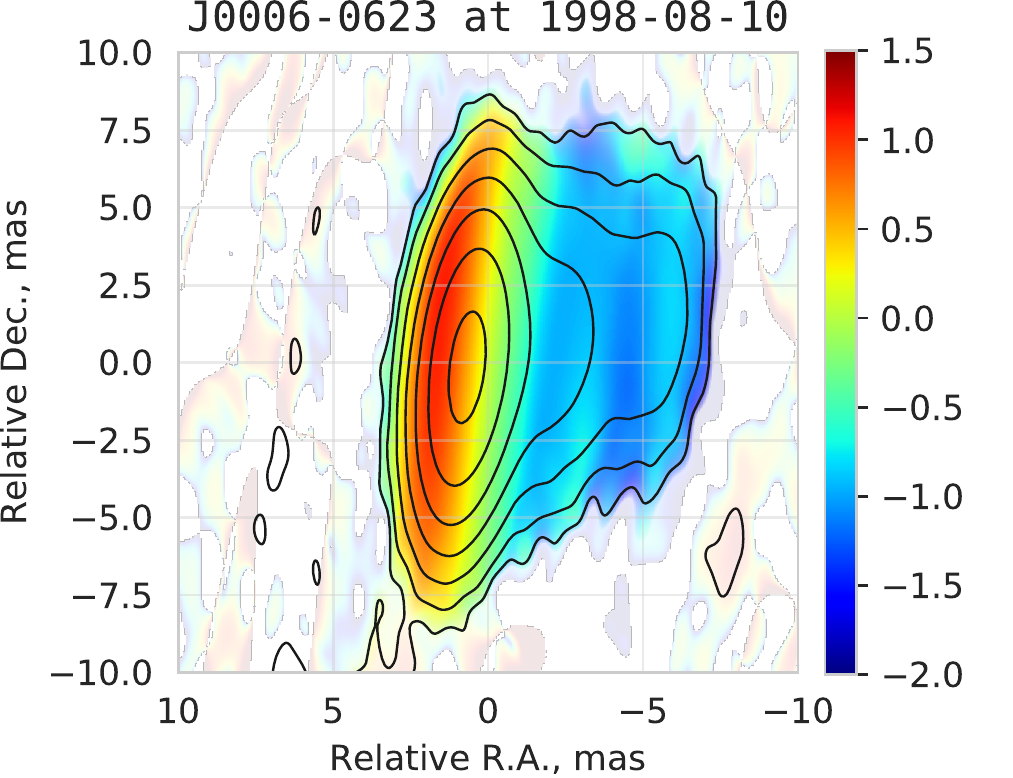}
\includegraphics[width=0.24\linewidth]{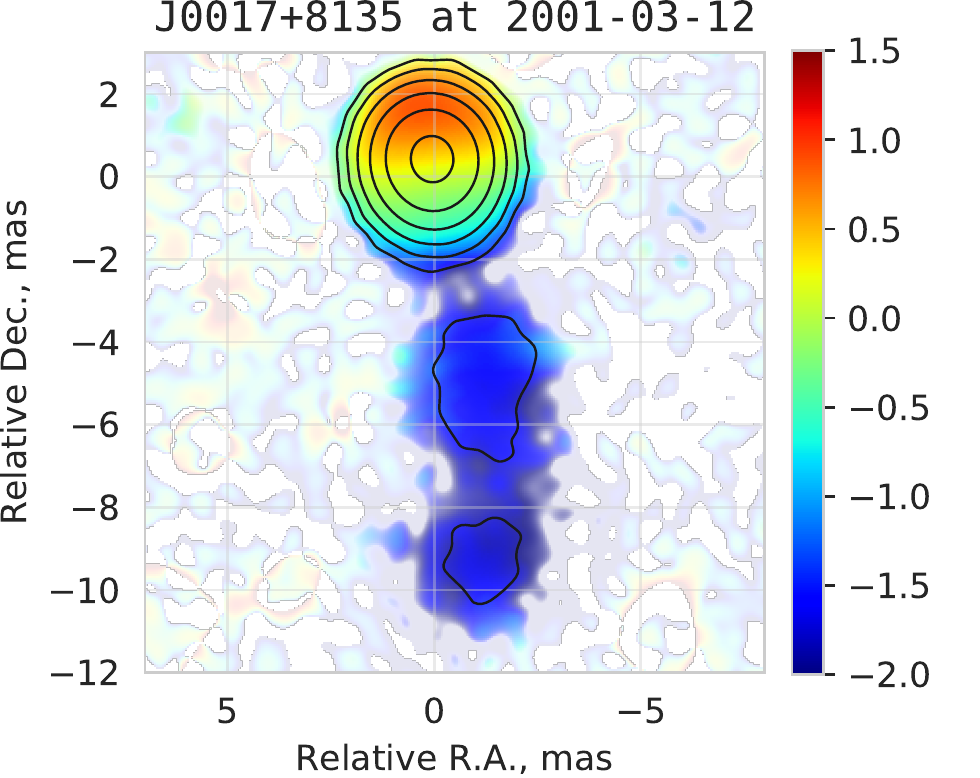}
\includegraphics[width=0.24\linewidth]{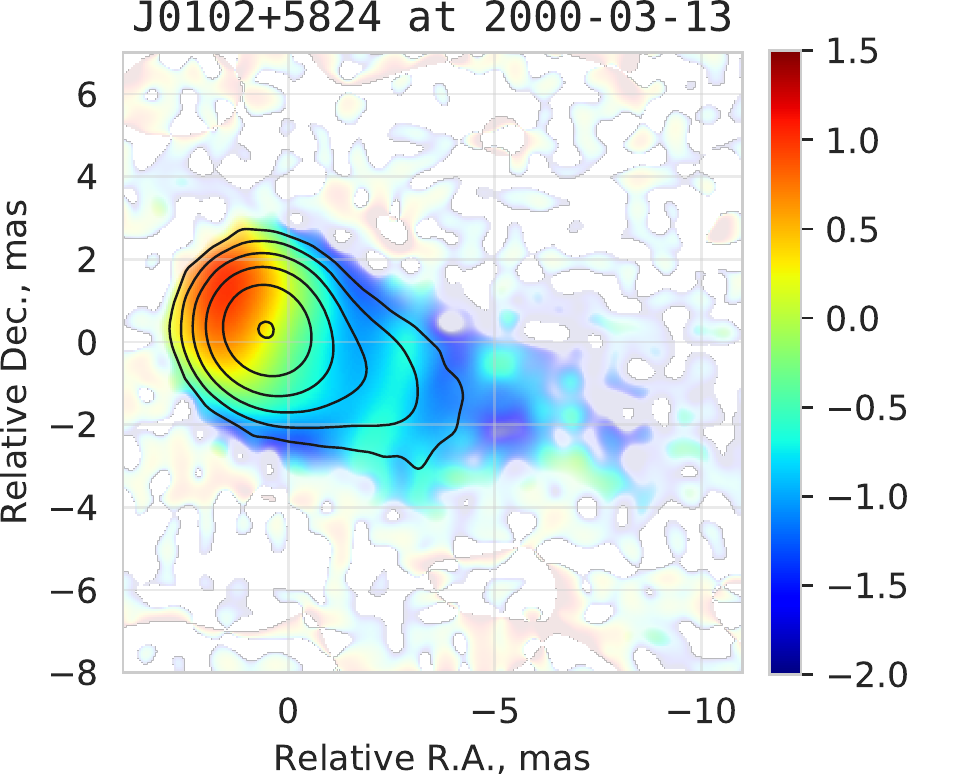}
\includegraphics[width=0.24\linewidth]{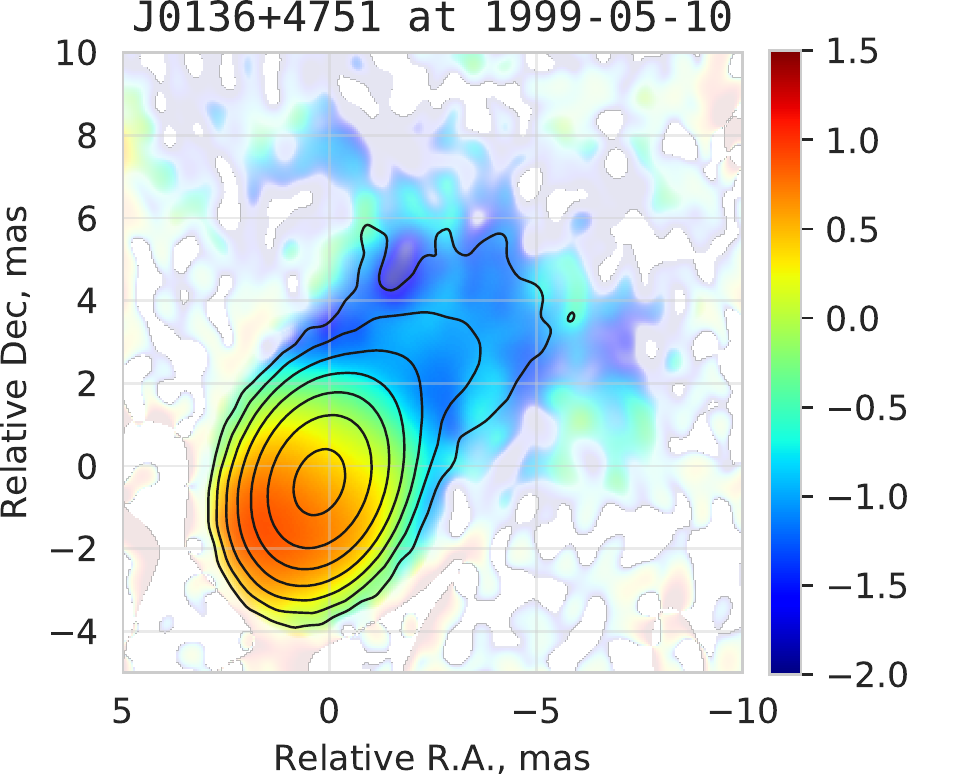}
\includegraphics[width=0.24\linewidth]{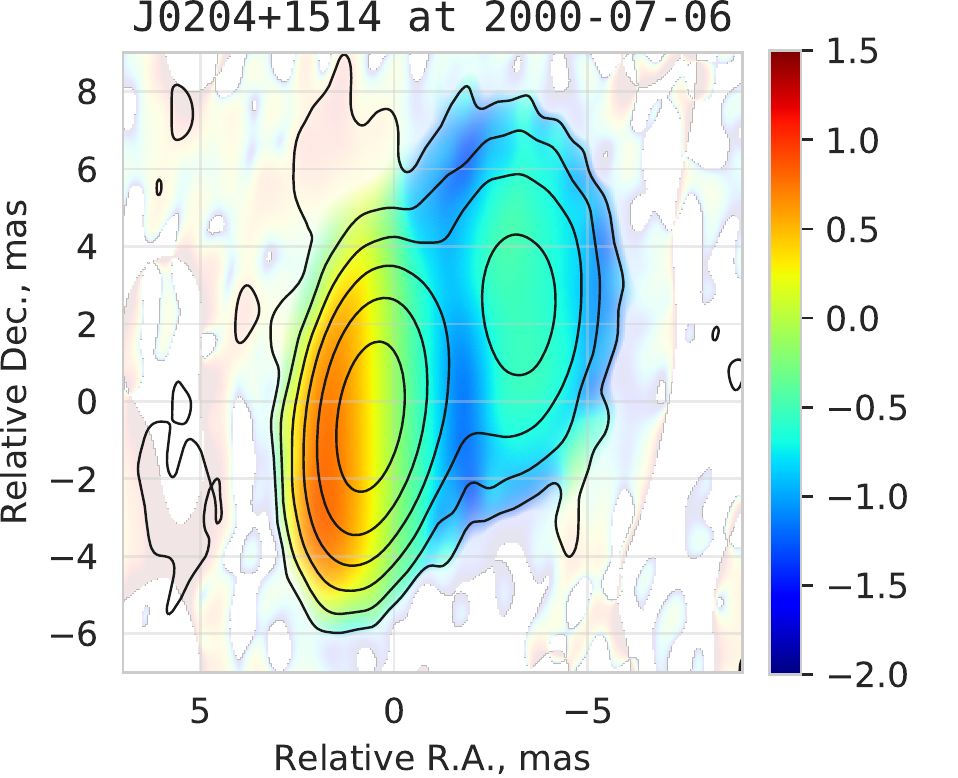}
\includegraphics[width=0.24\linewidth]{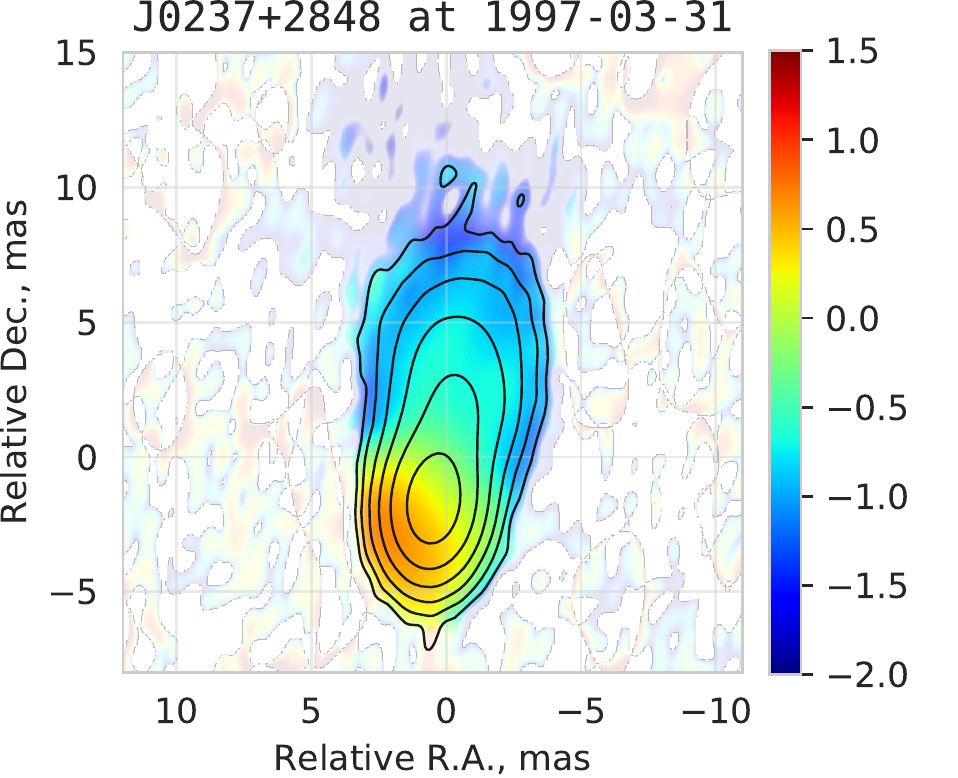}
\includegraphics[width=0.24\linewidth]{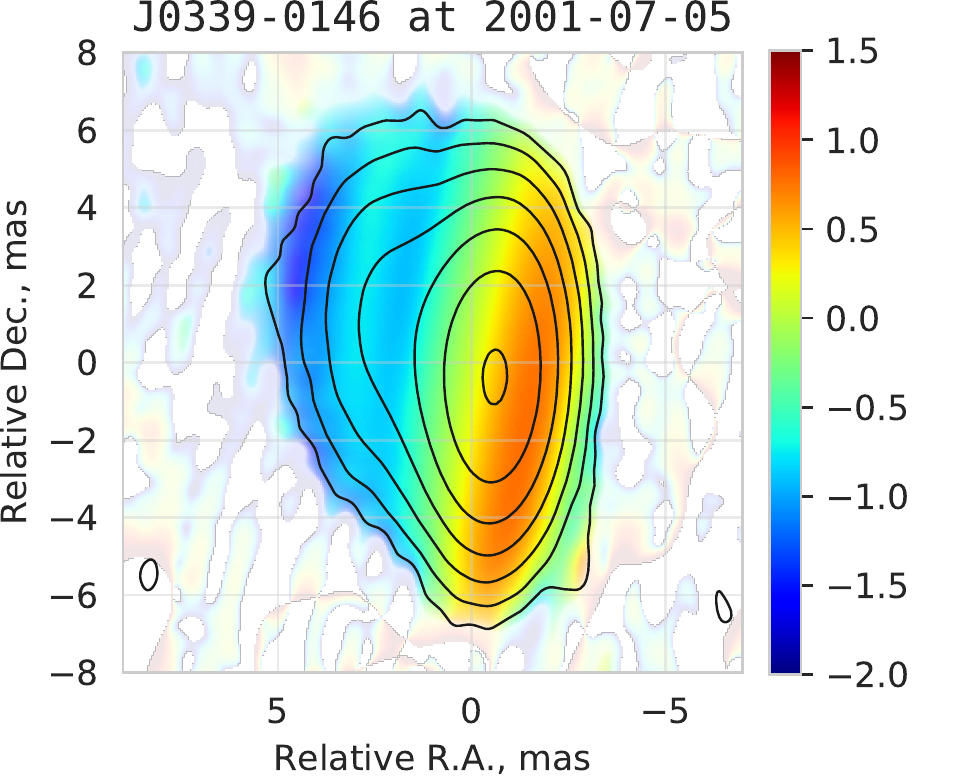}
\includegraphics[width=0.24\linewidth]{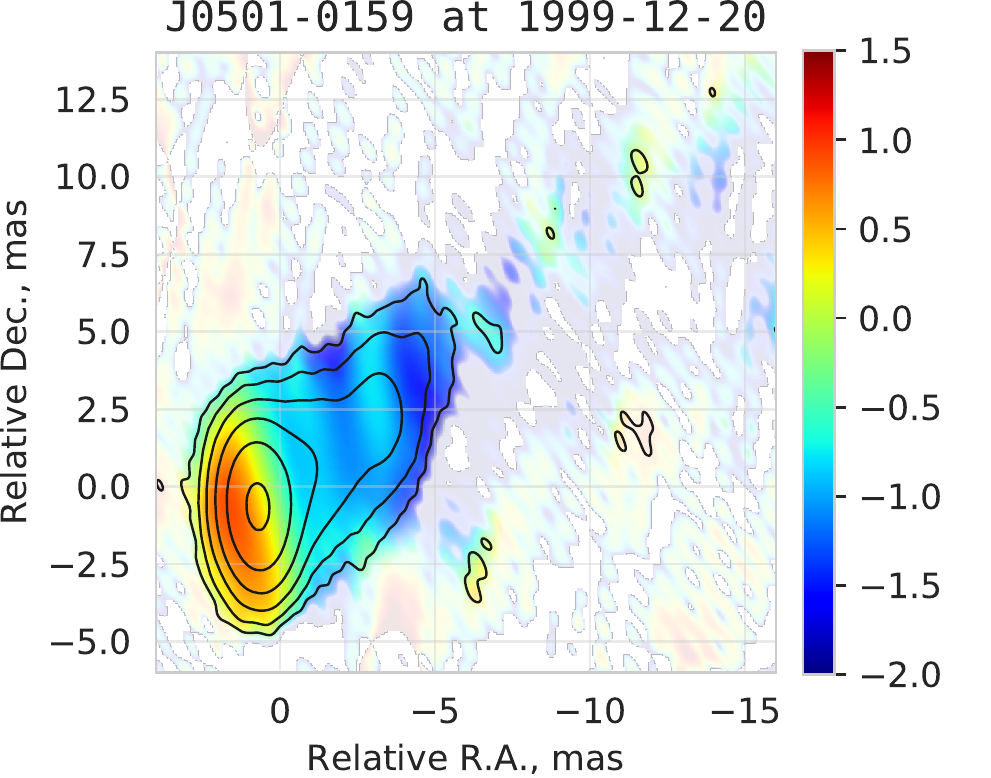}
\includegraphics[width=0.24\linewidth]{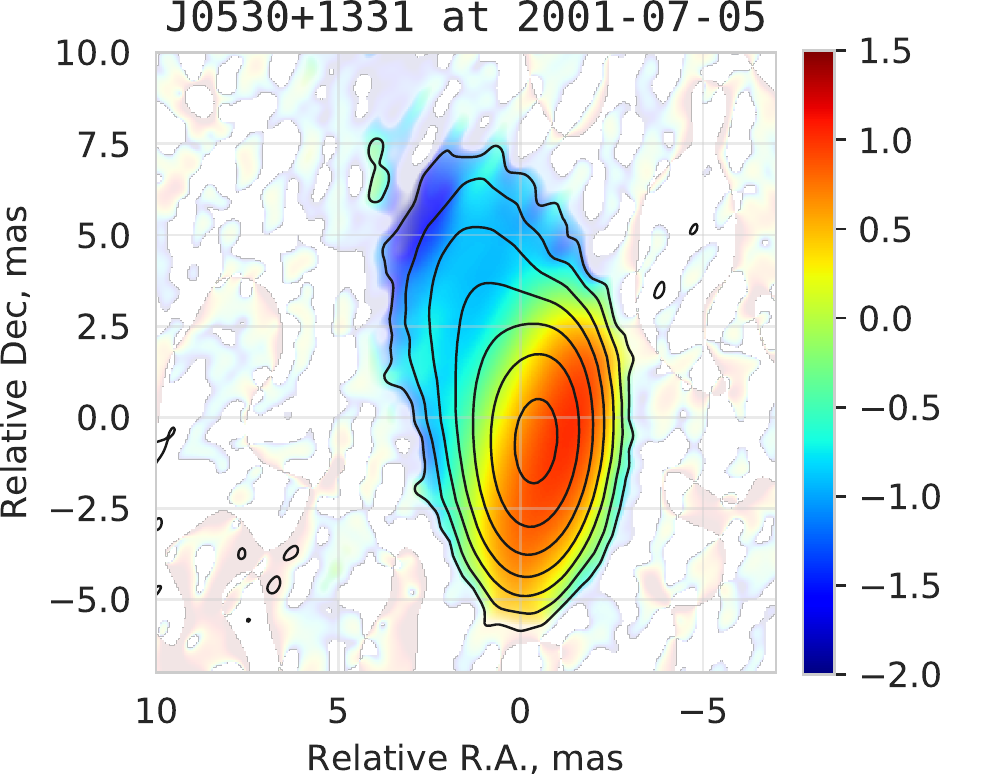}
\includegraphics[width=0.24\linewidth]{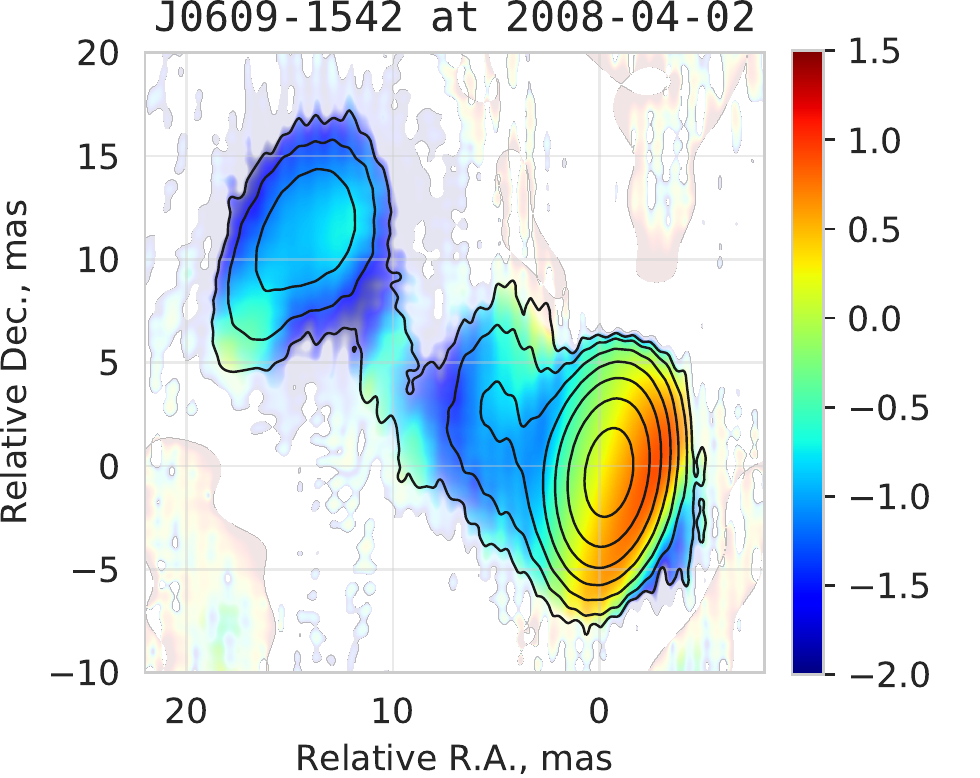}
\includegraphics[width=0.24\linewidth]{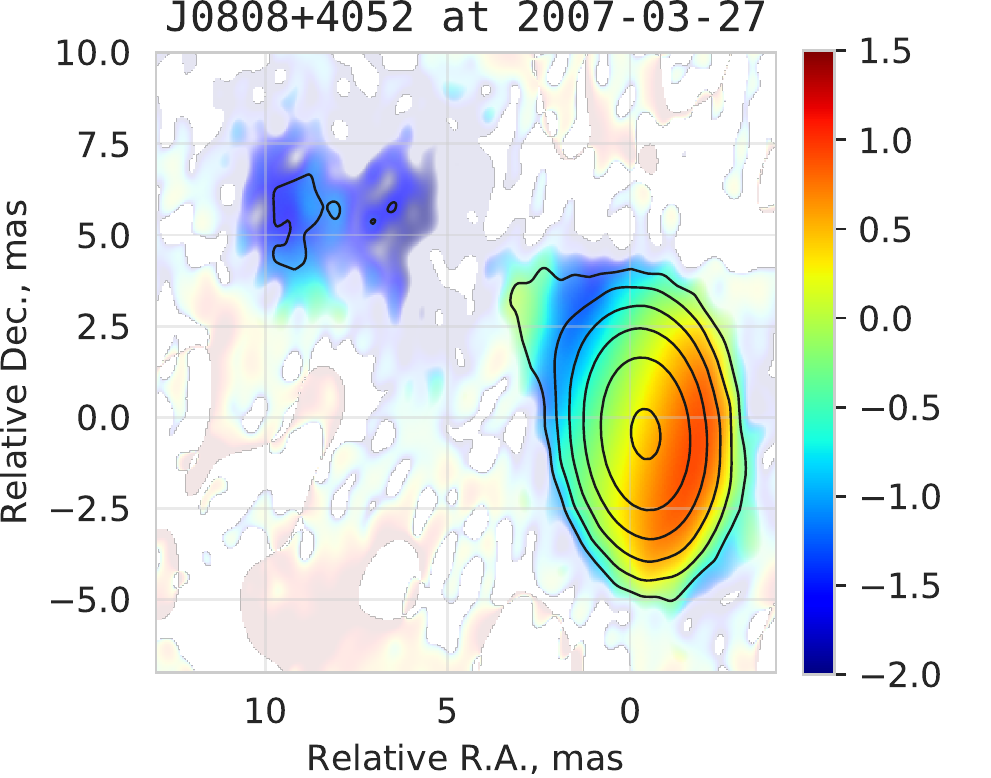}
\includegraphics[width=0.24\linewidth]{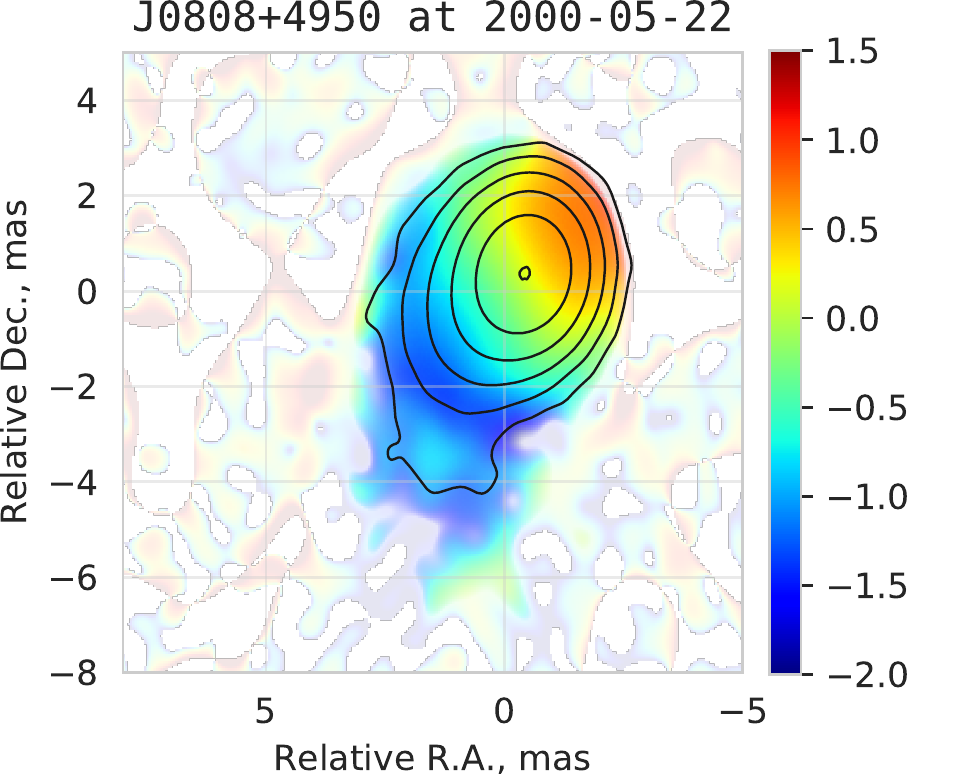}
\includegraphics[width=0.24\linewidth]{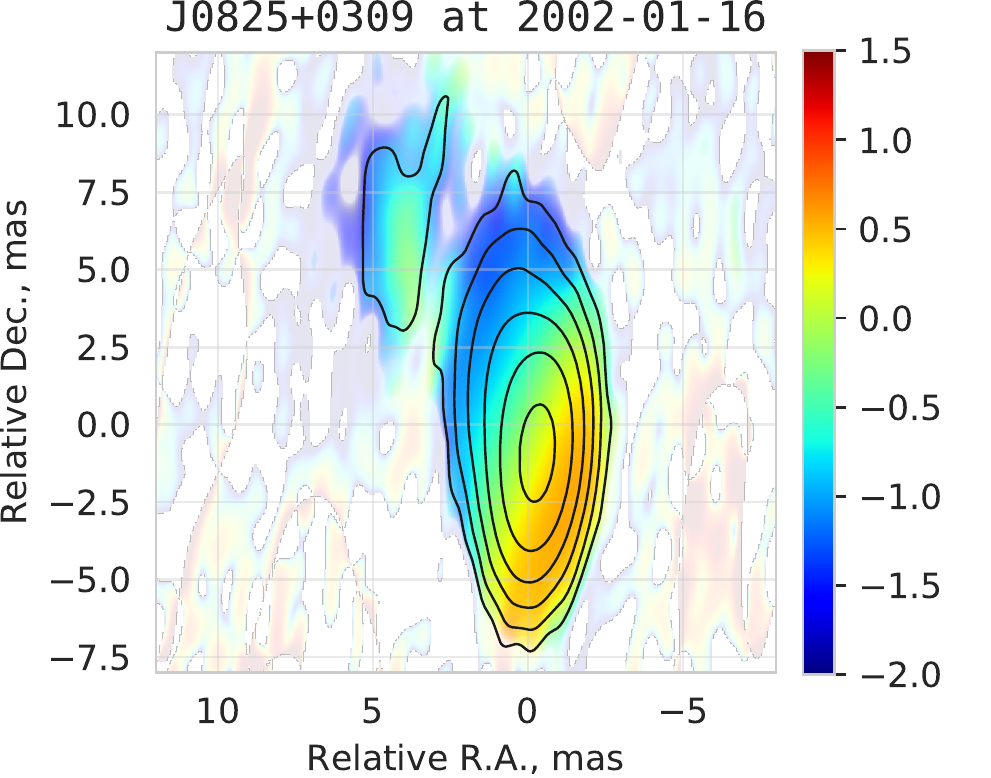}
\includegraphics[width=0.24\linewidth]{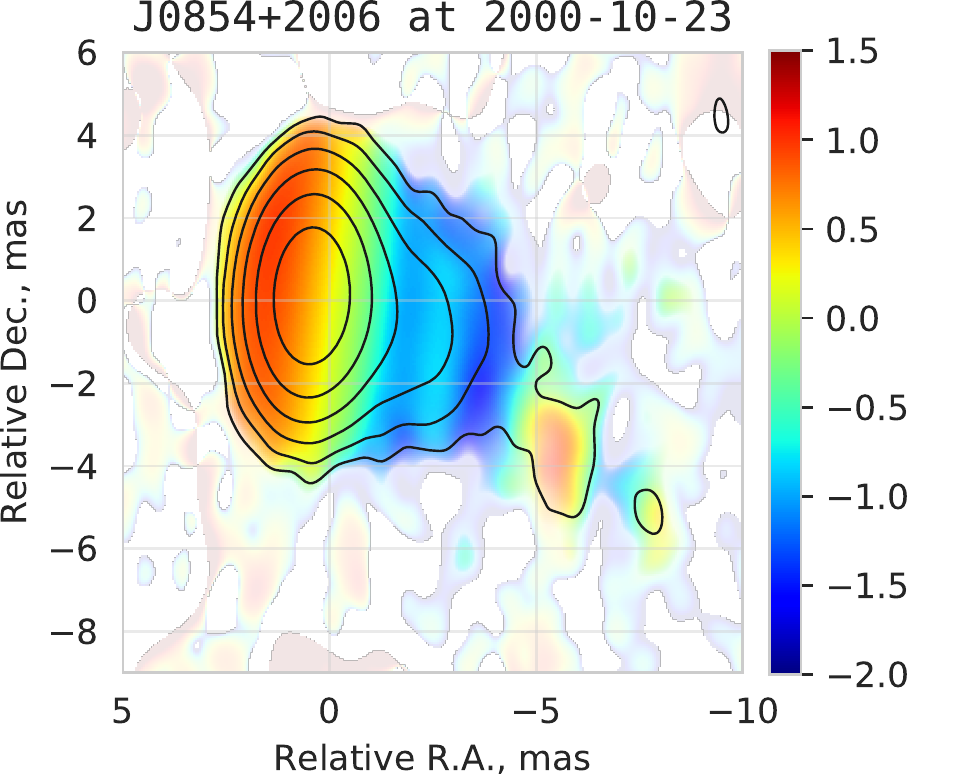}
\includegraphics[width=0.24\linewidth]{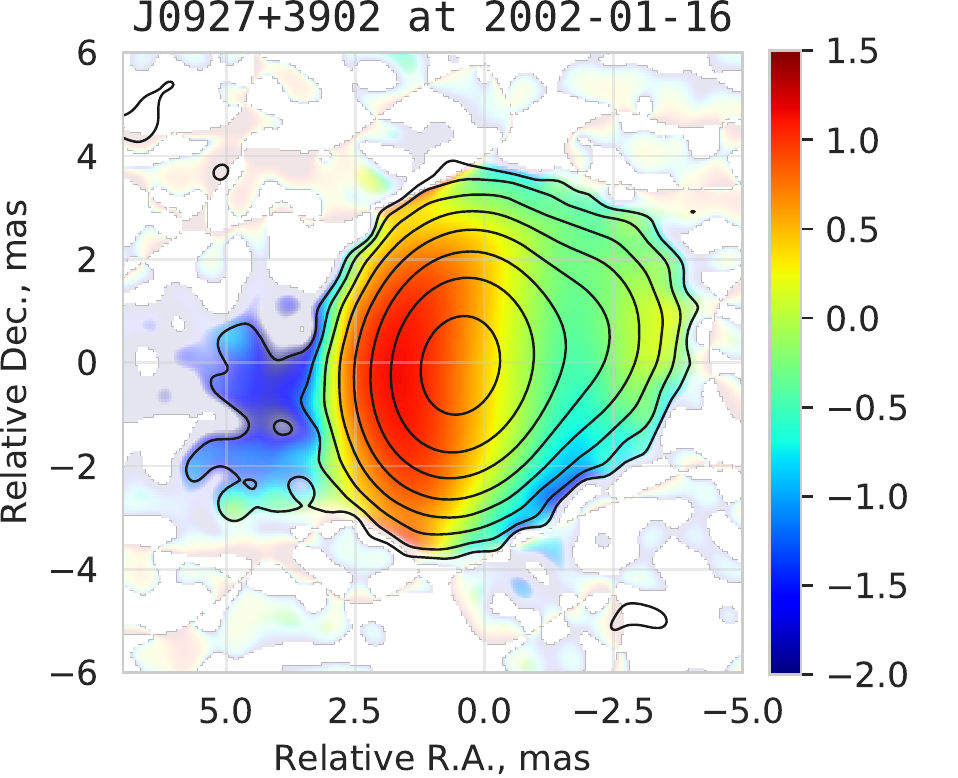}
\includegraphics[width=0.24\linewidth]{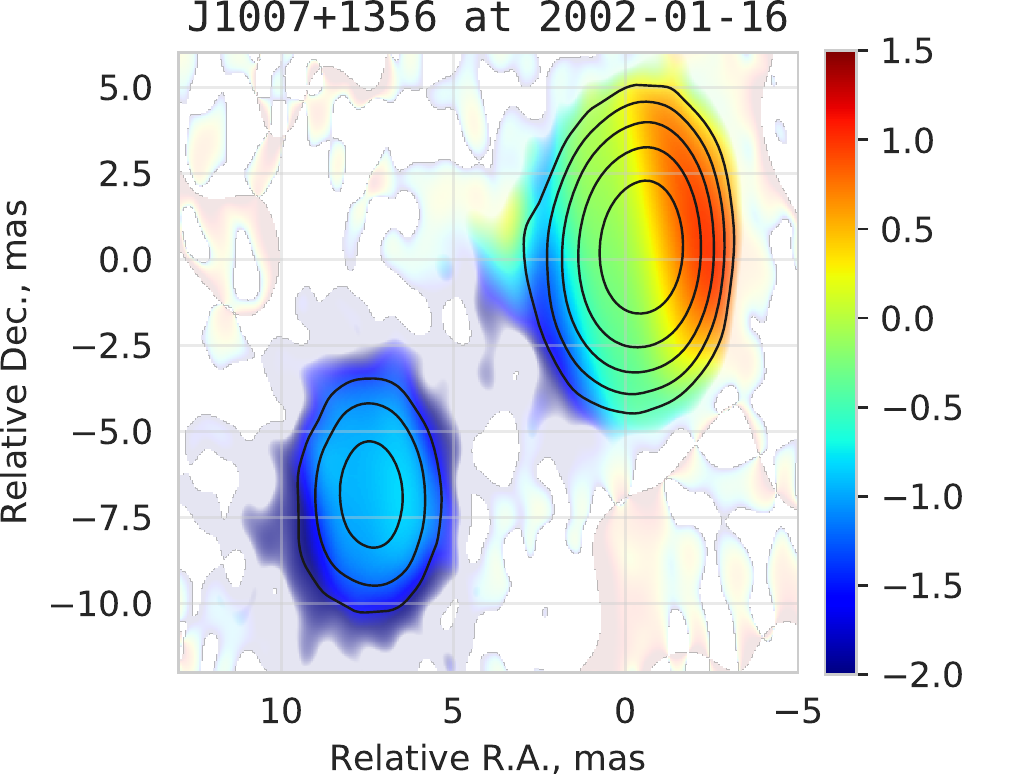}
\includegraphics[width=0.24\linewidth]{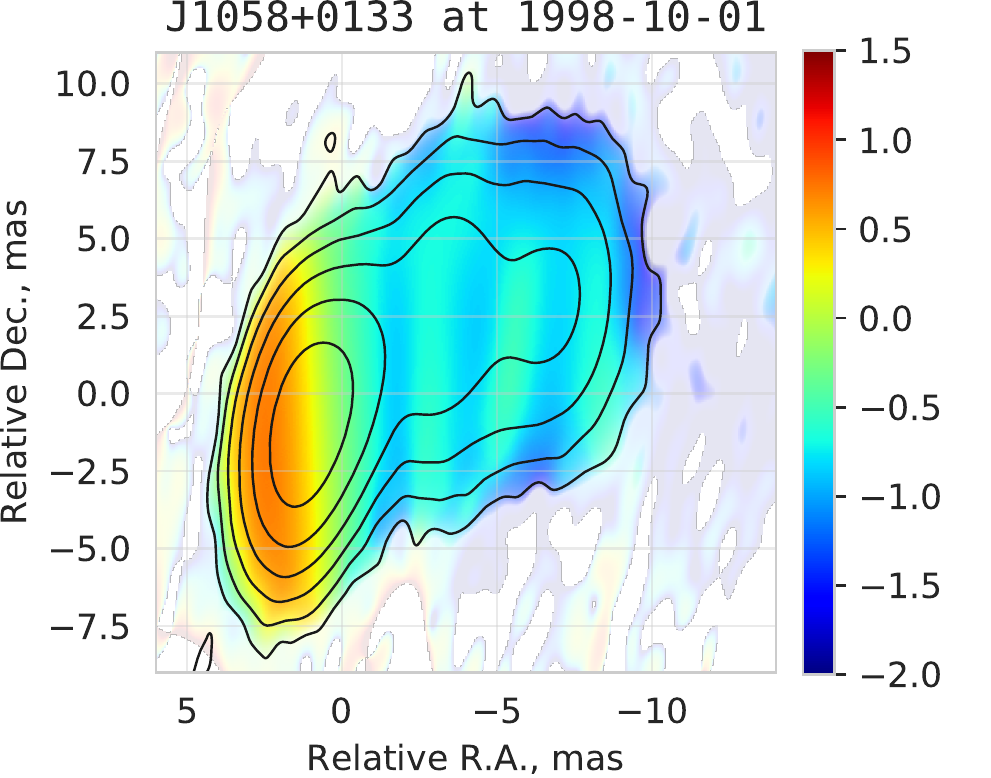}
\includegraphics[width=0.24\linewidth]{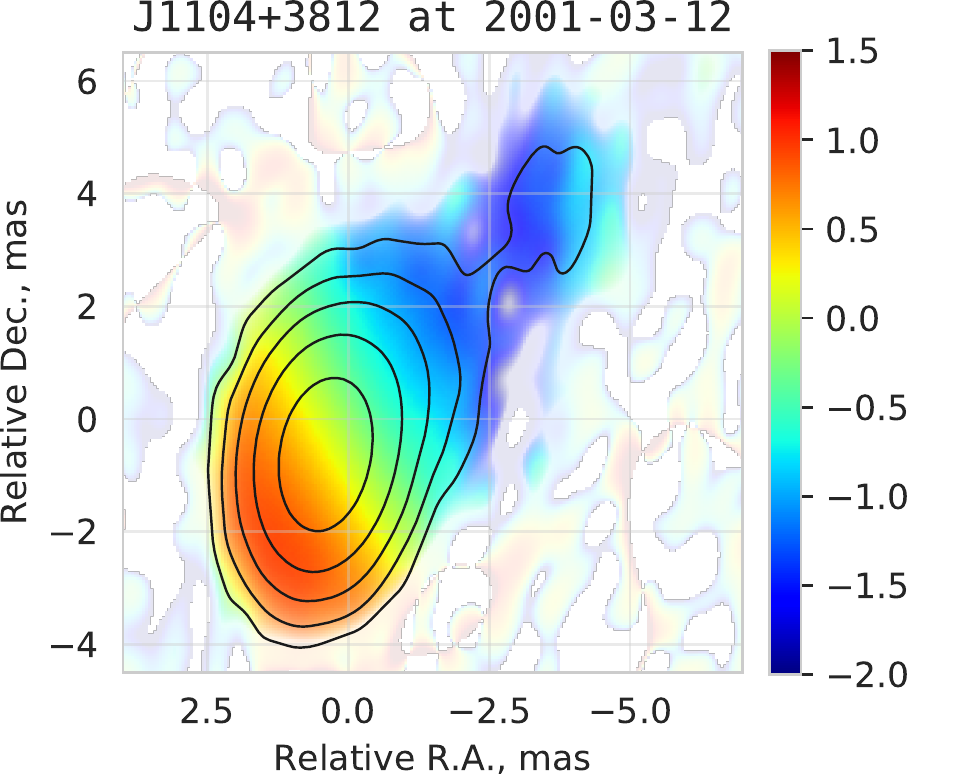}
\includegraphics[width=0.24\linewidth]{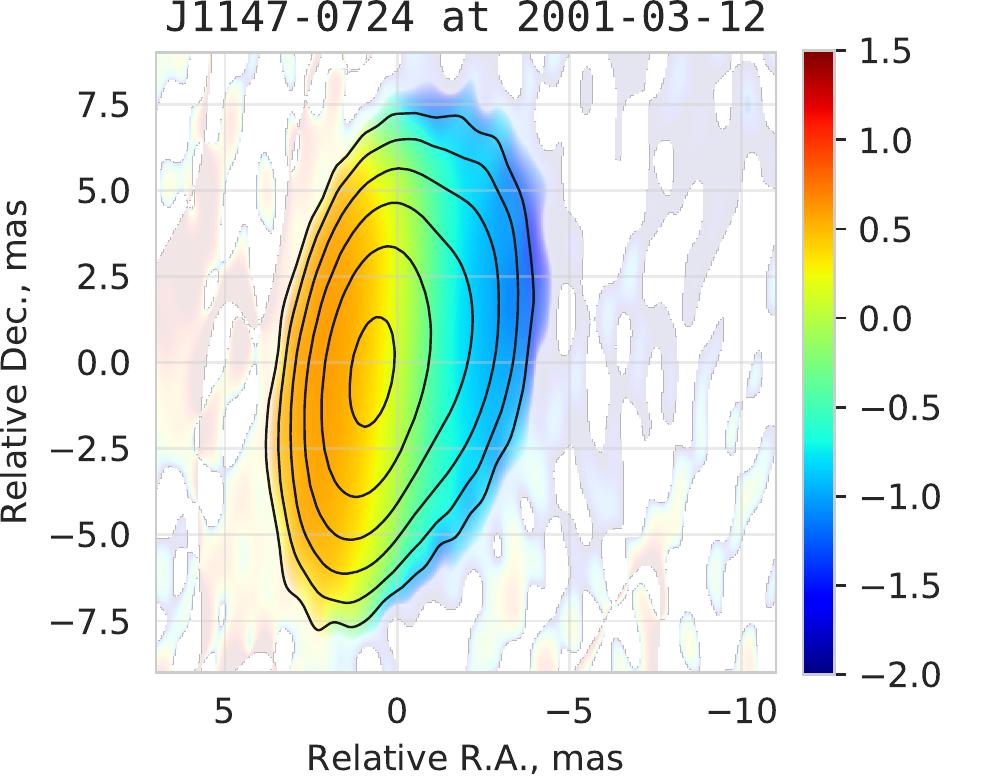}
\includegraphics[width=0.24\linewidth]{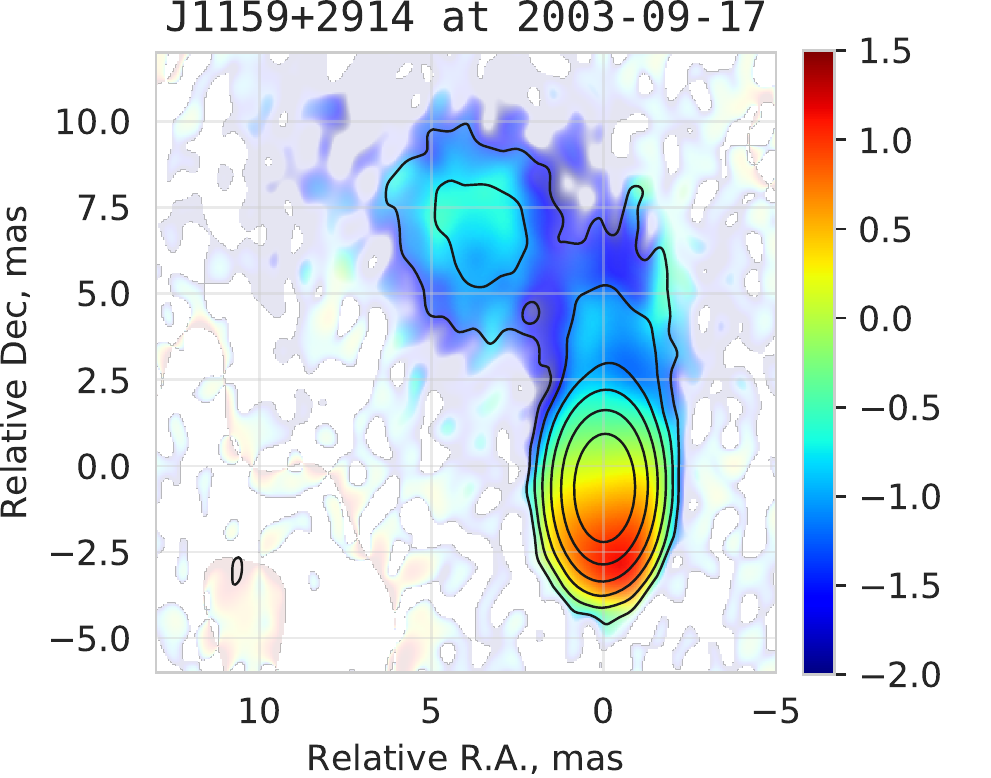}
\includegraphics[width=0.24\linewidth]{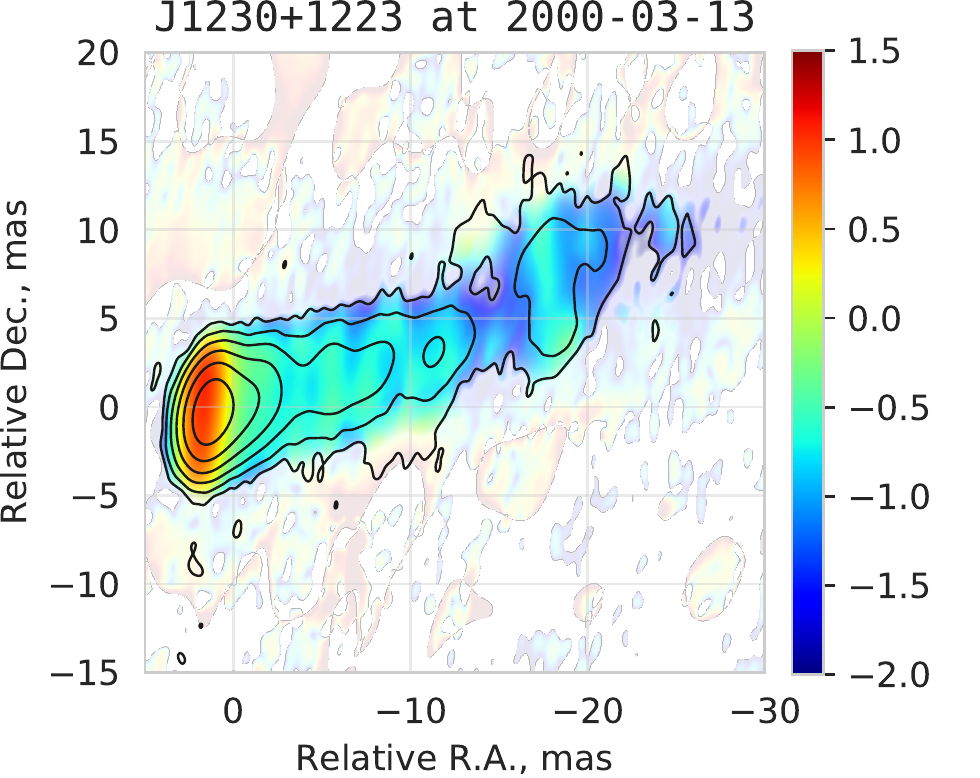}
\includegraphics[width=0.24\linewidth]{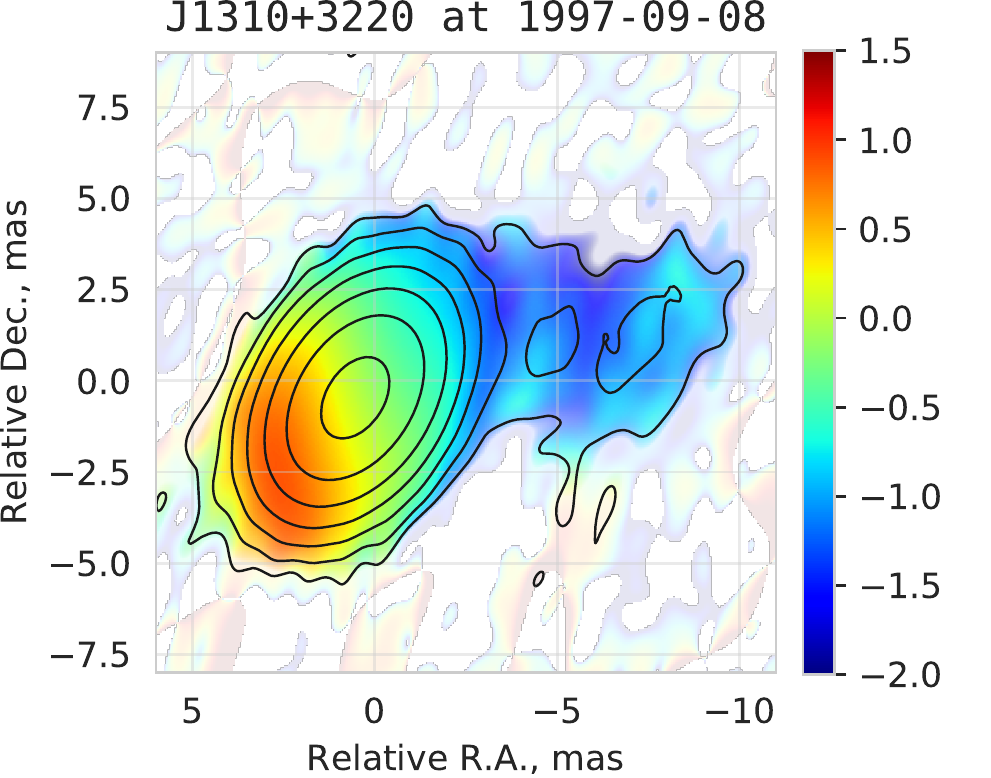}
\includegraphics[width=0.24\linewidth]{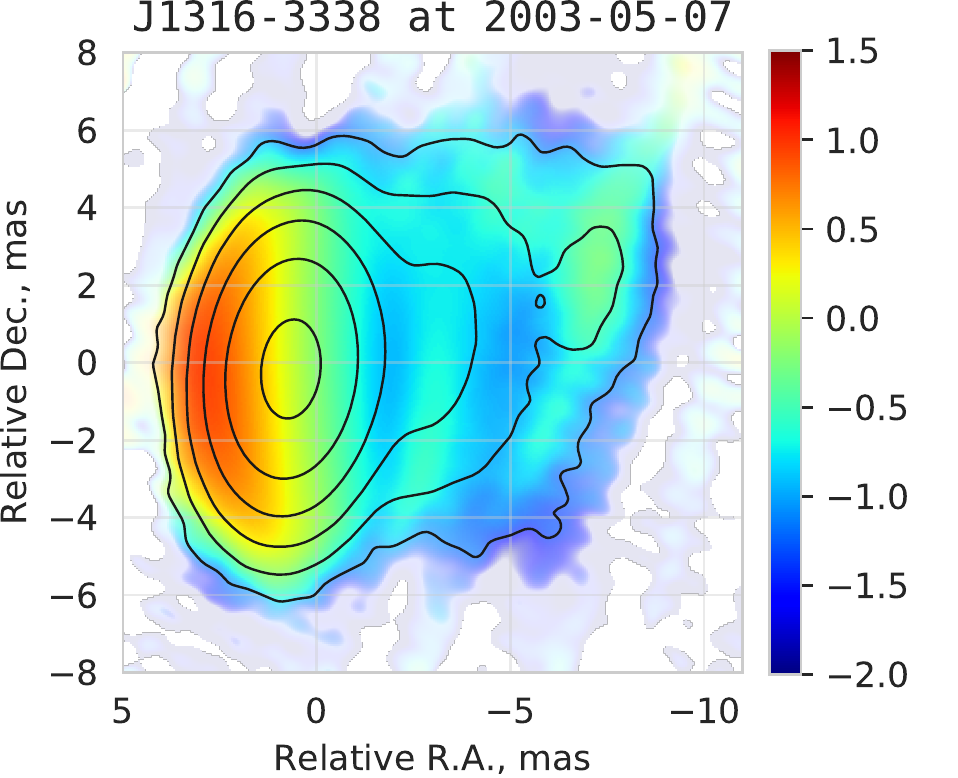}
\includegraphics[width=0.24\linewidth]{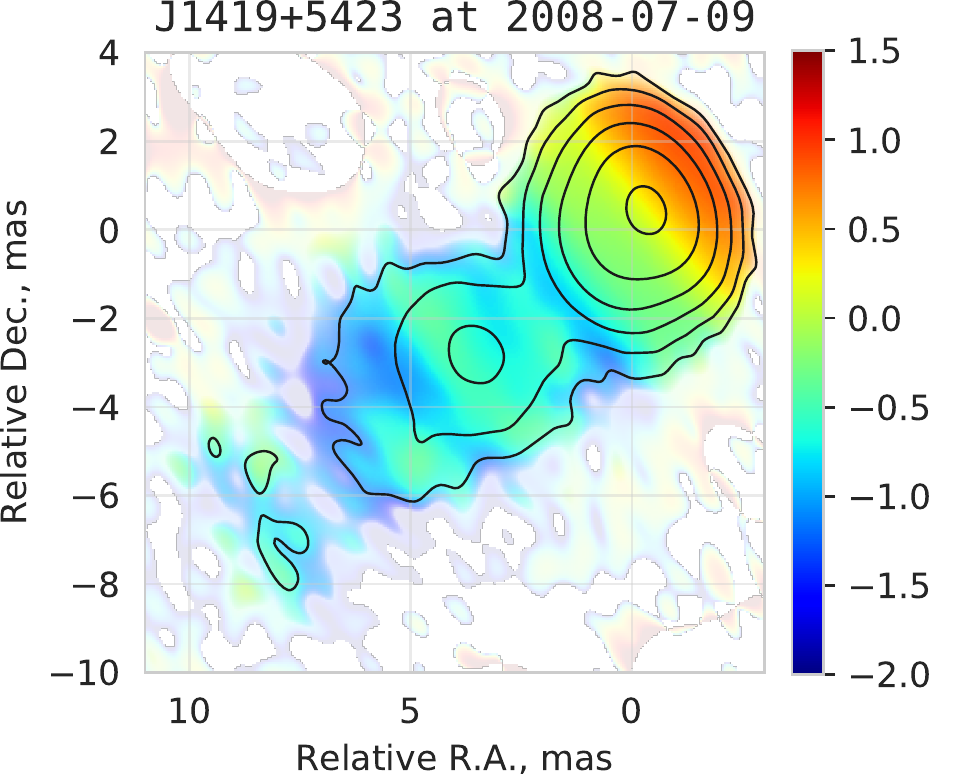}
\caption{One spectral index image per source corresponding to observations at given epochs. The 2-8~GHz spectral index value \corr{$\alpha$ defined as $S\propto\nu^\alpha$} is shown in colour, and 8~GHz intensity  contours are overlaid.}
\label{f:spectral_index_sample}
\end{figure*}

\begin{figure*}
\centering
\includegraphics[width=0.24\linewidth]{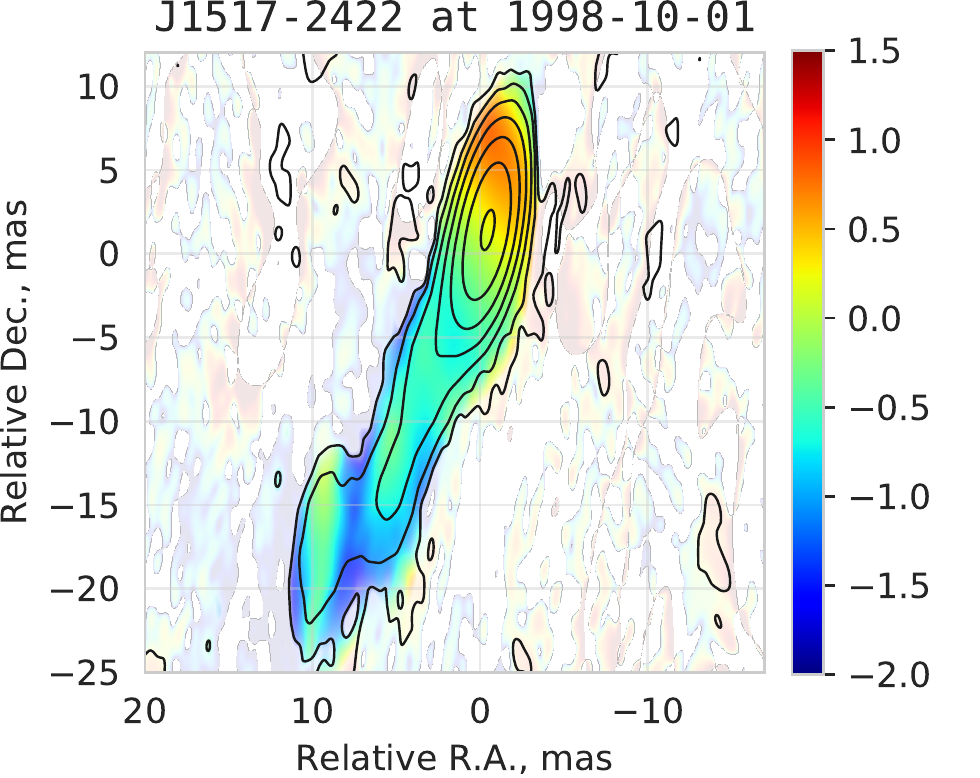}
\includegraphics[width=0.24\linewidth]{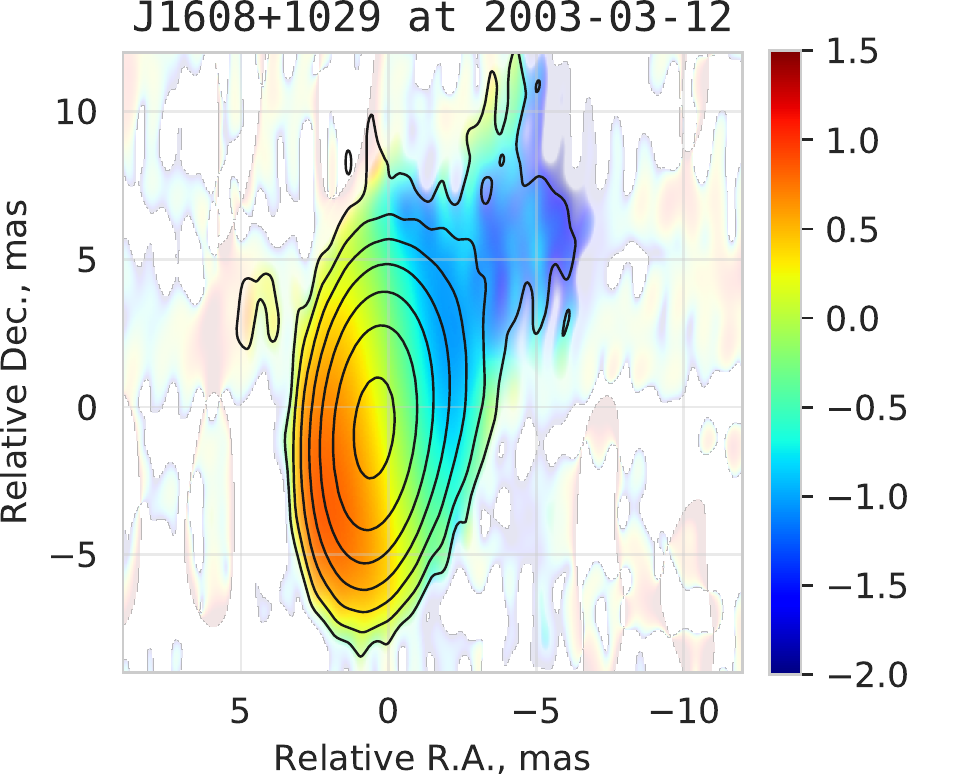}
\includegraphics[width=0.24\linewidth]{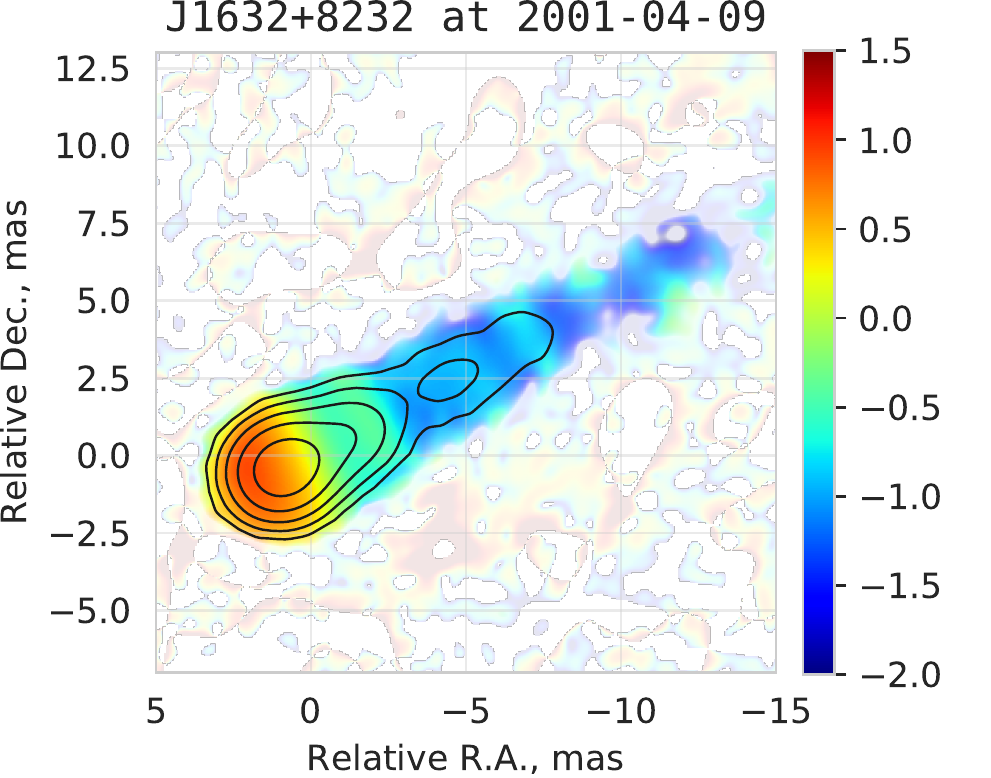}
\includegraphics[width=0.24\linewidth]{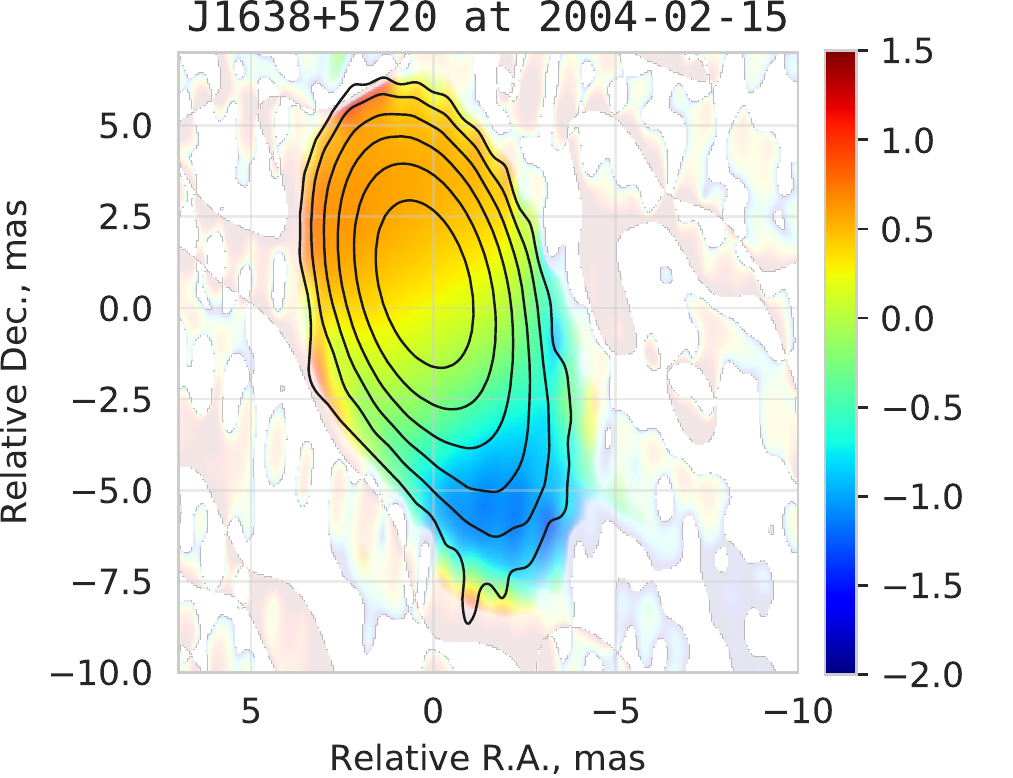}
\includegraphics[width=0.24\linewidth]{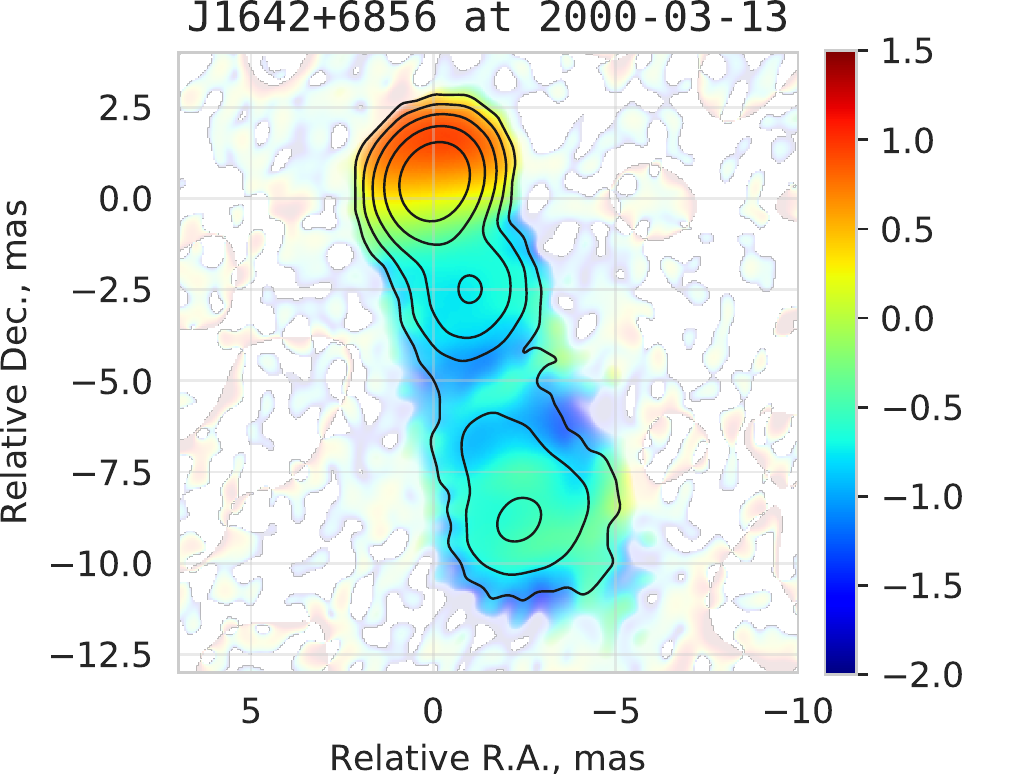}
\includegraphics[width=0.24\linewidth]{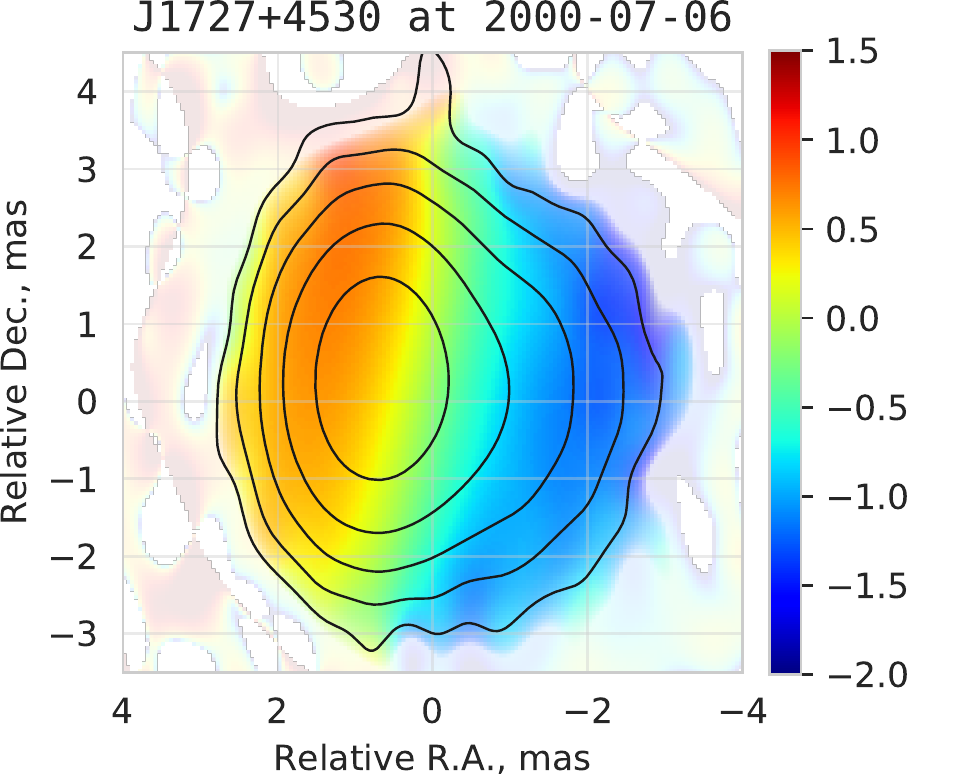}
\includegraphics[width=0.24\linewidth]{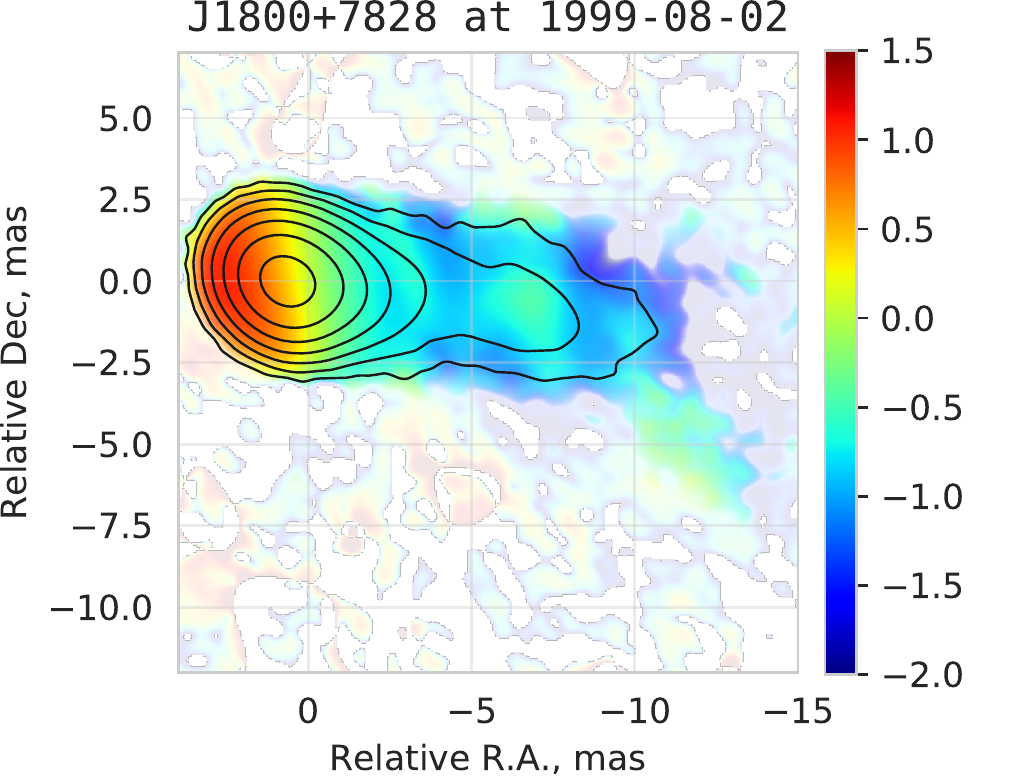}
\includegraphics[width=0.24\linewidth]{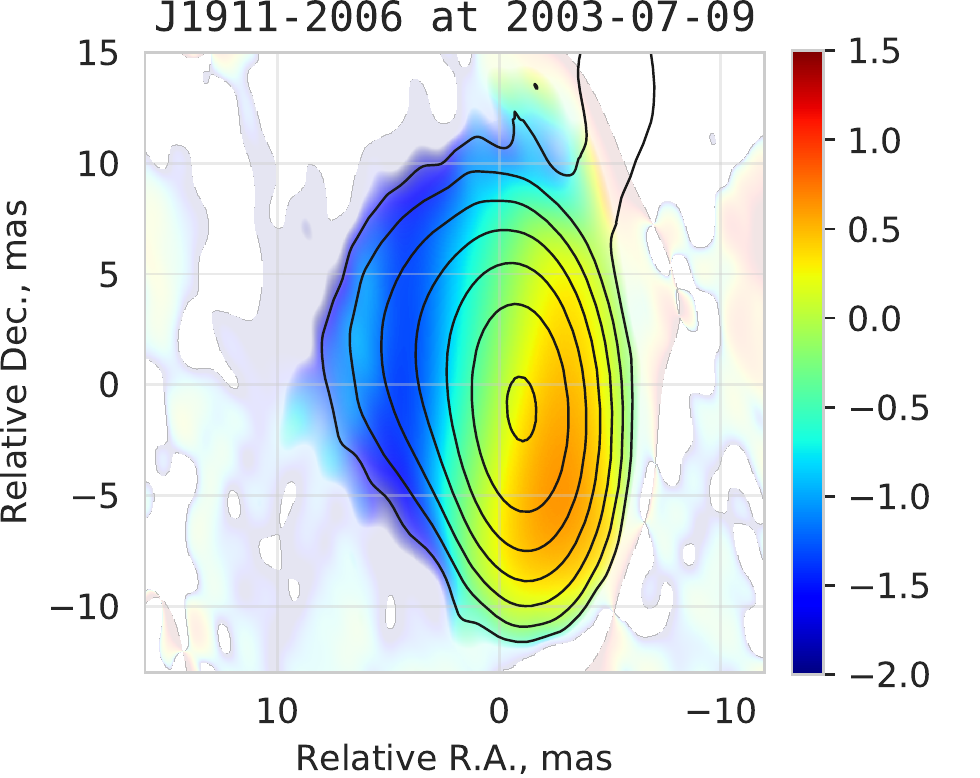}
\includegraphics[width=0.24\linewidth]{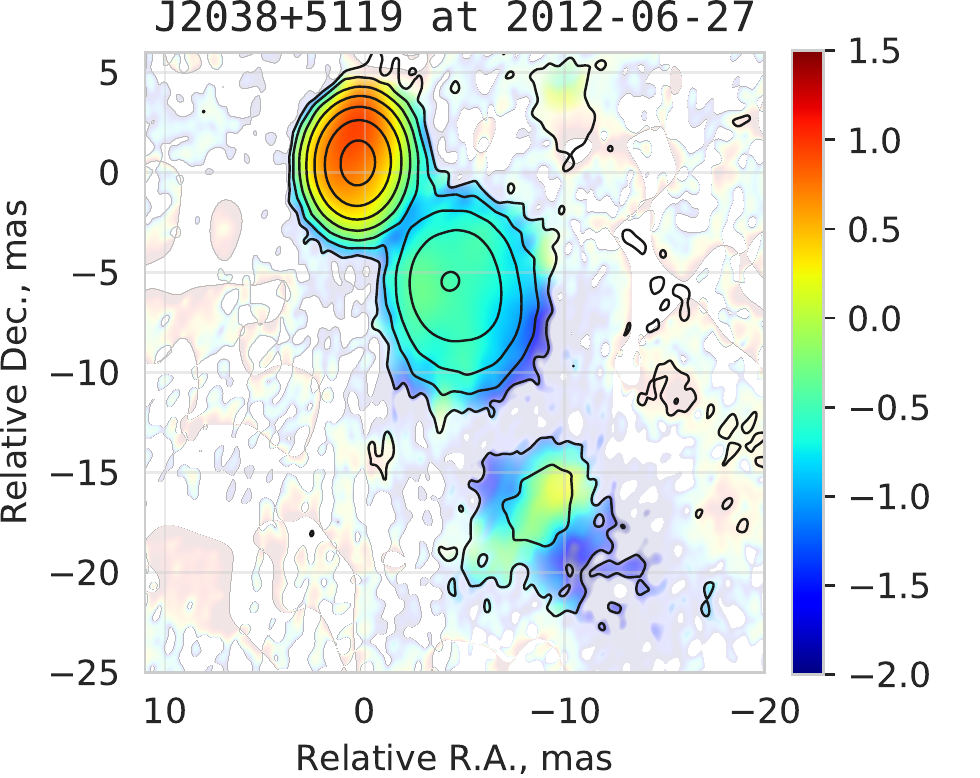}
\includegraphics[width=0.24\linewidth]{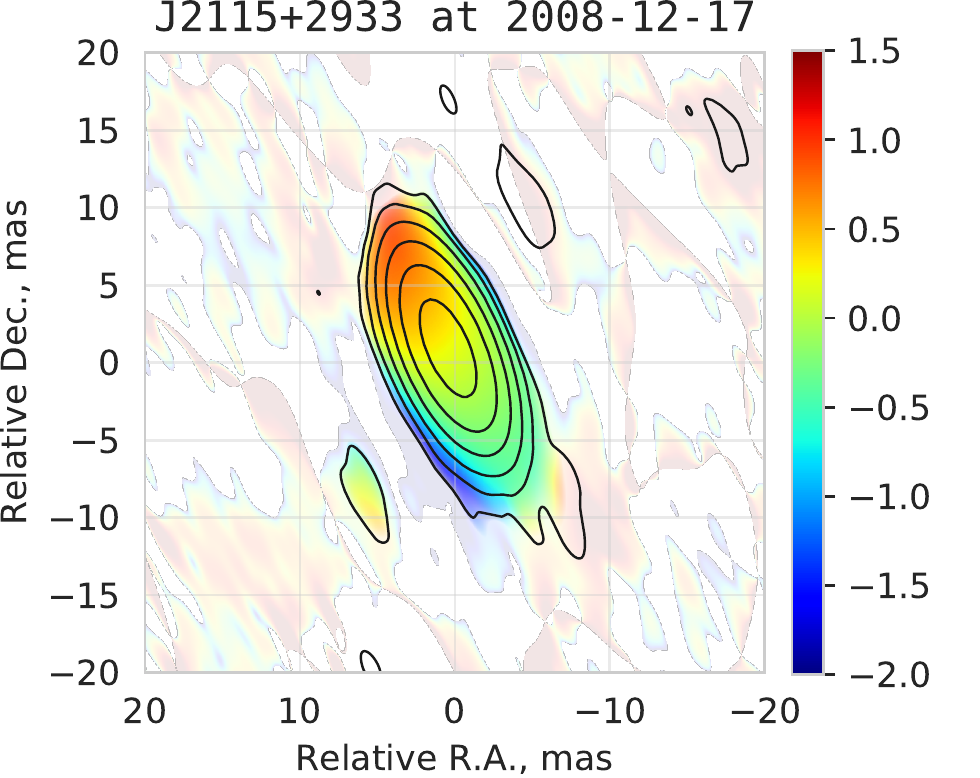}
\includegraphics[width=0.24\linewidth]{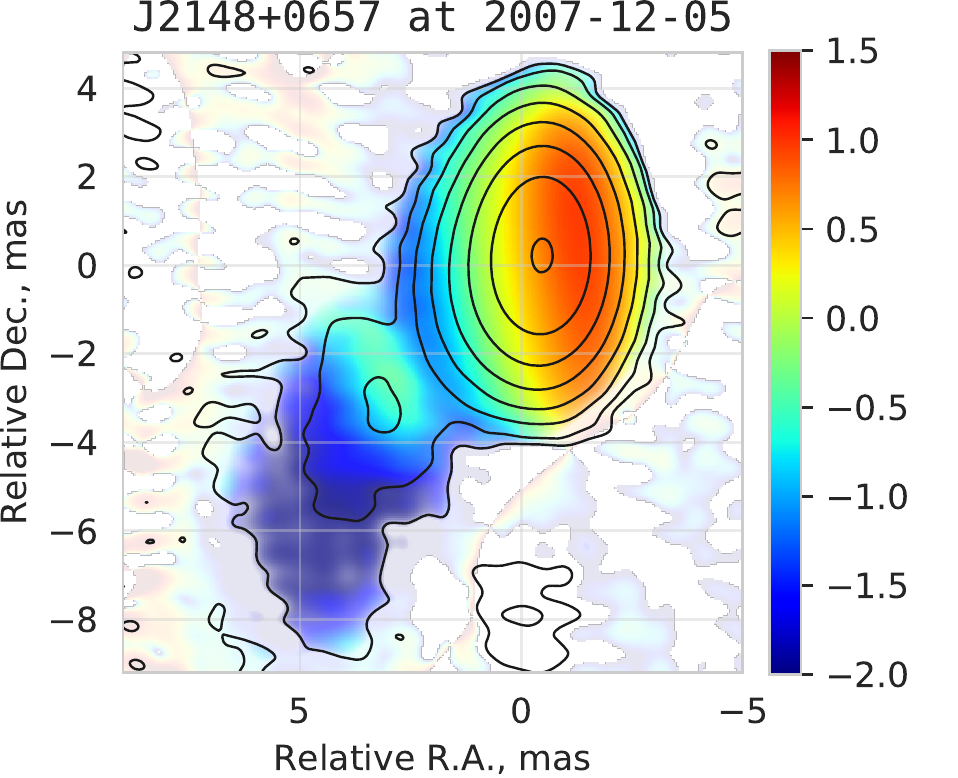}
\includegraphics[width=0.24\linewidth]{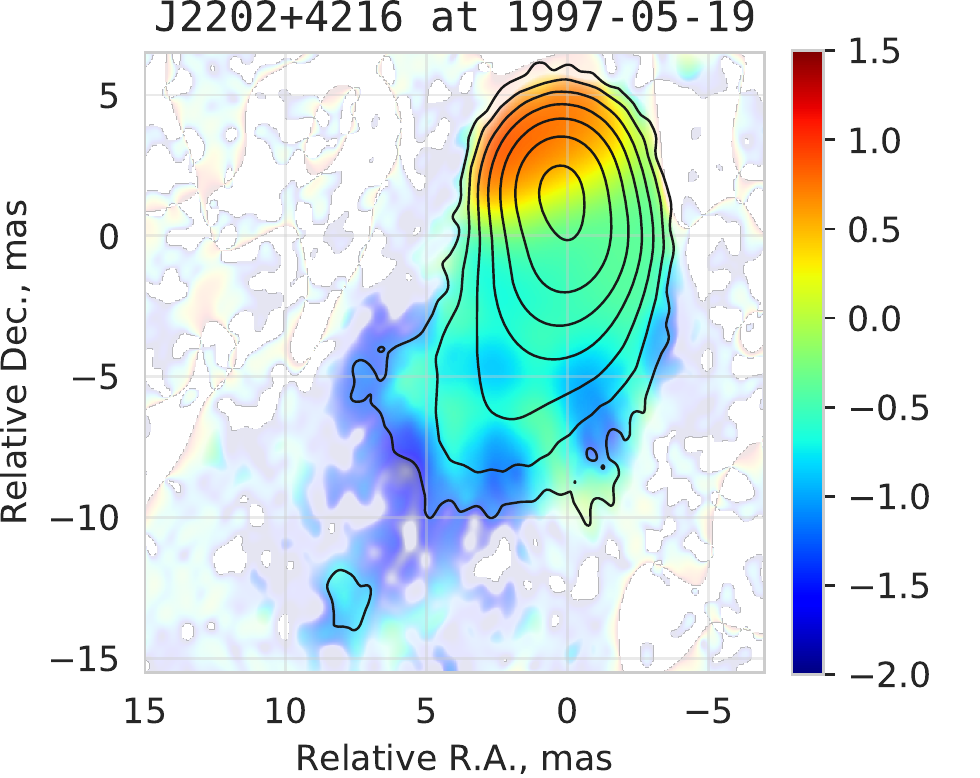}
\includegraphics[width=0.24\linewidth]{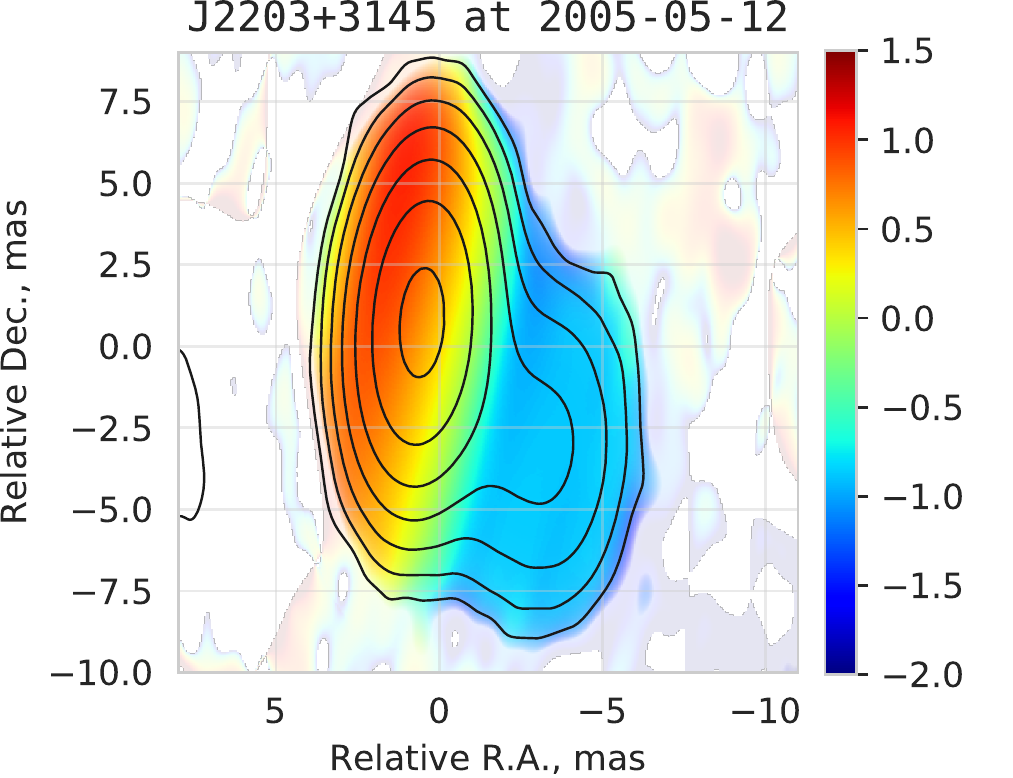}
\includegraphics[width=0.24\linewidth]{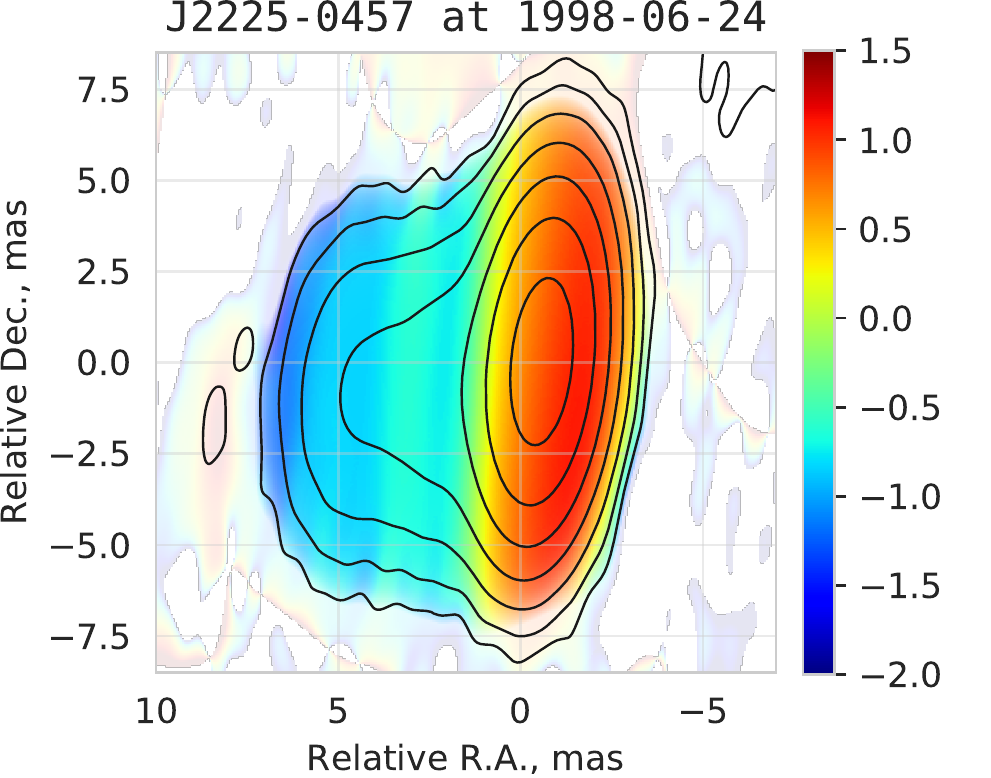}
\includegraphics[width=0.24\linewidth]{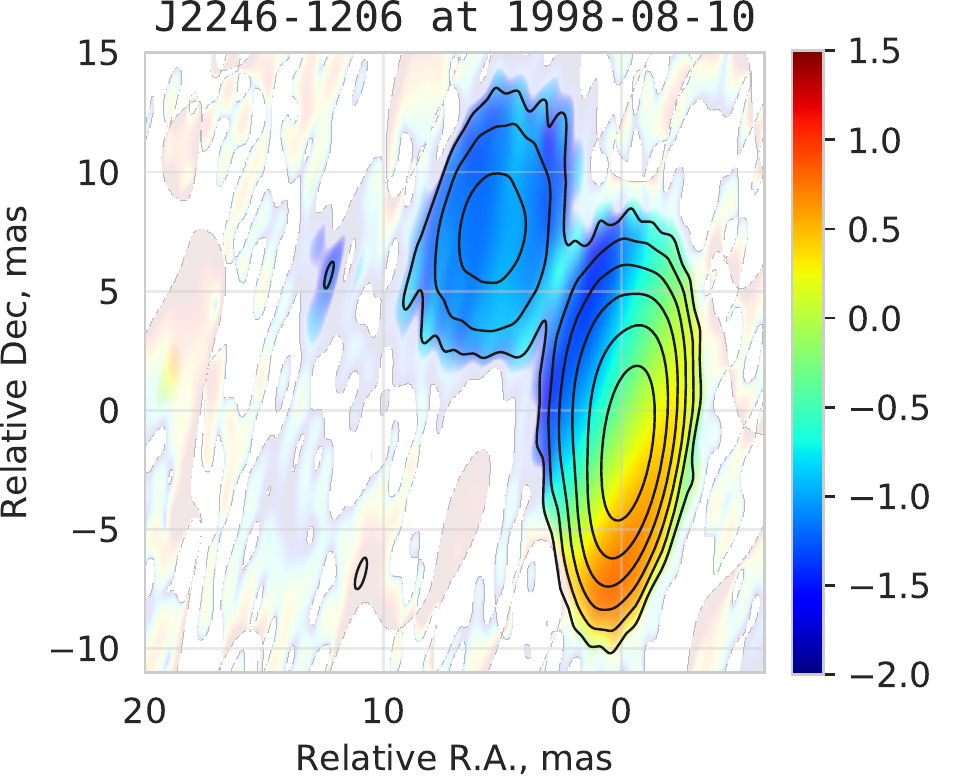}
\includegraphics[width=0.24\linewidth]{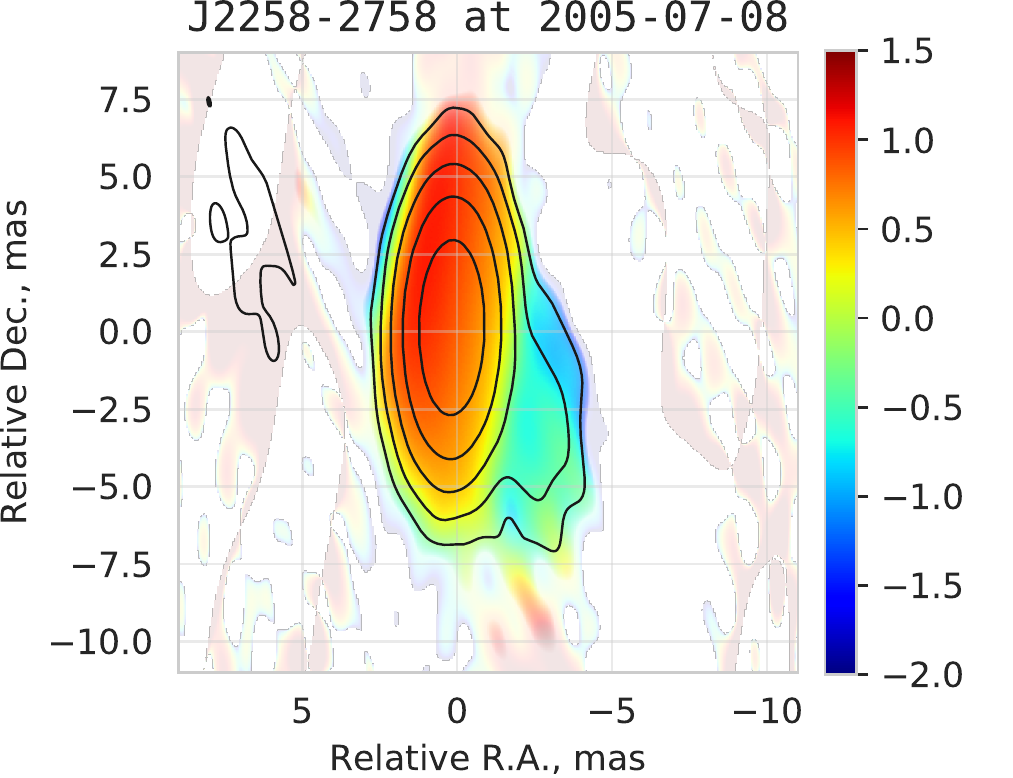}
\contcaption{}
\end{figure*}


\bsp    
\label{lastpage}
\end{document}